\PassOptionsToPackage{colorlinks=true,citecolor=blue,linkcolor=blue,urlcolor=scipostdeepblue}{hyperref}

\documentclass[preprint]{SciPost}

\definecolor{lightblue}{rgb}{0.68, 0.85, 0.9}
\definecolor{airforceblue}{rgb}{0.36, 0.54, 0.66}

\usepackage{hyperref}

\usepackage[backend=biber,url=false,eprint=true,doi=true,sorting=none,backref=true]{biblatex}
\usepackage[draft]{todonotes}

\usepackage{comment}

\usepackage{amsmath,amssymb,amsfonts,amsthm}

\usepackage{bbm}

\usepackage{enumerate}
\usepackage{enumitem}
\usepackage{latexsym}

\usepackage{longtable,booktabs}

\usepackage{array}

\usepackage{caption}
\usepackage{accents}

\usepackage{subcaption}

\usepackage{graphicx}

\newcommand{\bite}{\begin{itemize}}
\newcommand{\eat}{\end{itemize}}
\newcommand{\beq}{\begin{equation}}
\newcommand{\eeq}{\end{equation}}

\newcommand{\beqa}{\begin{align}}
\newcommand{\eeqa}{\end{align}}
\newcommand{\barr}{\begin{array}}
\newcommand{\earr}{\end{array}}

\newcommand{\ie}{\textit{i.e.}~}

\newcommand{\wrt}{\textit{w.r.t.}~}

\newcommand{\mb}[1]{\mathbf{#1}}
\newcommand{\mc}[1]{\mathcal{#1}}
\newcommand{\mbb}[1]{\mathbb{#1}}
\newcommand{\mf}[1]{\mathfrak{#1}}

\newcommand{\unit}[1]{\mathbf{\hat{#1}}}

\newcommand{\id}{\mathbbm{1}}

\newcommand{\vect}[1]{\boldsymbol{#1}}
\newcommand{\expect}[1]{\langle #1\rangle}
\newcommand{\innerp}[2]{\langle #1 \vert #2 \rangle}
\newcommand{\expectop}[3]{\langle #1 \vert #2 \vert #3 \rangle}
\newcommand{\bra}[1]{\langle #1 \vert}
\newcommand{\ket}[1]{\vert #1 \rangle}
\newcommand{\supersc}[1]{$^{\textrm{#1}}$}

\newcommand{\onehalf}{\frac{1}{2}}

\newcommand{\Tr}{\mathrm{Tr}}

\newcommand{\Cyl}{\mathrm{Cyl}}

\newtheorem{thm}{Theorem}[section] 

\theoremstyle{plain} 
\newcommand{\thistheoremname}{}
\newtheorem{genericthm}[thm]{\thistheoremname}

\newtheorem*{genericthm*}{\thistheoremname}
\newenvironment{namedthm*}[1]
{\renewcommand{\thistheoremname}{#1}%
	\begin{genericthm*}}
	{\end{genericthm*}}

\newcommand*{\addheight}[2][.5ex]{
	\raisebox{0pt}[\dimexpr\height+(#1)\relax]{#2}
}

\usepackage{parskip}
\graphicspath{{figures/}}

\begin{document}


\begin{center}{\Large \textbf{
Gauging Time Reversal Symmetry in Quantum Gravity: Arrow of Time from a Confinement--Deconfinement Transition
}}\end{center}

\begin{center}
Deepak Vaid\textsuperscript{*}
\end{center}

\begin{center}
Independent Researcher
\\
* dvaid79@gmail.com
\end{center}

\begin{center}
\today
\end{center}


\section*{Abstract}
{\bf
The question of the origin of time's arrow is a major outstanding problem in physics. Here we present a mechanism for the emergence of a cosmological arrow of time from a confinement--deconfinement transition in a $ Z_2 $ lattice gauge theory living on the spin-network states of Loop Quantum Gravity. Following Chen and Vishwanath \cite{Chen2015Gauging}, who showed that time-reversal symmetry can be gauged on tensor network states, and using the spin-network/tensor-network correspondence \cite{Qi2013Exact,Han2016Loop}, we introduce a $ Z_2 $ gauge field on spin networks encoding a local time-reversal symmetry. The effective theory of this gauge field contains a confined phase -- corresponding to a pre-geometric ``quantum gravitational foam'' with no coherent arrow of time -- and a deconfined phase -- corresponding to semiclassical spacetime with a uniform cosmological arrow. The emergence of the arrow of time is identified with the confinement--deconfinement transition, detected by the Wilson loop order parameter. The deconfined phase is further shown to correspond to a symmetry-protected topological (SPT) phase of the CZX type, whose topological order provides additional stability of the coherent time orientation against local perturbations. We conjecture that the topologically protected surface excitations of this SPT phase give rise to fermionic matter degrees of freedom.
}

\vspace{10pt}
\noindent\rule{\textwidth}{1pt}
\tableofcontents\thispagestyle{fancy}
\noindent\rule{\textwidth}{1pt}
\vspace{10pt}

\section{Introduction: Time's Arrow and Quantum Gravity}\label{sec:intro}

The question of ``time'' has perplexed humanity since the dawn of the civilization. Modern physics provides two conflicting roles for time. In quantum mechanics, time is simply an external parameter representing external clocks \wrt which the evolution of a quantum system takes places. In general relativity, on the other hand, time is a dimension on par with the three spatial dimensions and taken together, the four dimensions constitute a dynamical entity known as ``spacetime'' which serves as the arena on which all physics processes take place. However, unified though it may be in a certain geometrical sense with the spatial dimensions, time retains a character distinct from them. The Universe we live in appears to have a distinct \textit{arrow of time}\footnote{see \cite{Zeh1989The-Physical} for a comprehensive review of the arrow of time in classical, quantum and gravitational contexts.}, a sense in which time \emph{flows} from the ``past'' and towards the ``future''. Spatial dimensions do not appear to display such behavior, with all directions being ``equally likely'' - in the sense that the matter distribution in the Universe does not appear to have a preferred spatial orientation on the largest observable scales\footnote{Whether or not the Universe is spatially isotropic on the largest scales is a matter of some debate \cite{Land2005Examination}.}.

The equations of motion of physical systems, in classical and quantum mechanics and in general relativity are manifestly time-symmetric \cite{Ardakani2018Time,Roberts2012Time}. Yet, we know very well that macroscopic systems \emph{appear} to be endowed with a preferred direction for time evolution. In classical physics this is commonly understood as a consequence of the thermodynamic second law for systems with large number of degrees of freedom. According to the second law, \emph{isolated systems} always evolve from a state with low entropy to a state with high entropy. Thus the direction of increasing entropy can be identified with the direction of the flow of time. In quantum mechanics the situation was unclear until recently when advances in the field of many body localization and the eigenstate thermalization hypothesis \cite{Popescu2006Entanglement, Linden2009Quantum,Nandkishore2015Many-Body,Goold2016The-role} have clarified that many body quantum systems coupled to an environment with a sufficiently large Hilbert space will evolve such that their final state is the equilibrium or highest entropy state, regardless of the initial starting state of the combined system. In other words even if the Universe started in a pure state, individual subsystems within the Universe would \emph{almost} always evolve from states of low entropy to states of high entropy. There are also several proposals for how an arrow of time can spontaneously arise even in closed systems \cite{Maccone2009Quantum,Barbour2014The-Solution,Barbour2013A-Gravitational,Barbour2014Identification}\footnote{see, however, \cite{Jennings2010Entanglement,Jennings2009Comment,Maccone2009Quantum} for comments on and alternatives to the proposed resolution of the arrow of time given in \cite{Maccone2009Quantum}}. The above considerations, however, apply only to closed systems. The universe is not, to the best of our knowledge, a closed system.

Our universe appears to have a well-defined arrow of time on the largest cosmological scales. In particular, the initial state of the universe - at the so-called ``Big-Bang'' - appears to have been a low entropy state and over ``time'' the overall entropy of the Universe has increased and more and more complex structures have formed leading eventually to the present day epoch where systems have become complex enough to allow for the emergence of life\footnote{One should, in general, be careful not to confuse the two separate issue of ``complexity'' and ``entropy'', though they clearly are intertwined in many cases. Entropy is commonly associated with ``disorder'' and complex systems are typically thought of as being ``ordered'', low entropy configurations. However, as Jeremy England has shown \cite{England2012Statistical}, in open systems which are exposed to a constant energy flux, the formation of complex self-replicating structures which dissipate energy efficiently is inevitable. From this perspective complexity and entropy are mutually competing aspects of any system far from equilibrium.}. One could argue that from the perspective of Popescu, Short and Winter \cite{Popescu2006Entanglement} that the emergence of an arrow of time is inevitable in any subsystem of the Universe and that any such system coupled to a sufficiently large environment will always evolve towards a thermal state even if the combined system - \ie the Universe - starts in a pure state. The question is, to what extent does this apply to the Universe as a whole? Our Universe, on the largest observable scales, appears to exhibit a uniform sense of the flow of time. Individual isolated subsystems of the Universe might evolve towards a thermal state, but there is no reason, \emph{a priori}, for why the sense of time-evolution in these different subsystems should be the same. This raises the question why does the observable Universe not consist of regions where time runs in opposite directions? 

Ultimately, any question about the arrow of time must find their resolution in a theory of quantum gravity - the long sought after unification of quantum mechanics and general relativity which would permit us to understand the structure of spacetime at the Planck scale ($ l_p \sim 10^{-34}m $) and in regions of high matter density and strong gravitational fields such as in and around black holes and near the Big Bang. One proposed route to quantum gravity is known as Loop Quantum Gravity (\cite{Vaid2016LQG-for-the-Bewildered}) in which states of quantum geometry are modeled by graphs whose edges are labeled by representations of $ SU(2) $. For obvious reasons, such graphs are also known as ``spin networks''. In the framework of LQG several authors have discussed the role of time-reversal symmetry \cite{Christodoulou2012How-to-detect,Rovelli2012Discrete} and how one could possibly detect an ``anti-spacetime'' - a region of spacetime where the local arrow of time points in the opposite direction to that of the overall global arrow in that spacetime.

Several authors \cite{Qi2013Exact,Han2016Loop,Chirco2018Group,Colafranceschi2021Holographic,Chirco2022Quantum} have shown that there exists a mapping between spin-networks and ``tensor networks'' which are a tool used to study the properties of quantum spacetime in approaches to quantum gravity inspired by the AdS/CFT correspondence. In \cite{Chen2015Gauging} Chen and Vishwanath have shown how time-reversal symmetry can be made from a global to a local symmetry and described its action on tensor networks states.

In this work we build upon these developments to present a resolution of the question of the origin of a cosmological arrow of time by introducing a $ Z_2 $ gauge field representing a local time-reversal symmetry living on spin-networks. We show that the effective theory of this gauge field is a $ Z_2 $ lattice gauge theory which admits a confinement-deconfinement transition. The confined phase corresponds to a pre-geometric state with no coherent arrow of time, while the deconfined phase corresponds to semiclassical spacetime with a uniform cosmological arrow. We further conjecture that the topologically protected surface excitations of the resulting SPT phase give rise to fermionic matter degrees of freedom. This microscopic approach is complemented by prior work \cite{Christodoulou2012How-to-detect,Rovelli2012Discrete} which permits a macroscopic, continuum description of the $ Z_2 $ gauge field as a symmetry which changes the sign of the determinant of the tetrad field.

The original contributions of this work are as follows. (i) We introduce a local $ Z_2 $ gauge field on spin networks as the microscopic representation of time-reversal symmetry in Loop Quantum Gravity, and show that its effective theory is a $ Z_2 $ lattice gauge theory (\autoref{sec:z2-action}). (ii) We identify the cosmological arrow of time with the confinement-deconfinement transition of this gauge theory, with the Wilson loop as the order parameter, fully consistent with Elitzur's theorem (\autoref{sec:z2-action}). (iii) We establish a precise structural correspondence between 4-valent $ j=1/2 $ spin-network intertwiners and the CZX model, showing that the deconfined phase realizes a symmetry-protected topological (SPT) order of the CZX type (\autoref{sec:spt-lqg}). (iv) We establish a structural correspondence between the LQG intertwiner qubit and the CZX code subspace as $Z_2$-invariant effective qubits, with the $ \hat{Q} $-sign flip on the LQG side identified with the on-site $ U_{CZX} $ symmetry on the CZX side (\autoref{sec:spt-lqg}). (v) We conjecture that the topologically protected surface excitations of this SPT phase give rise to fermionic matter degrees of freedom, supported by three independent lines of argument (\autoref{subsec:edge-modes}).

The layout of this paper is as follows. In \autoref{sec:topo-order} we explain the crucial differences between Symmetry Protected Topological (SPT) order, Topological Order (TO) and Symmetry Enriched Topological (SET) order and outline why we expect our classical spacetime to arise as a SPT phase of some underlying fundamental theory. In \autoref{sec:mps-all} we explain the concept of matrix product states and tensor network states and show how \emph{local} time reversal symmetry can be implemented in tensor network states (TNS). In \autoref{sec:spin-networks} the correspondence between spin-network states of LQG and tensor networks is elaborated. In \autoref{sec:volume-z2} we introduce the volume operator of LQG and explain how in the simplest case -- when all spin network edges are labeled by the $ j=1/2 $ representation of $ \mf{su}(2) $ -- the \emph{local} action of time-reversal symmetry causes the volume of a given region of space to change sign. In \autoref{sec:spt-czx} we discuss the CZX model which describes a SPT phase with gapless edge excitations in a many body system. In \autoref{sec:spt-lqg} we establish the mapping between the CZX model and spin-network states, justify the restriction to the $ j = 1/2 $ sector, and identify the CZX SPT phase with the deconfined phase of the $ Z_2 $ gauge theory. In \autoref{sec:z2-action} we formalize the $ Z_2 $ gauge field on the spin network, derive the effective lattice gauge theory action, analyze its phase structure, and identify the emergence of the arrow of time with the confinement-deconfinement transition. Finally we end with a discussion of our results, address possible criticisms and suggest problems for future research in \autoref{sec:discussion}. For those not from the quantum gravity community necessary background on LQG and spin-networks states is provided in \autoref{sec:adm-tetrad} and \autoref{sec:quantum-geometry}. Time reversal in the context of the connection formulation in classical GR is discussed in \autoref{sec:gr-time-reversal}. For completeness, in \autoref{sec:time-reversal} we give a brief outline of how time-reversal symmetry is implemented in quantum mechanical systems. 

\section{Topological Order and Spacetime Geometry}\label{sec:topo-order}

There exist many materials in nature which display transitions between phases which cannot be identified by measuring the change in the expectation value of some \emph{local} order parameter field. These systems fall outside the Landau-Ginzburg paradigm for classifying the phase space of many body systems. The order (or disorder) in these systems is characterized by the expectation values of \emph{non-local} operators such as the Wilson loop operator:
\begin{equation}\label{eqn:wilson-loop-expectation}
	\expect{\mc{O}_\gamma} = \expectop{\Psi}{\exp\left(-i\oint_{\gamma} dx^\mu \mb{A}_\mu\right)}{\Psi}
\end{equation}
where $ \gamma $ is a closed path. Such phases are characterized by expectation values of topological operators. By now, a bewildering variety of topological phases have been discovered in condensed matter systems and sometimes it gets difficult to keep track of the terminology associated with the different phases. Senthil \cite{Senthil2015Symmetry-Protected} provides an accessible description of some of the common types of topological phases as shown below. This listing is not exhaustive but is indicative of the diversity of different types of topological orders. A systematic classification beyond group cohomology, using cobordism theory, is developed in \cite{Kapustin2014Symmetry,Freed2021Reflection}.

\begin{figure}[htbp]\label{fig:topo-phases}
	\begin{tabular}{|>{\raggedright\arraybackslash}p{4cm}|p{6cm} |p{4cm} |}
		\hline
		\centering \textbf{Species} & \centering \textbf{Description} & \textbf{Examples} \\ \hline
		SPTs - Symmetry Protected Topological states & Short-range entangled states, which can be deformed to a trivial product state using local unitary transformations, at the cost of breaking a protected symmetry. In other words the non-trivial topological order of the state is \emph{protected} by the existence of an unbroken symmetry. & Topological Insulators, AKLT State, Haldane phase \\ \hline
		TOs - Topological Orders & Long-range entangled states which cannot be transformed into a trivial product state by local unitary transformations even by breaking all global symmetries & FQHE, Spin Liquids     \\ \hline
		SETs - Symmetry Enriched Topological States  & These are TOs, which in addition to long-range entanglement, also have some global symmetry.                                                                            & FQHE, Spin Liquids     \\ \hline
	\end{tabular}%
	\caption{Topological orders: terminology, characteristics and examples}
\end{figure}

Significant conceptual and technical breakthroughs \cite{Van-Raamsdonk2010Building,Swingle2012Constructing} in the last decade lead us to believe that the geometry of spacetime arises from the entanglement between the various subsystems of some underlying pre-geometric strata. As first argued \footnote{In earlier work \cite{Swingle2009Entanglement} Swingle had suggested that the emergence of spacetime could be seen as a consequence of applying the renormalization group to the tensor network representation of the quantum state of a many body system. While this work was deeply significant in and of itself, it was van Raamsdonk who first explained in a clear manner how the precise connection between geometry and entanglement should arise.} in a simple and elegant fashion by Van Raamsdonk in 2010 \cite{Van-Raamsdonk2010Building}, the amount of entanglement between two different regions of spacetime determines the geometric connectivity of the points within those regions with each other. This argument builds upon the earlier groundbreaking work of Ryu and Takayanagi \cite{Ryu2006Holographic,Ryu2006Aspects} who showed that the entropy of entanglement between two different regions of spacetime - in the restricted setting of AdS spacetime - is proportional to the area of the \emph{minimal} surface which separates the two regions:
\begin{equation}\label{eqn:rt-formula}
	S_{EE} \propto A_{min}.
\end{equation}
In the context of loop quantum gravity, recent work \cite{Colafranceschi2022Holographic,Livine2018Intertwiner,Baytas2018Gluing} has shown that similar holographic scaling relations arise from the intertwiner entanglement in spin networks, providing a robust microscopic basis for \eqref{eqn:rt-formula}.
Now, imagine disconnecting the two bulk regions from each other, by ``pinching'' off the boundary surface joining the two regions. As the area of the surface shrinks to zero, points in one region become increasingly disconnected from points in the other, until when the area reaches zero, the two regions become completely geometrically separate. At this stage, as per the RT formula \eqref{eqn:rt-formula} the entanglement entropy between the two regions also goes to zero. This simple argument - anchored in the well understood physics of the AdS/CFT correspondence - is sufficient to suggest that an apparently smooth, extended macroscopic geometry arises when smaller regions of quantum spacetime are ``stitched'' together by entanglement.

There is a flip side to this argument. If the metric distance between two neighboring regions of space is inversely proportional to the entanglement entropy between the two regions, then as the entanglement entropy between any two regions increases the metric distance between the two should decrease. Now, the spacetime we observe around us exhibits the property of \emph{locality}. The metric geometry changes smoothly as we increase the scales and distances between different systems. For instance, we ``know'' that Saturn is closer to Earth than is Alpha Centauri. This means that the geometric degrees of freedom around our planet have a greater amount of entanglement with the degrees of freedom around Saturn than they do with the degrees of freedom around Alpha Centauri. If this were not the case, then it would not be possible a to have a \emph{sense} of the separation between the Earth and Alpha Centauri.

The same structure is observed on the largest scales in the universe. This suggests that the underlying many-body system from which our macroscopic geometry emerges exhibits short range entanglement (SRE) rather than long range entanglement (LRE). In a state with LRE the macroscopic geometry would be ill-defined as points arbitrarily far apart from each other with respect to the microscopic geometry could be arbitrarily close to each other with respect to the emergent macroscopic geometry. This leads us to the following ansatz:

\noindent\begin{minipage}{\linewidth}
	\centering
	\begin{quote}
		\emph{Emergent macroscopic geometry exhibits (approximate) smoothness and locality when the underlying quantum many body system is in a SRE phase.}
	\end{quote}
\end{minipage}

However, only SRE is not sufficient to ensure the emergence of a connected, bulk geometry. This is because states with SRE can, in general, be deformed to trivial product states by local unitary operations. In order to have a state with \emph{non-trivial} topological order which cannot be deformed into trivial product states by local unitaries the existence of an unbroken symmetry is required. As long as the symmetry is preserved the SRE state will have non-trivial topological order and cannot be deformed into a trivial state. Such a state is therefore said to possess \emph{symmetry protected topological} (SPT) order. In light of this fact the statement of the previous ansatz is modified as follows:

\noindent\begin{minipage}{\linewidth}
	\centering
	\begin{quote}
		\emph{Emergent macroscopic geometry exhibits (approximate) smoothness and locality when the underlying quantum many body system is in a SPT phase.}
	\end{quote}
\end{minipage}

\section{Matrix Product States and All That}\label{sec:mps-all}

\emph{Readers familiar with matrix product states and tensor network states may skip directly to \autoref{subsec:mps-local-gauge}, where the local gauge action relevant for the remainder of the paper is introduced.}

Consider a many body system consisting of $ N $ quantum degrees of freedom. The most general many body state of such as system can be written as:
\begin{equation}\label{eqn:many-body-state}
	\ket{\psi} = \sum_{i_1, i_2, \ldots, i_N} C_{i_1, i_2, \ldots, i_N} \ket{i_1, i_2, \ldots, i_N}
\end{equation}
where $ \ket{i_k} $ is the state of the $ k^\text{th} $ degree of freedom. For instance, in a spin-chain, the single body Hilbert space would be $ \mc{H}^2 = \{ \ket{\uparrow}, \ket{\downarrow} \} $ and the sum in the above equation would be over all the possible spin configurations of the sites in the chain. It is easy to see that the total Hilbert space of the spin-chain is:
\begin{equation}\label{eqn:n-body-hilbert-space}
	\mc{H}^{2^N} = \underset{i = 1 \ldots N}{\bigotimes} \mc{H}^2_i
\end{equation}
and the dimensionality of this Hilbert space is $ 2^N $. As $ N $ increases, the amount of information required to specify a given many-body state increases exponentially. Moreover in the representation \eqref{eqn:many-body-state}, there is no way to distinguish short-range correlations from long-range ones. The co-efficients $  C_{i_1, i_2, \ldots, i_N} $ are defined over the entire Hilbert space and cannot be decomposed into local contributions in any simple manner.

Most importantly for our purposes, it is not clear what the local action of the time-reversal symmetry would be on a state of the form \eqref{eqn:many-body-state}. One can define the global action:
\begin{equation}\label{eqn:trs-many-body-state}
	\mc{T} \ket{\psi} = \sum_{i_1, i_2, \ldots, i_N} C^*_{i_1, i_2, \ldots, i_N} U_1 \otimes \ldots \otimes U_N \ket{i_1, i_2, \ldots, i_N}
\end{equation}
Here the action of $ \mc{T} $ on a state is expressed as a product of a unitary transformation $ U $ followed by complex conjugation $ K $: $ \mc{T} = KU $. Each factor of $ U_i $ acts on the $ i^\text{th} $ site, while complex conjugation acts on the coefficients $ C^*_{i_1, i_2, \ldots, i_N} $. It is precisely this complex conjugation which makes it difficult to understand how to apply time reversal \emph{locally}, \ie to a subset of the system $ \{i_m, \ldots, i_{m+n}\} $, rather than to the whole sysem. What is needed is a way to express \eqref{eqn:trs-many-body-state} in a way that allows us to act with the complex conjugation \emph{locally}, and only on select parts of the system, rather than on the global state. It turns out it is possible to express \eqref{eqn:trs-many-body-state} in such a way using the technique of matrix product states (MPS) or their generalization to dimensions greater than one, tensor network states (TNS).

\subsection{Matrix Product States}\label{subsec:mps}

A typical MPS is written as \cite[Sec 3]{Bridgeman2016Hand-waving}:
\begin{equation}\label{eqn:mps-ansatz}
	\ket{\psi} = \sum_{i_1, i_2, \ldots, i_N} \Tr\left[ A^{(1)}_{i_1} A^{(2)}_{i_2} \ldots A^{(N)}_{i_N} \right] \ket{i_1, i_2, \ldots, i_N}
\end{equation}
where $ A^{(k)}_{i_k;\alpha\beta} $ is a set of matrices associated with the $ k^\text{th} $ lattice site; with $ i_k \in \{1,2,\ldots,D\} $, where $ D $ is the dimension of the single site Hilbert space and $ \alpha, \beta \in \{ 1, 2, \ldots, d\} $, where $ d $ is known as the ``bond dimension''. As an example, for a chain of spin $ 1/2 $ particles, each site can be one of two states $ \{\ket{\uparrow}, \ket{\downarrow} \} $ and thus with each site we associate two matrices: $ A^{(k)}_{\uparrow;\alpha\beta}  $ and $ A^{(k)}_{\downarrow; \alpha\beta} $. From now on for ease of notation, instead of using the up/down notation $ \{\ket{\uparrow}, \ket{\downarrow} \} $ we will use $ \{ \ket{1}, \ket{0} \} $ to represent the two states of a spin $ 1/2 $ system. Typically, we will consider systems invariant under translations, which allows us to use a single set of matrices $ A_{i_k;\alpha\beta} $ for each site. We will do so in the following examples.

Working with the spin half chain, consider, for instance the matrices:
\begin{equation}\label{eqn:mps-state-1}
	A_0 = (1); \quad A_1 = (0)
\end{equation}
Inserting these into the MPS expression \eqref{eqn:mps-ansatz} - and keeping in mind that the subscripts $ \{0,1\} $ on the matrices correspond to the allowed states $ \{\ket{1}, \ket{0} \} $ at each site, rather than to the sites themselves - we can see that whenever the state at a given site is $ \ket{1} $, the coefficient of that term will vanish. Thus we are left with the result that \eqref{eqn:mps-state-1} represents the simple state: $ \ket{00\ldots0} $. A less trivial example is that of the GHZ (Greenberger-Horne-Zeilinger) \cite{Greenberger1990Bells} state:
\begin{equation}\label{eqn:ghz-state}
	\ket{\psi_{GHZ}} = \ket{00\ldots0} + \ket{11\ldots1}
\end{equation}
for which the associated MPS matrices are:
\begin{equation}\label{eqn:ghz-matrices}
	A_0 = \begin{pmatrix} 1 & 0 \\ 0 & 0 \end{pmatrix}; \quad A_1 = \begin{pmatrix} 0 & 0 \\ 0 & 1 \end{pmatrix}
\end{equation}
One can explicitly evaluate the expression \eqref{eqn:mps-ansatz} with the choice of above matrices for, say, a two-site system to find that it does indeed correspond to \eqref{eqn:ghz-state}. It is important at this point to mention that it has been proven in \cite{Vidal2004Efficient} that \emph{any} many-body state can be written in the form \eqref{eqn:mps-ansatz} as long as the bond dimension $ d $ is large enough (though for generic states $d$ grows exponentially with system size; MPS are efficient only for states satisfying an area law). Therefore the MPS ansatz can be used to study local application of time-reversal, among other, symmetries to an arbitrary many-body system.

\begin{figure}[tbph]
	\centering
	\includegraphics[width=0.2\linewidth]{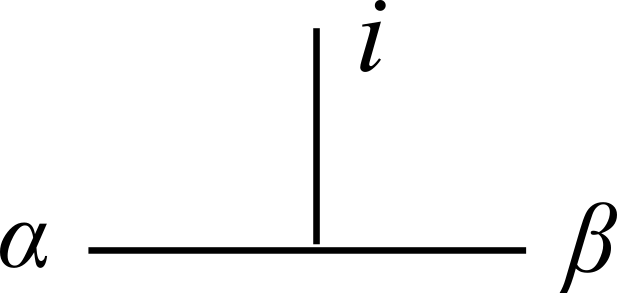}
	\caption{Single vertex of a matrix product state. Internal indices are denoted by $ \alpha, \beta $. Physical indices by $ i,j,\ldots $}
	\label{fig:mps-vertex}
\end{figure}

There is an illuminating diagrammatic notation \cite{Bridgeman2016Hand-waving} for MPS. The key component of this notation is the vertex diagram shown in \autoref{fig:mps-vertex}. The node labelled $ i $ represents the physical degree of freedom corresponding to the state $ \ket{i} $ in the single site Hilbert space. The site index itself is not shown.

\begin{figure}
\centering
		\begin{minipage}[t]{0.45\textwidth}
		\centering
		\includegraphics[width=0.6\linewidth]{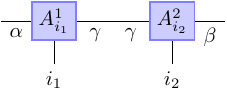}
		\caption{A connection between two adjacent vertices corresponds to multiplying the associated matrices.}
		\label{fig:mps-vertex-product}
	\end{minipage}
	\hspace*{5mm}
	\begin{minipage}[t]{0.45\textwidth}
		\centering
		\includegraphics[width=0.6\linewidth]{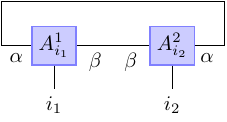}
		\caption{To take the trace we complete the loop by connecting the first and last vertices}
		\label{fig:mps-vertex-product-trace}
	\end{minipage}
\end{figure}

\autoref{fig:mps-vertex-product} shows how two connect two such vertices. In terms of matrices, such a connection corresponds to multiplying two site matrices: $ A^1_{i_1;\alpha\gamma} A^2_{i_2;\gamma\beta} $ (where the sum over the repeated index $ \gamma $ is implied). In order to obtain the numerical value of the wavefunction coefficient in \eqref{eqn:mps-ansatz}, the trace has to be taken after connecting all the vertices together. This operation is illustrated in \autoref{fig:mps-vertex-product-trace}, by the line connecting $ A^1_{i_1} $ and $ A^2_{i_2} $ from the outside.

\begin{figure}[h]
	\centering
	\includegraphics[width=0.5\linewidth]{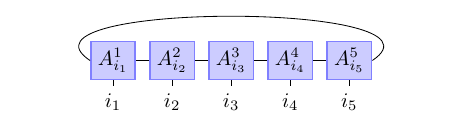}
	\caption{Diagrammatic representation of a matrix product state (MPS). The matrix $ A_{i} $ lives at the $ i^\text{th} $ lattice site. Indices $ i_1, \ldots, i_5 $ represent the \emph{physical} degrees of freedom at each site. The trace operation is represented by the line connecting by the first and last sites (figure courtesy of \cite{Bridgeman2016Hand-waving})}
	\label{fig:mps-state}
\end{figure}

An illustration of use of this notation to represent a MPS is shown in \autoref{fig:mps-state} for a system with five spins.

\subsection{Tensor Network States}\label{subsec:tns}

\begin{figure}[tbph]
	\centering
	\includegraphics[width=0.3\linewidth]{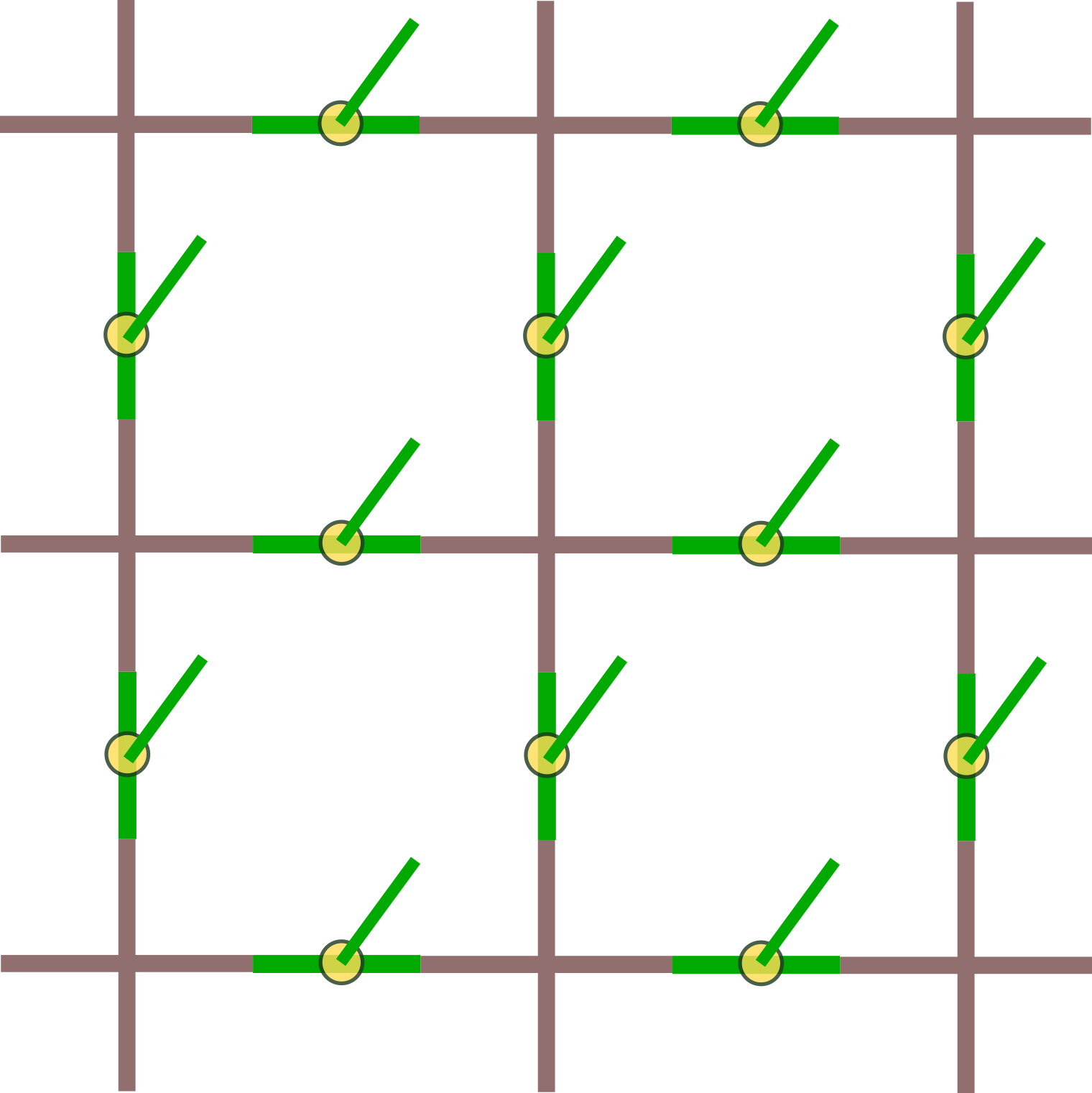}
	\caption{Tensor network state (TNS) representation of the toric code wavefunction. Each edge (green) has a physical index $ i $ and two internal indices $ \alpha, \beta $. The corresponding tensor is $ g^e_{i_e;\alpha\beta} $ where $ e $ labels the edge and $ i_e $ labels the state of the edge. The vertices have only internal indices and the corresponding tensor is written as $ A^v_{\alpha\beta\gamma\delta} $ where $ v $ labels the given vertex.}
	\label{fig:toric-code}
\end{figure}

One can easily extend this notation to arbitrary lattices in arbitrary number of dimensions. When dealing with graphs with degree greater than two, the term ``tensor network states'' or ``tensor product states'' is used instead of ``matrix product states'', because each matrix now has more than two internal indices. An example is shown in \autoref{fig:toric-code} which shows the diagrammatic representation for the state for a system defined on a two-dimensional square lattice, where the physical degrees of freedom live on the edges. This state is defined by two sets of tensors: a four index object $ A_{\alpha\beta\gamma\eta} $ for the vertices and a two index object $ g^e_{i_e;\alpha\beta} $ for the edges. Since the physical degrees of freedom only live on the edges, only the edge tensor has the corresponding index $ i_e $, where $ e $ represents the given edge and $ i_e \in \{1,\ldots,D\} $, with $ D $, as before, being the dimension of the Hilbert space of the physical degree of freedom.

In general given an arbitrary graph $ \Gamma $, the full wavefunction for such a graph is given by:
\begin{equation}\label{eqn:tensor-net-state}
	\ket{\Psi} = \sum_{\{i_1,\ldots,i_{n_v}\}, \{ j_1,\ldots,j_{n_e} \} } \Tr\left[ \prod_{v \in \Gamma} A^v_{i_v} \prod_{e \in \Gamma} g^e_{j_e} \right] \ket{i_1,\ldots,i_{n_v}; j_1, \ldots, j_{n_e} }
\end{equation}

\begin{table}
	\centering
	\begin{tabular}{p{6cm} c p{6cm}}
			\toprule
			Number of degrees of freedom & $ N $    &                          \\ \hline
			Site Hilbert space          & $ \mc{H} $   &                      \\ \hline
			Number of site degrees of freedom & $ d = \dim\mc{H} $	&		  \\ \hline
			Bond dimension				& $ D \in \mbb{N}$		&						  \\ \hline
			Vertex Tensors & $ A^v_{i_v; \alpha_1 \ldots \alpha_{v_k}} $ & $ v: $ vertex \newline $i_v$: vertex state/physical index \newline $\alpha_1 .. \alpha_k \in \{1,\ldots,D\}$: internal (``bond'') indices \\
			\midrule
			Edge Tensors & $ g^e_{i_e; \alpha\beta} $ &  $ e $: edge label \newline $i_e:$ edge state/physical index \newline $\alpha, \beta \in \{1,\ldots,D\}$: bonding indices \\
			\bottomrule
	\end{tabular}
	\caption{Data required to specify the state of a tensor network}
	\label{tbl:tns-data}
\end{table}

Here $ n_v, n_e $ are the total number of vertices and edges in the graph respectively; $ i_k, j_k $ are the physical degrees of freedom for vertices and edges respectively. The products are over all vertices and edges of the graph and the sum is over all possible configurations of the physical degrees of freedom on the graph. We have suppressed the internal indices on the tensors for simplicity. In the next section we will see that \eqref{eqn:tensor-net-state} is the precise form of the expression for the wavefunction of a spin-network state in LQG with the only difference being that the physical (or ``bulk'') degrees of freedom are located at the vertices instead of on the edges as in the case of the toric code.

\subsection{Dangling Indices: Bulk vs Boundary Degrees of Freedom}

We will also need to consider states with \emph{\textbf{dangling indices}}, \ie ``internal'' or bonding degrees of freedom which are not traced over. As we will see in the next section, spin-network states have two different types of such free indices. One kind comes from the intertwiner degrees of freedom which encode the bulk geometry of the spacetime and the second correspond to the boundary states of quantum geometry of a bulk region. Consider the MPS for a two-site system as shown in \autoref{fig:mps-vertex-product}. Such as state, before would be written as:
\begin{equation}\label{eqn:two-index-mps}
\ket{\Psi_{\alpha\beta}}= \sum_{\{i_1, i_2\}} A^1_{i_1;\alpha\gamma} A^2_{i_2;\gamma\beta} \ket{i_1, i_2}
\end{equation}
Now the state is a two-index object. Likewise if the graph has $ n $ dangling edges, then the resulting state will also have $ n $ indices of size equal to the bond dimension. Consider a graph $ \Gamma $, which has a subset of $ n $ vertices which are connected to only a single edge, each. These vertices and their accompanying edges $ \{ v_1,\ldots,v_n; e_1, \ldots, e_n \} $ form a subset $ \partial \Gamma \subset \Gamma $. Let $ \bar{\Gamma} = \Gamma/\partial \Gamma$, be the ``bulk'' of the graph which does not have any dangling edges. Resulting state is $ n $ index object.
\begin{equation}\label{eqn:n-index-mps}
\ket{\Psi_{\alpha_1,\ldots,\alpha_n}} = \ket{\Psi_\text{bulk}}\ket{\alpha_1,\ldots,\alpha_n}
\end{equation}
where:
\begin{equation}\label{eqn:bulk-mps}
\ket{\Psi_\text{bulk}} = \underset{\{i_v,j_e\} \in \bar{\Gamma}}{\sum} \left[ \underset{v \in \bar\Gamma}\prod A^v_{i_v} \underset{e \in \bar\Gamma}\prod g^e_{j_e} \right] \ket{ \{ i_v \}, \{ j_e \} }
\end{equation}


\subsection{Global Gauge Symmetry of MPS}\label{subsec:mps-gauge}

The central thrust of this work is the idea that gauge symmetry breaking in spin-network states can lead to the emergence of an arrow of time. For this to work, we have to understand how gauge symmetry can act on MPS - and thereby on spin-network states. Now any MPS is given by data as shown in \autoref{tbl:tns-data}. Any such state has a \emph{global} gauge symmetry under which all the matrices are transformed by a constant matrix, as in:
\begin{equation}\label{eqn:mps-global-gauge}
	A^{(k)}_{i_k} \rightarrow M A^{(k)}_{i_k} M^{-1}
\end{equation}
where $ M $ is a constant matrix. Then it is easy to see that the resulting state will be the same as before, because all factors of $ M, M^{-1} $ will cancel when taking the trace:
\begin{equation}\label{eqn:mps-ansatz-gauge}
	\ket{\psi} \rightarrow \sum_{i_1, i_2, \ldots, i_N} \Tr\left[ M A^{(1)}_{i_1} M^{-1} M A^{(2)}_{i_2} M^{-1} \ldots M A^{(N)}_{i_N} M^{-1 }\right] \ket{i_1, i_2, \ldots, i_N} = \ket{\psi}
\end{equation}
Diagrammatically the effect of a such a gauge symmetry can be illustrated as follows:


\begin{figure}[h]
	\[
		\vcenter{\hbox{\includegraphics[width=0.40\linewidth]{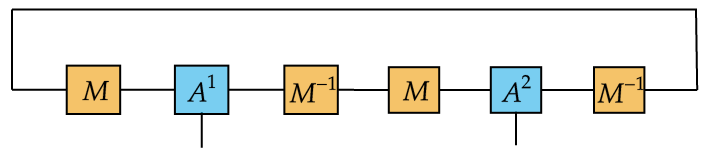}}}
	\Rightarrow
		\vcenter{\hbox{\includegraphics[width=0.3\linewidth]{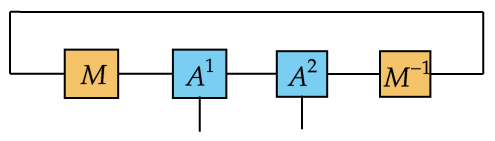}}}
	\Rightarrow
		\vcenter{\hbox{\includegraphics[width=0.2\linewidth]{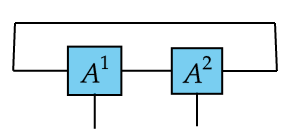}}}
	\]
	\caption{Action of a global gauge transformation on a matrix product state. After tracing out all the degrees of the freedom, the resulting state is unchanged.}
\end{figure}

\subsection{Local Gauge Action}\label{subsec:mps-local-gauge}

For our purposes what is needed is to be able to understand how to convert the global symmetry into a local one. We know how to do this for symmetries under groups such as $ U(1) $ or $ SU(2) $. For a system of spin, for instance, endowing the system with a global $ SU(2) $ symmetry implies that the system should remain unchanged under a simultaneous rotations of \emph{all} the spins under the same $ SU(2) $ rotation. In order to make this into a local symmetry, we simply restrict the rotations to individual spins such that instead of a single element of $ SU(2) $ there now exists a gauge \emph{field} $ g(\vec{x}) $ (where $ \vec{x} $ is the co-ordinate of individual sites either in a discrete or continuous system) which changes from point to point in the system. If we want to know how the state of a given spin changes we apply the value of $ g(\vec{x} $ at the location of \emph{that} spin only.

For time-reversal the situation becomes more subtle. This is because time-reversal is not a unitary operation, but an anti-unitary one, \ie in addition to transformation of the degrees of freedom by some group element we must also take the complex conjugate of the state. For global time-reversal of a many body state of the form \eqref{eqn:many-body-state} this can be accomplished as follows:
\begin{align}\label{eqn:psi-time-reversal}
	\mc{T} \ket{\psi} & = \mc{T} \sum_{i_1, i_2, \ldots, i_N} C_{i_1, i_2, \ldots, i_N} \ket{i_1, i_2, \ldots, i_N} \nonumber \\
	& = \sum_{i_1, i_2, \ldots, i_N} C^*_{i_1, i_2, \ldots, i_N} U_1 \otimes \ldots \otimes U_N  \ket{i_1, i_2, \ldots, i_N}
\end{align}
where the complex conjugation acts on the coefficient $ C_{i_1, i_2, \ldots, i_N} $ of the many-body state. It is clear that this is a global effect. It is not clear how to make complex conjugation into a \emph{local} operation, at least as long as the state is written in the form \eqref{eqn:many-body-state}. This is where the utility of the MPS or TNS representation becomes manifest. The action of (global) time-reversal on the MPS state \eqref{eqn:mps-ansatz} is given by:
\begin{equation}\label{eqn:mps-time-reversal}
	\mc{T} \ket{\psi} =  \sum_{i_1, i_2, \ldots, i_N} \Tr\left[ A^{(1)}_{i_1} A^{(2)}_{i_2} \ldots A^{(N)}_{i_N} \right]^* U_1 \otimes \ldots \otimes U_N  \ket{i_1, i_2, \ldots, i_N}
\end{equation}
From this expression one can see the effect of (global) time-reversal is equivalent to replacing the matrices in the above representation:
\begin{equation}\label{eqn:trs-site-action-v1}
	\tilde A^{(k)}_{i_k} = U^T{}_{i_k}{}^{j_k} (A^{(k)}_{j_k})^*
\end{equation}
Recall, that in the MPS representation \eqref{eqn:mps-ansatz} the expression $ A^{(k)}_{i_k} $ is actually short-form for the full matrix: $ A^{(k)}_{i_k;\alpha\beta} $. Here $ \alpha, \beta $ are ``bond'' indices and connect matrices of neighboring sites with each other, $ k $ is the site index and $ i_k \in \{1,2,\ldots,D\} $ indexes the basis states of the (physical) site Hilbert space. For simplicity ignoring all other sites we focus only on a single site in the MPS. The action of the unitary $ U $ on a single site $ k $ can then be expressed as\footnote{The placement of the indices, lower and upper, on the unitary matrix $ U_i{}^j $ is used purely for notational elegance and has no physical significance here.}:
\[ A^{(k)}_{i_k} \ket{i_k} \rightarrow \left(A^{(k)}_{i_k}\right)^* U_{i_k}{}^{j_k} \ket{j_k} = \left( U^T \right)_{j_k}{}^{i_k} \left(A^{(k)}_{i_k}\right)^* \ket{j_k} = \tilde A^{(k)}_{j_k} \ket{j_k}  \]
with $ \tilde A^{(k)}_{j_k} $ defined as in \eqref{eqn:trs-site-action-v1}.

\begin{figure}
	\centering
	\begin{subfigure}[t]{0.48\textwidth}
		\centering
		\includegraphics[width=0.8\linewidth]{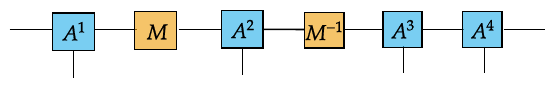}
		\caption{Insertion of a symmetry flux at a single site}
		\label{fig:site-symmetry-flux}
	\end{subfigure}
	\hfill
	\begin{subfigure}[t]{0.48\textwidth}
		\centering
		\includegraphics[width=0.8\linewidth]{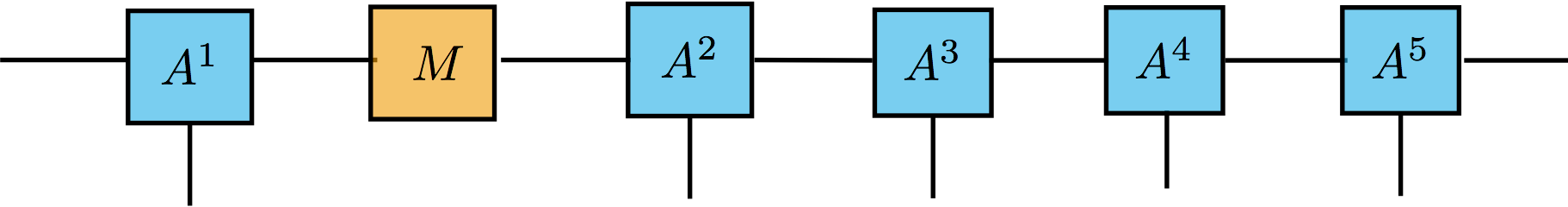}
		\caption{Insertion of a symmetry flux on a number of sites}
		\label{fig:local-symmetry-flux}
	\end{subfigure}
	\caption{Insertion of local gauge symmetry fluxes in a MPS}
\end{figure}

Now, for a time-reversal symmetric and short-range entangled (SRE) state the relation \eqref{eqn:trs-site-action-v1} can equivalently be written as \cite{Chen2015Gauging,Pollmann2010Entanglement}:
\begin{equation}\label{eqn:trs-site-action-v2}
	\tilde A^{(k)}_{\alpha\beta} = M_{\alpha\gamma} A^{(k)}_{\gamma\eta} M^{-1}_{\eta\beta},
\end{equation}
where now the matrix $ M $ acts on the bond indices of $ A $, rather than on the physical state index as in \eqref{eqn:trs-site-action-v1}. This is illustrated below in \autoref{fig:site-symmetry-flux}. This can be used to extend the application of TRS to a group of neighboring sites rather than to only a single site. In this case each member of the neighborhood will undergo a transformation given by \eqref{eqn:trs-site-action-v2}. The factors of $ M $ and $ M^{-1} $ which are located in the interior of the region will cancel each other out leaving only a factor of $ M $ and $ M^{-1} $ on either side of the boundary of the region as shown in \autoref{fig:local-symmetry-flux}.

\begin{figure}[h]
    \centering
    \includegraphics[width=0.75\columnwidth]{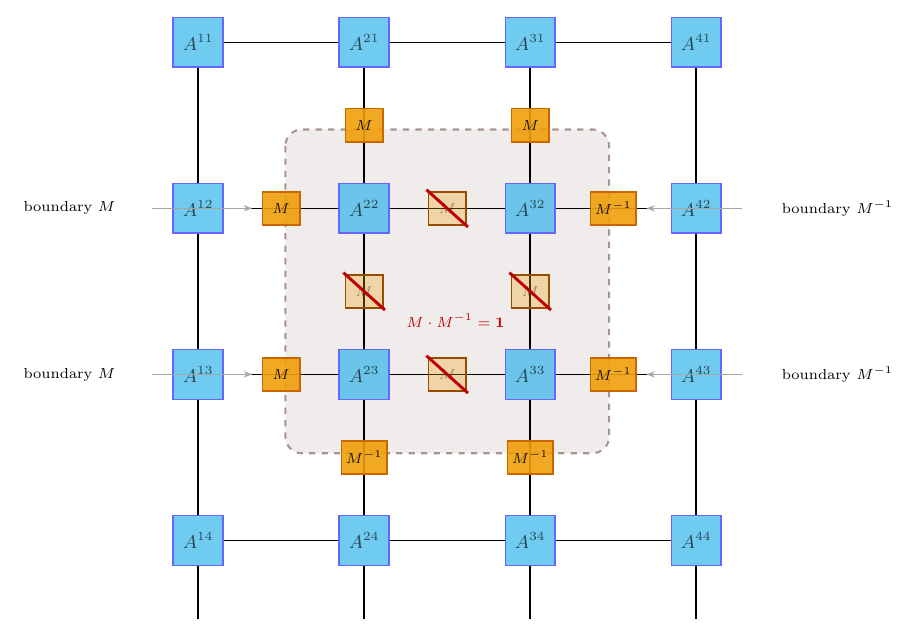}
    \caption{Matrix insertion on a subregion of a 2D tensor network state. Applying the symmetry $M$ to the bonds of the shaded subregion (cols 2--3, rows 2--3) inserts $M$ and $M^{-1}$ on every bond touching the region. Interior bonds carry both $M$ and $M^{-1}$ which cancel ($M \cdot M^{-1} = \mathbf{1}$, shown with strikethrough). Only the boundary bonds retain a net $M$ or $M^{-1}$ insertion.}
    \label{fig:tns-matrix-insertion-2d}
\end{figure}

\begin{figure}[h]
	\centering
	\begin{subfigure}[t]{0.48\textwidth}
		\centering
		\includegraphics[width=0.7\columnwidth]{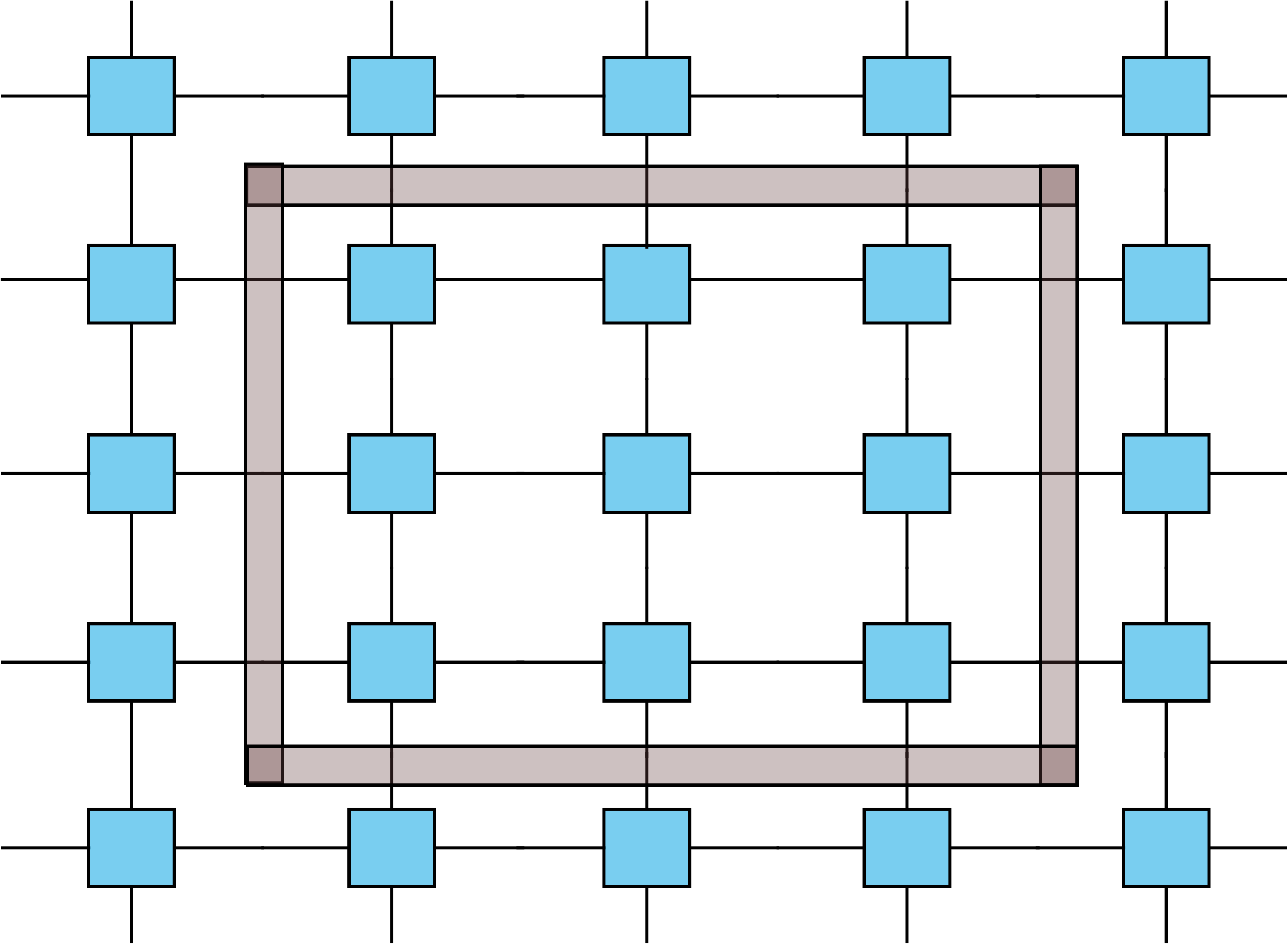}
		\caption{Applying constant gauge transformation to given region, leaves symmetry flux insertions along boundary of region.}
		\label{fig:symmetry-flux-line}
	\end{subfigure}
	\hfill
	\begin{subfigure}[t]{0.48\textwidth}
		\centering
		\includegraphics[width=0.7\columnwidth]{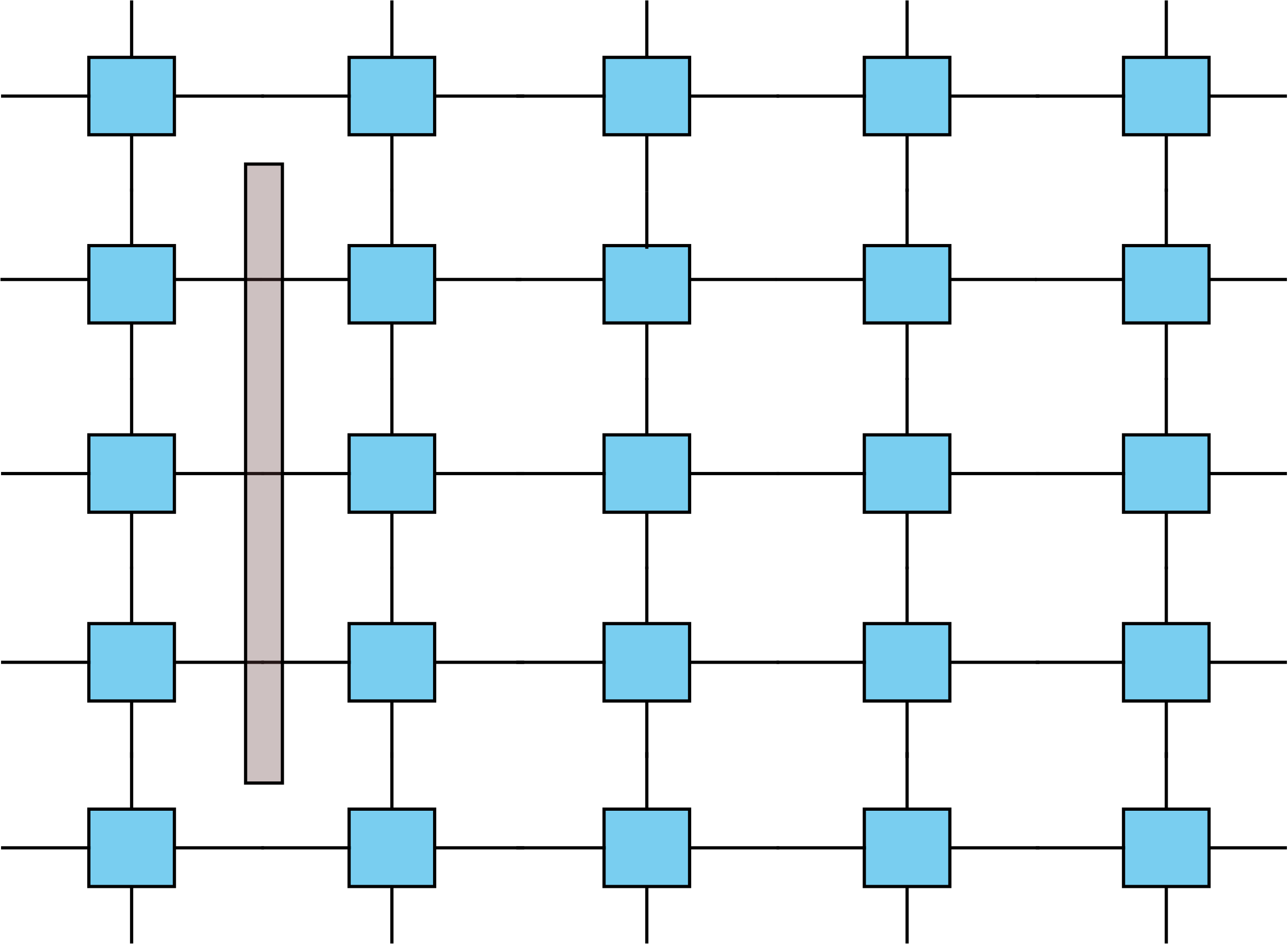}
		\caption{Insertion of a symmetry flux twist only along one line}
		\label{fig:symmetry-flux-twist}
	\end{subfigure}
 	\caption{Local symmetry flux insertions in a 2D MPS}
	\label{fig:symmetry-flux-2d}
\end{figure}

We can extend this reasoning to two and three dimensional systems. In this case the site tensors will have three or more indices and a matrix insertion will be required on every leg. As illustrated in \autoref{fig:tns-matrix-insertion-2d}, when $M$ is inserted on every bond of a subregion in a 2D tensor network, the $M$ and $M^{-1}$ factors on interior bonds cancel pairwise, leaving non-trivial matrix insertions only on the boundary bonds of the region. When applied to a subregion of a given lattice, the result will be that matrices on the interior will multiply to give the identity and we will be left with the non-trivial unitaries only on the boundary of the region as shown in \autoref{fig:symmetry-flux-line}.

\section{Spin Networks}\label{sec:spin-networks}

\begin{figure}[h]
	\centering
	\begin{subfigure}[t]{0.48\textwidth}
		\centering
		\includegraphics[width=0.6\linewidth]{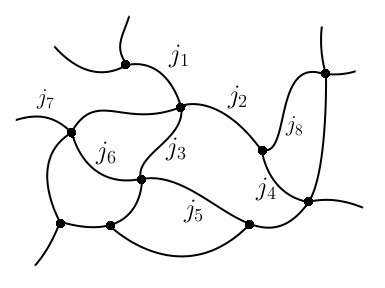}
		\caption{A graph with edges labelled by representations of $ SU(2) $. Each $ j_k $ is an $ SU(2) $ spin taking values in the set $ \{1/2,1,3/2,2,\ldots \} $}
		\label{fig:spin-network}
	\end{subfigure}
	\hfill
	\begin{subfigure}[t]{0.48\textwidth}
		\centering
		\includegraphics[width=0.5\linewidth]{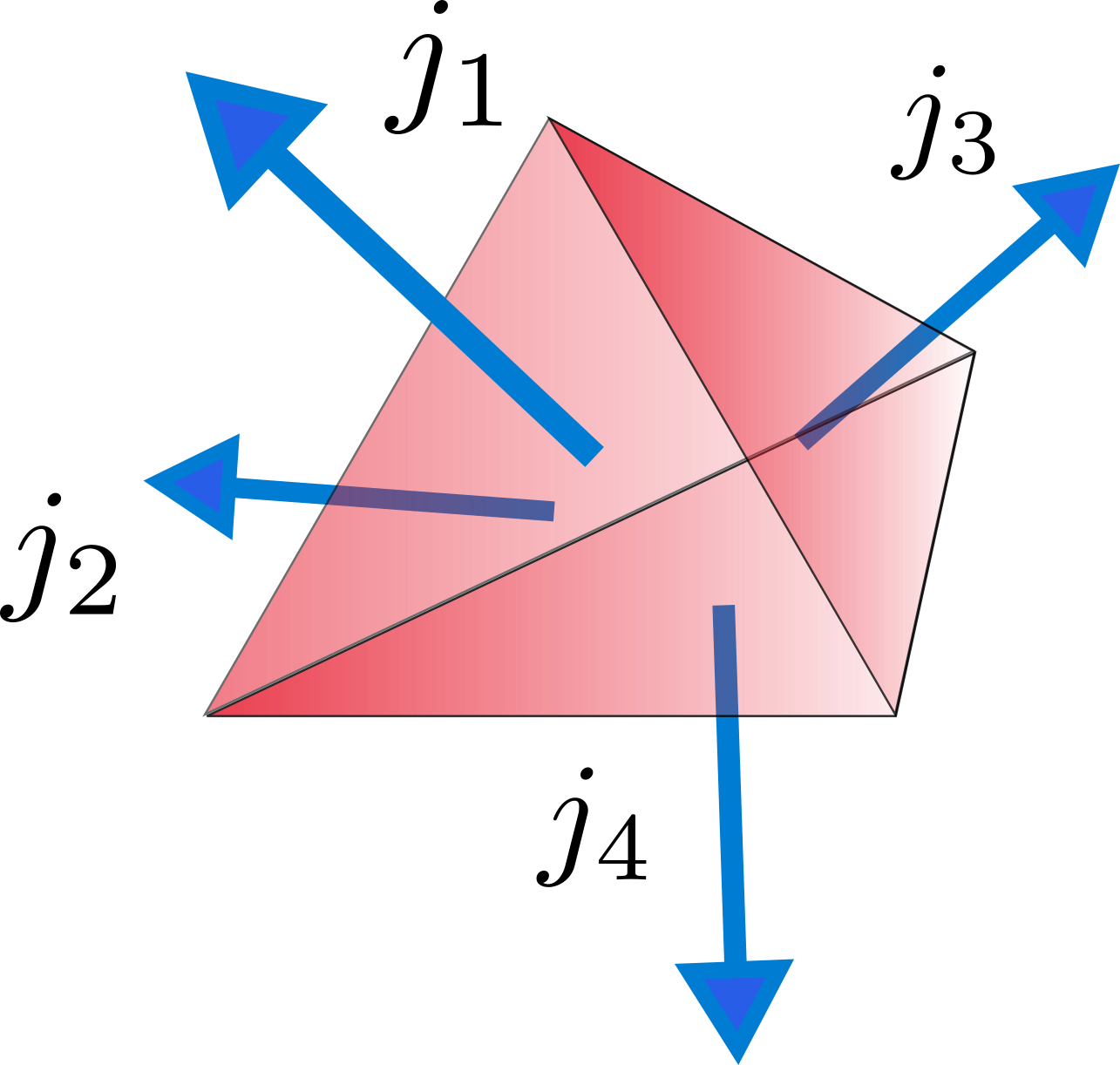}
		\caption{A single vertex connected to four edges, with each edge labeled by a representation $ j_i $ of $ SU(2) $. The vertex is represented schematically by a tetrahedron. The spin on each edge can then be thought of as representing the normal to the correspond face of the tetrahedron.}
		\label{fig:spin-net-vertex}
	\end{subfigure}
 	\caption{States of quantum geometry represented by spin-networks}
\end{figure}

Spin networks are very simple to describe physically. Consider an arbitrary graph $ \Gamma $, as shown in \autoref{fig:spin-network}, whose edges are labelled by representations $ j_e \in \mbb{Z}/2 $ of $ SU(2) $. This is very similar to the tensor networks defined in the previous section, with the only difference being that now the edge Hilbert spaces can all be of different dimensionality $ D_e = \text{dim}[\mc{H}_e] = 2j_e + 1 $ depending on the representation $ j_e $ living on the respective edges. The tensor associated with each edge will then be a matrix $ g^{e}_{i_e;\alpha\beta} $ of rank $ 2 $ and dimension $ D_e $, with $ i_e \in \{ \ket{-j_e}, \ket{-j_e + 1}, \ldots, \ket{j_e - 1}, \ket{j_e} \} $ being the state living on the edge\footnote{Recall that given a particle of spin $ j$, the Hilbert space is spanned by a set of vectors $ \{ \ket{j;m}  \} $, where $ m \in \{ -j, -j+1,\ldots,j-1,j \} $. Here $ m $ is referred to as the ``magnetic'' quantum number. In the main text we have used just $ \ket{m} $ for these states, rather than the full $ \ket{j;m} $, for notational symplicity.} taking values in $ \mc{H}_e $.


To complete the definition of a spin-network we need to assign tensors $ T^v_{j_v} $ to the vertices also. Now, as opposed to the edge tensors, the rank of the vertex tensors will vary, being given by the degree $ k_v $ of each vertex\footnote{In graph theoretical language, the ``degree'' a vertex is the number of edges which meet at that vertex.} Given a vertex $ v $, let $ \{ e^v_m; m \in \{ 1, \ldots, k_v \} \} $ be the set of edges which meet at $ v $. Then the tensor $ T^v_{j_v} $ will be an element of:
\begin{equation}\label{eqn:tensor-hilbert-space}
\mc{H}_v  = \left[ \underset{m = 1 .. k_v}{\bigotimes} \mc{H}_{e^v_m} \right]
\end{equation}
which is the tensor product space of the edge Hilbert spaces $ \mc{H}_{e^v_m} $. Later will we have to restrict further to the subspace $ \mc{H}_v^{Inv}$ invariant under the action of $ SU(2) $ gauge transformations.

As a simple illustration of this consider the case of single vertex with four edges \autoref{fig:spin-net-vertex} . For simplicity, we set $ j = 1/2 $ for all edges. Now the state space $ \mc{H}^2 $ of a single edge is spanned by the states $ \{\ket{\uparrow}, \ket{\downarrow} \} $, or equivalently $ \{ \ket{1}, \ket{0} \} $. The state space of all edges taken together is then: $ \mc{H}^v = \{\mc{H}^2\}^{\otimes 4} $ and is spanned by states of the form $ \ket{m_1, m_2, m_3, m_4} $ where the coefficients $ m_k $ take values in $\{0,1 \}$. An arbitrary state of the edge system can then be written as:
\begin{equation}\label{eqn:tetra-edge-wf}
\ket{\Psi_v} = \sum_{\{m_k\}} C_{m_1, m_2, m_3, m_4} \ket{m_1, m_2, m_3, m_4}
\end{equation}
where the sum is over all allowed spin states of the edges: $ m_i \in \{-j_i, -j_1 + 1, \ldots, j_1 - 1, j_i \} $ and $ C_{m_1, m_2, m_3, m_4} $ is the weight for the corresponding state vector. Not all states of the form \eqref{eqn:tetra-edge-wf} are physically allowed. We want quantum states which corresponds to classical states of three geometry. Physical states in the classical theory are required to satisfy the constraints discussed in the previous section. There are three constraints: the Hamiltonian constraint $ \mc{H} $, the diffeomorphism constraint $ \mc{C}^a $ and the Gauss constraint $ \mc{G}^i $. We are interested in the integral form of the Gauss constraint \eqref{eqn:gauss-constraint-v3}. When translated into the language of operators acting on spin-networks the \emph{quantum} version of the Gauss constraint can be written as:
\begin{equation}\label{eqn:quantum-gauss-constraint}
	\sum_i J_i = 0	
\end{equation}
where the Gaussian surface now encloses a single vertex of our spin-network and the quantum version of the Gauss constraint says that the sum of the angular momenta carried by the edge spins should be zero. Physical states should satisfy this constraint. In other words:
\begin{equation}\label{eqn:physical-wf}
	(J_1 + J_2 + J_3 + J_4) \ket{\Psi_v}_{phys} = 0
\end{equation}
This relationship has a simple geometric interpretation. It is the statement that the faces of the tetrahedra in \autoref{fig:spin-net-vertex} should close. Of course, \eqref{eqn:quantum-gauss-constraint} applies to all vertices, not just the 4-valent ones. Given a node with $ n $ edges attached to it, the resulting state is a quantum version of a classical polyhedron with $ n $ faces. The set of all face normals $ \vec{N}_i $ must satisfy $ \sum_i \vec{N}_i = 0 $ if the polyhedron is to close. The angular momentum vector is the quantum analog of this normal vector and hence the classical closure condition:
\begin{equation}\label{eqn:classical-closure-condition}
\sum_i \vec{N}_i = 0
\end{equation}
implies the quantum closure condition \eqref{eqn:quantum-gauss-constraint}.

Now as shown in \autoref{sec:quantum-geometry} the wavefunction for a spin-network defined on an arbitrary graph can be written in the form:

\begin{equation}\label{eqn:spin-net-state}
\ket{\Psi} = \underset{\{i_v\}, \{ j_e \}}{\sum} \Tr\left[ \underset{v \in \Gamma}\prod I^v_{i_v} \underset{e \in \Gamma}\prod D^e_{j_e} \right] \ket{i_1,\ldots,i_{n_v}; j_1, \ldots, j_{n_e} } 
\end{equation}
with the various terms in this equation being defined in \autoref{tbl:spin-net-data}.

\begin{table}[h]
	\centering
	\begin{tabular}{c c p{6cm}}
		\toprule
		Vertex Tensors & $ I^v_{i_v; \alpha_1 \ldots \alpha_{v_k}} $ & $ v: $ vertex \newline $i_v$: vertex state/physical index \newline $\alpha_i \in \{ 1,..,d_i \} $; $ d_i = 2 j_i + 1 $: contracts with the state on the $ i^\text{th} $ edge joined to the vertex. \\
		\midrule
		Edge Tensors & $ D^e_{j_e; \alpha\beta} $ &  $ e $: edge \newline $j_e:$ $ SU(2) $ label of edge \newline $\alpha, \beta \in \{1,\ldots,(2 j_e + 1) \} $: bonding indices \\
		\bottomrule
	\end{tabular}
	\caption{Data required for specifying the state of a spin network}
	\label{tbl:spin-net-data}
\end{table}

We can see from the form of \eqref{eqn:spin-net-state} and \autoref{tbl:spin-net-data} that the expression for the state vector living on a spin network is identical to that of a tensor network state given in \eqref{eqn:tensor-net-state}.

\begin{figure}[h]
	\centering
	\includegraphics[width=0.3\linewidth]{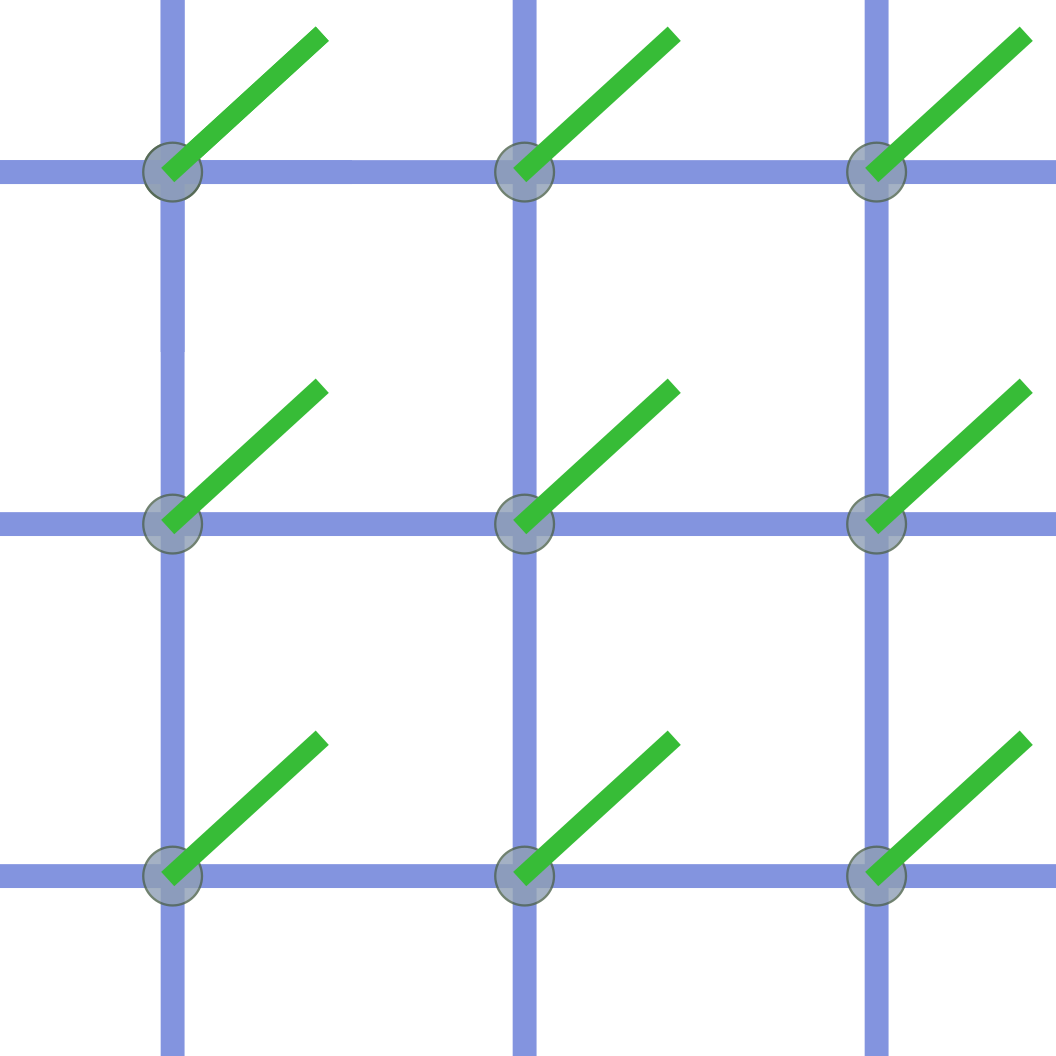}
	\caption{A spin-network as a tensor network whose ``physical'' degrees of freedom correspond to the allowed intertwiner states of four-valent vertices.}
	\label{fig:spin-tensor-network}
\end{figure}

It is worth noting that the condition \eqref{eqn:physical-wf} is \emph{exactly} the same as the ``branching rules'' for so-called string-net states as first described by Levin and Wen in \cite{Levin2004String-net}.


\subsection{Microscopic Structure of Spacetime}\label{subsec:quantum-geometry}

\begin{figure}[h]
	\centering
	\begin{subfigure}[t]{0.48\textwidth}
		\centering
		\includegraphics[width=0.7\textwidth]{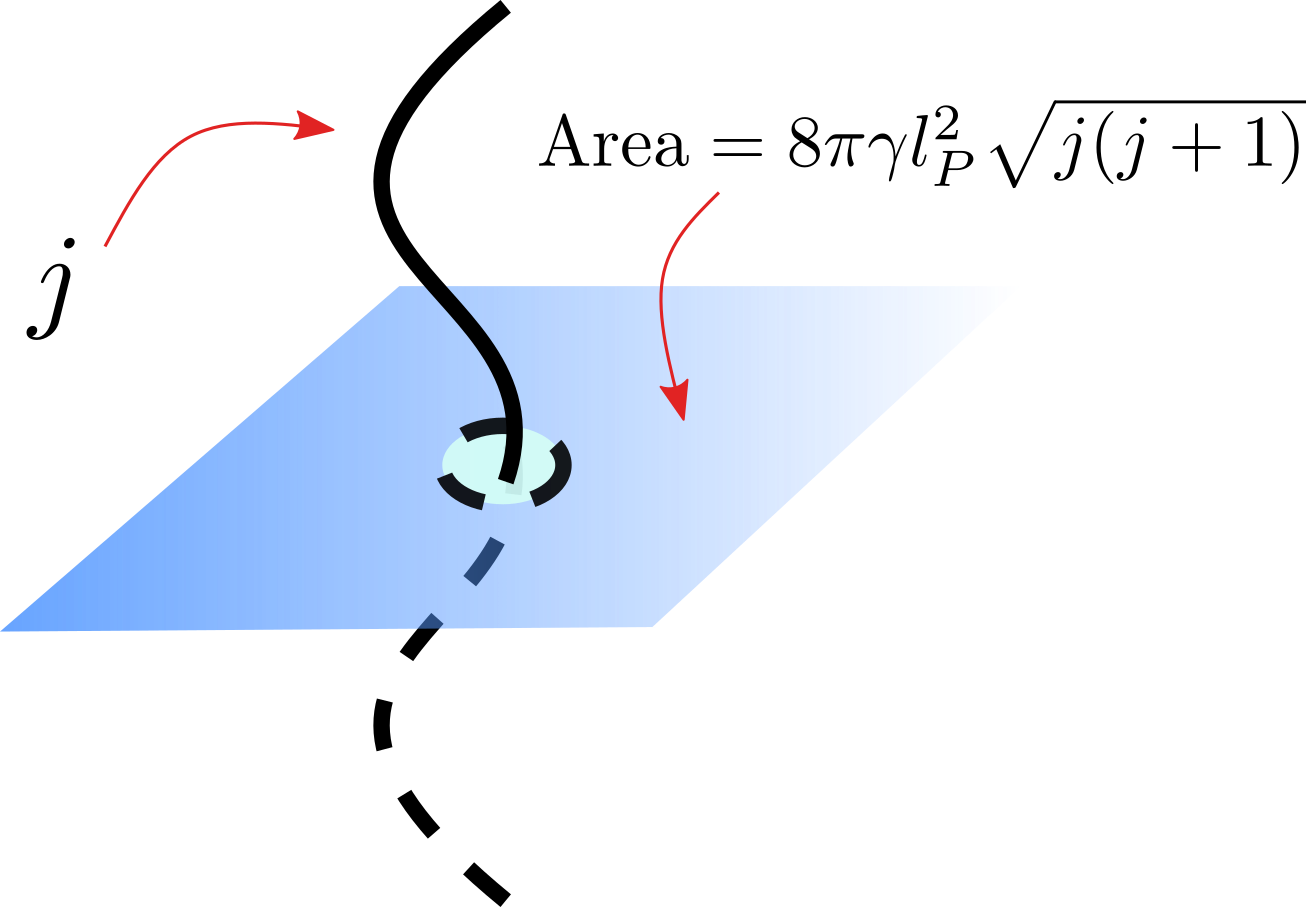}
		\caption{The area associated with a single spin puncture}
		\label{fig:area-operator}
	\end{subfigure}
	\hfill
	\begin{subfigure}[t]{0.48\textwidth}
		\centering
		\includegraphics[width=0.7\textwidth]{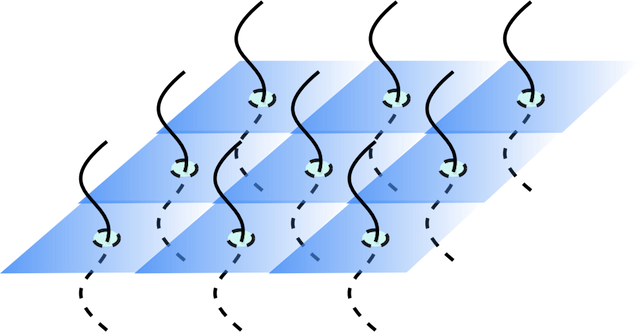}
		\caption{A graph state with many edges puncturing a given surface}
		\label{fig:many-punctures}
	\end{subfigure}
	   \caption{How macroscopic geometry emerges from quantum geometry}
       \label{fig:area-quanta}
\end{figure}

Spin-networks have a beautiful interpretation in terms of geometric observables. As such spin-networks are abstract graphs which do not require any background manifold. The whole idea is that the notion of a smooth manifold with some geometry \emph{emerges} from considering superpositions or ensembles of large numbers of such graphs. The first step towards achieving this goal is to relate the physical variables living on the graph with geometric quantities. Since spin-networks arise as the quantization of Einstein's equations, which are themselves built out of purely geometric quantities such as the metric and Christoffel connection, it is natural that they would encode spacetime geometry. The manner in which this happens is as follows. Consider a single edge of a graph carrying a spin $ j $. Whenever this edge intersects a surface, as shown in \autoref{fig:area-operator}, it endows the surface with a quantum of area at the location of the puncture. The area arises as the eigenvalue of the Casimir operator $ \unit{j}^2 $ acting on the state of the spin-network \emph{at that edge}. These eigenvalues thus take the form:
\begin{equation}\label{eqn:area-eigenvalue}
	A = 8 \pi \gamma l_p^2 \sqrt{j(j+1)}
\end{equation}
where $ l_p $ is the Planck length, $ \gamma $ is known as the Barbero-Immirzi parameter and $ j $ is the spin on that edge. From this equation it is clear that the area operator is gapped, \ie it has smallest eigenvalue which is greater than zero. This occurs for the smallest value of spin $ j = 1/2 $,
\begin{equation}\label{eqn:area-minimum}
	A_{min} = 4 \sqrt{3} \, \pi \, \gamma l_p^2
\end{equation}


\section{Volume Operator and $Z_2$ Symmetry}\label{sec:volume-z2}

Having established the physical correspondence between spin-networks states and gauged tensor networks, we now return to the physics of the problem under consideration. We wish to argue that the \emph{local} action of time-reversal symmetry on a spin network leads to a change in the sign of the eigenvalues of the LQG volume operator acting in that region. This effect can be modeled at the level of the Palatini action by introducing a $ Z_2 $ field $ \phi $ which couples non-trivially to the tetrad.

There are two commonly used versions of the volume operator in LQG. They are named after the respective authors as the Rovelli-Smolin (RS) \cite{Rovelli1994Discreteness} or the Ashtekar-Lewandowski (AL) \cite{Ashtekar1996Quantum,Ashtekar1997Quantum} versions. These have the form:
\begin{align}\label{eqn:volume-ops}
	\hat{V}_{S}^{R S} \ket{\Psi_{\Gamma}} & =\gamma^{3 / 2} l_{p}^{3} \sum_{v \in S \cap \Gamma} \sum_{i, j, k}\left|\frac{i C_{r e g}}{8} \epsilon^{i j k} \hat{J}_{i} \hat{J}_{j} \hat{J}_{k}\right|^{1 / 2} \ket{\Psi_{\Gamma}} \nonumber \\
	\hat{V}_{S}^{A L} \ket{\Psi_{\Gamma}} & =\gamma^{3 / 2} l_{p}^{3} \sum_{v \in S \cap \Gamma}\left|\frac{i C_{r e g}}{8} \epsilon_{v}\left(i,j,k\right) \epsilon^{i j k} \hat{J}_{i} \hat{J}_{j} \hat{J}_{k}\right|^{1 / 2} \ket{\Psi_{\Gamma}}
\end{align}
where $ \gamma $ is the so-called Barbero-Immirzi parameter whose precise value or even physical relevance is a matter of some debate. Its actual value will not affect our arguments so for convenience we will set $ \gamma = 1 $; $ l_p $ is the Planck length; $ \Gamma $ is the spin-network; $ S $ is the subset of the graph whose volume we wish to determine; $ v \in S $ is the set of all vertices contained within $ S $; $ \{\hat J_i \} $ are the angular momentum operators living on each edge attached to $ v $.

\begin{figure}[h]
	\centering
	\begin{subfigure}[t]{0.48\textwidth}
		\centering
		\includegraphics[width=0.7\textwidth]{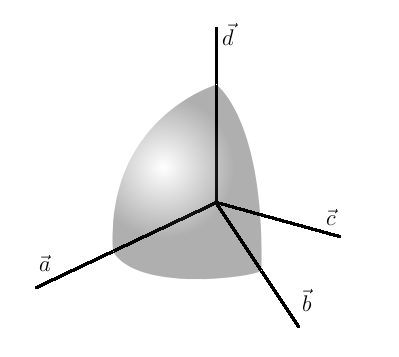}
		\caption{The volume associated with the solid angle spanned by three classical vectors is given by $ \vect{a}\cdot(\vect{b} \times \vect{d}) $ }
		\label{fig:solid-angle-classical}
	\end{subfigure}
	\hfill
	\begin{subfigure}[t]{0.48\textwidth}
		\centering
		\includegraphics[width=0.7\textwidth]{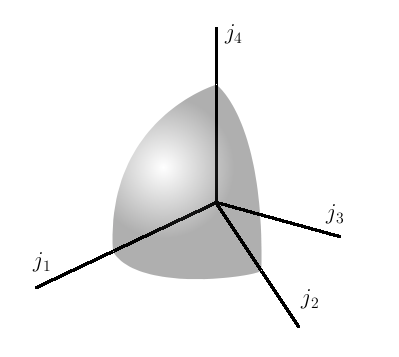}
		\caption{The volume associated with the solid angle spanned by three quantum vectors is given by $  \hat{J}_1 \cdot (\hat J_2 \times \hat J_4) $}
		\label{fig:solid-angle-quantum}
	\end{subfigure}
	\caption{Mapping classical volume to a quantum expression involves replacing classical vectors with angular momentum operators. Figure credit \cite{Vaid2016LQG-for-the-Bewildered}.}
	\label{fig:volume-operators}
\end{figure}

These two forms are identical except for the factor of $ \epsilon_v(i,j,k) $ which occurs in the AL version. The $ \epsilon^{i j k} \hat{J}_{i} \hat{J}_{j} \hat{J}_{k} \equiv \hat{J}_i \cdot (\hat J_j \times \hat J_k) $ term (\autoref{fig:solid-angle-quantum}) is essentially the quantum mechanical version of the classical geometric expression for the volume of a region spanned by three vectors $ \vect{a}\cdot(\vect{b} \times \vect{c}) $ (\autoref{fig:solid-angle-classical}) . The factor of $ 1/8 $ ensures that for each set of three edges only the volume of the quadrant spanned by those edges is counted. 

\begin{figure}[h]
	\centering
	\includegraphics[width=0.3\linewidth]{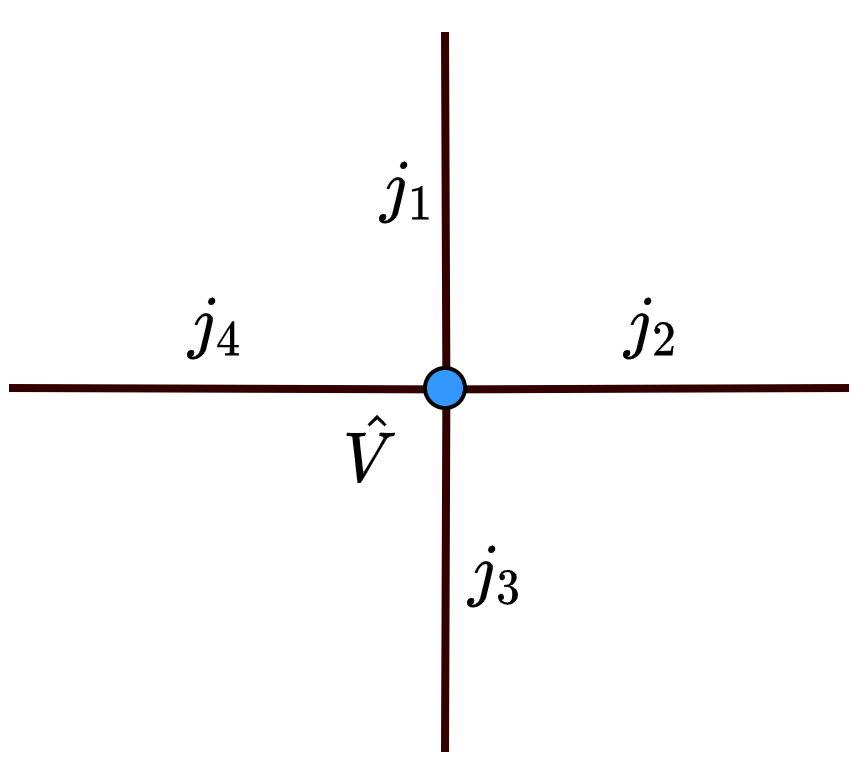}
	\caption{A 4-valent vertex with the vertex intertwiner shown as a four-index tensor}
	\label{fig:volume-op-2}
\end{figure}

Using these expression we can now ask what is the ``volume'' of the quantum tetrahedron shown in \autoref{fig:spin-net-vertex}. To make matters simple let us assume that all of the angular momentum operators are of the spin $ 1/2 $ representation. In this case the Hilbert space on which the volume operator will act is given by the tensor product space:
\begin{equation}\label{eqn:tetrahedron-hilbert-space}
	\mc H = \mc H_\onehalf \otimes \mc H_\onehalf \otimes \mc H_\onehalf \otimes \mc H_\onehalf
\end{equation}

The requirement of gauge invariance restricts the allowed space of states to the subspace $ \mathit{Inv}(\mc H) $ invariant under $ SU(2) $ rotations at each of the vertices. From the representation theory of $ SU(2) $ we know that the tensor product of four spin $ 1/2 $ representations can be written as a direct sum of irreducible representations as follows:

\begin{equation}\label{eqn:vol-irreps}
	1/2 \otimes 1/2 \otimes 1/2 \otimes 1/2 = 0 \oplus 0 \oplus 1 \oplus 1 \oplus 1 \oplus 2
\end{equation}

\begin{figure}[h]
	\centering
	\includegraphics[width=0.6\linewidth]{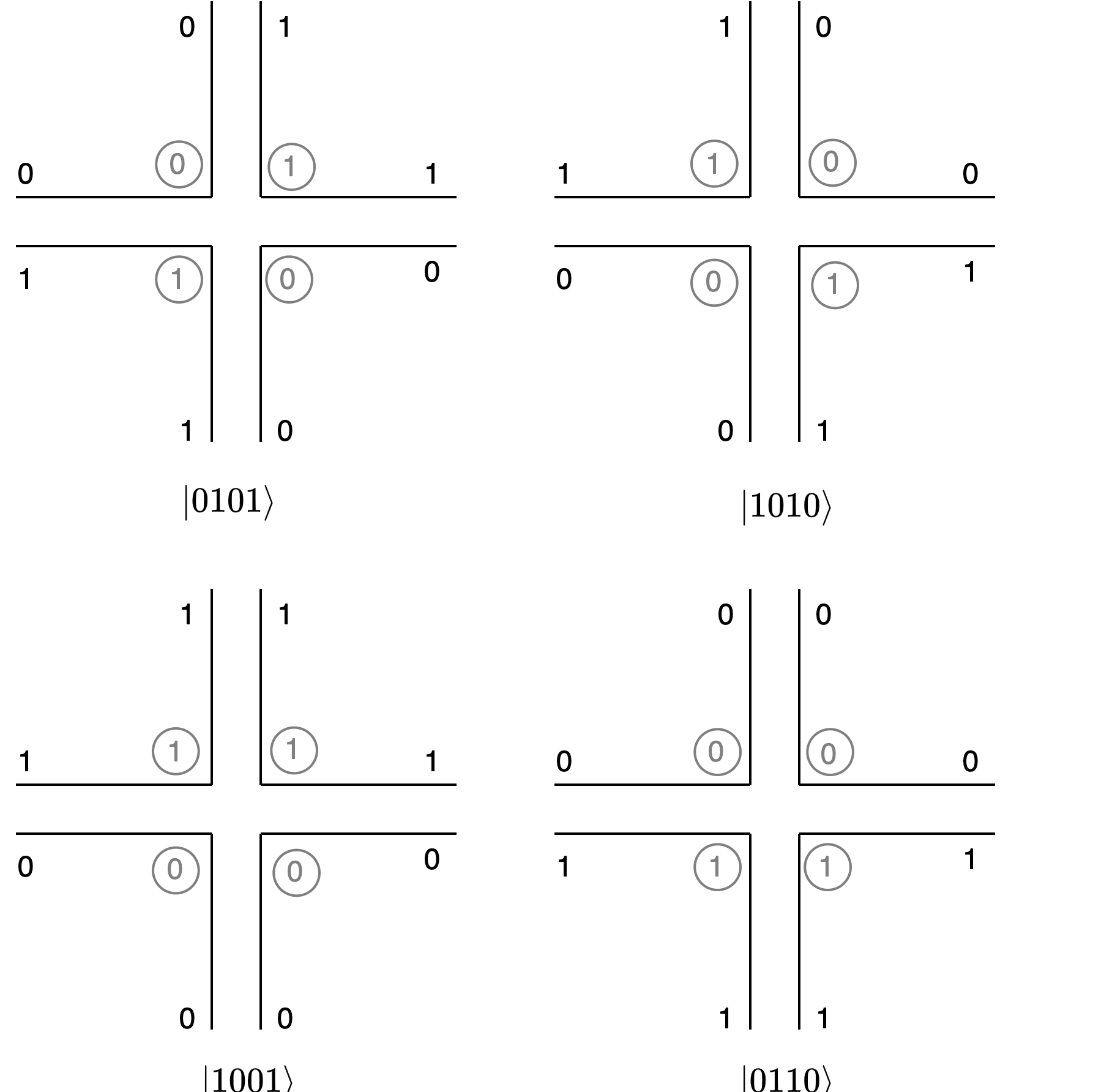}
	\caption{Tensor network representation of the 4-valent intertwiner state. The ``first'' qubit of each state lives on the top edge and then we proceed counter-clockwise around the vertex.}
	\label{fig:4-valent-state}
\end{figure}

The physical states live in the two copies of scalar (``s-wave'') states. As derived in the supplementary calculations, the correct orthonormal basis for the $SU(2)$ singlet subspace of four spin-1/2s consists of the states:
\begin{align}\label{eqn:vol-states}
	\ket{\Phi_1} & = \frac{1}{\sqrt{3}} \left( \ket{0101} + \ket{1010} - \frac{1}{2}(\ket{1001} + \ket{0110} + \ket{0011} + \ket{1100}) \right) \nonumber \\
	\ket{\Phi_2} & = \frac{1}{2} \left( \ket{1001} + \ket{0110} - \ket{0011} - \ket{1100} \right)
\end{align}
where $ \{\ket{\Phi_1}, \ket{\Phi_2}\} $ form an orthonormal basis for the 2D singlet subspace. These states are eigenvectors of the signed interior operator $\hat{Q}$ with eigenvalues $\pm q$, corresponding to the two possible time orientations. Note that the standard volume operator $\hat{V} \propto \sqrt{|\hat{Q}|}$ is degenerate on this subspace, but the time-orientation is captured by the sign of $\hat{Q}$.
Here the $ \ket{0}, \ket{1} $ are the two possible states of each edge carrying a spin $ 1/2 $ label. One can then write down the effective degrees of freedom carried by a four-valent intertwiner in terms of a single spin $ 1/2 $ state $ \ket{\tilde\Psi} $ spanned by the invariant basis states $ \{\ket{\Phi_1}, \ket{\Phi_2}\} $:
\begin{equation}\label{key}
	\ket{\tilde \Psi} = \alpha \ket{\Phi_1} + \beta \ket{\Phi_2}
\end{equation}.
The states $ \{\ket{\Phi_1}, \ket{\Phi_2}\} $ can also be expressed in a manner which makes the connection with tensor network states apparent. Any state of $ N $ spins can be written as:
\begin{equation}\label{eqn:spin-n-state}
	\ket{\Psi} = i^{A_1, \ldots, A_N}\ket{i_1 i_2 \ldots i_N}
\end{equation}
where $ \{\ket{i_k}\}; ~ k \in {1,\ldots, D} $, with $ D $ being the dimension of the site Hilbert space. Thus we can write:
\begin{align}\label{eqn:vol-states-3}
	\ket{\Phi_1} & = i_1^{ABCD}\ket{i_A i_B i_C i_D} \nonumber \\
	\ket{\Phi_2} & = i_2^{ABCD}\ket{i_A i_B i_C i_D}
\end{align}
where the two 4-index tensors $ i_1^{ABCD} $ and $ i_2^{ABCD} $ live on the vertex. These states can be represented using the diagrammatic tensor network notation as shown in \autoref{fig:4-valent-state}.

\section{CZX State}\label{sec:spt-czx}

\begin{figure}[h]
     \centering
     \begin{subfigure}[t]{0.3\textwidth}
         \centering
         \includegraphics[height=4cm]{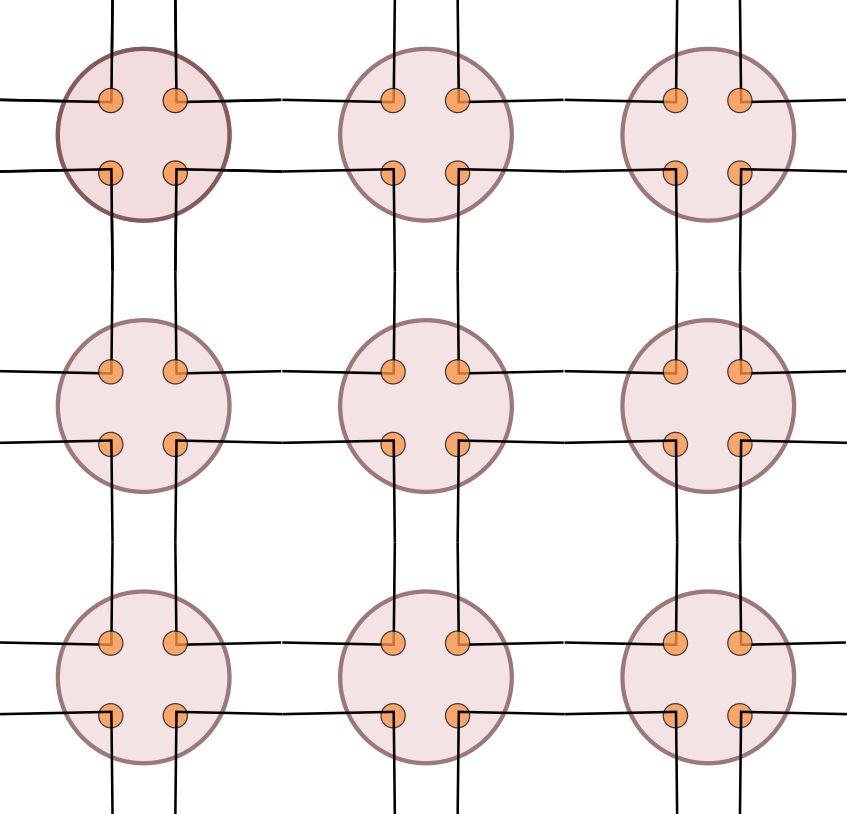}
         \caption{CZX model \cite{Chen2011Symmetry,Chen2010Local} is example of 2D SPT phase with $ Z_2 $ symmetry. four spins at each site; four spins in neighboring sites form a plaquette.}
         \label{fig:czx-lattice}
     \end{subfigure}
     \hfill
     \begin{subfigure}[t]{0.3\textwidth}
         \centering
         \includegraphics[height=4cm]{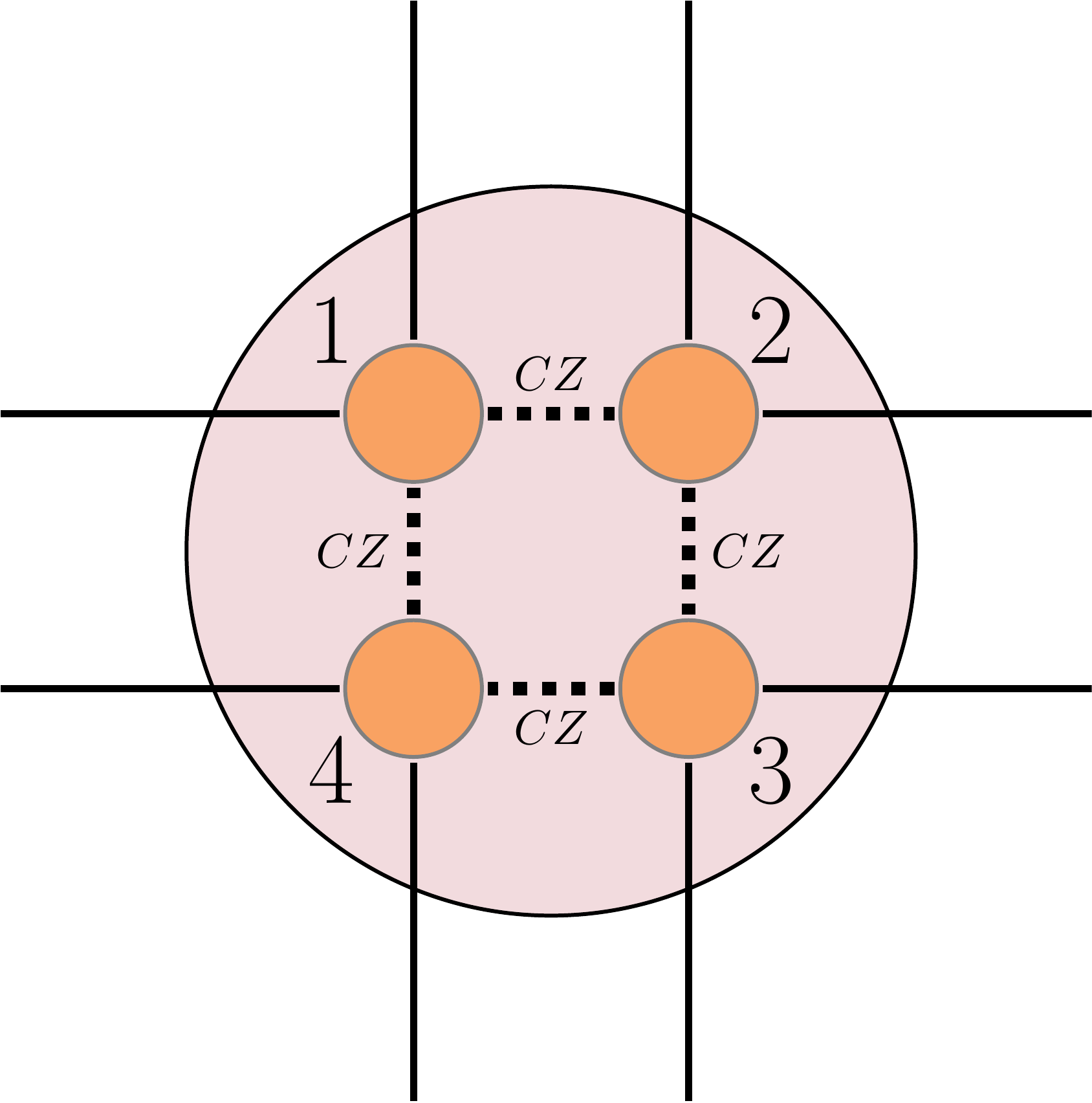}
         \caption{Action of unitary on vertices is given by $ U_{CZX} = U_X U_{CZ} $, where
         		$ U_X = X_1 \otimes X_2 \otimes X_3 \otimes X_4 $. The $ CZ $ gate acts between pairs of neighboring spins - connected by the dashed lines - in a single site in a clockwise fashion.}
         \label{fig:czx-site}
     \end{subfigure}
     \hfill
     \begin{subfigure}[t]{0.3\textwidth}
         \centering
         \includegraphics[height=4cm]{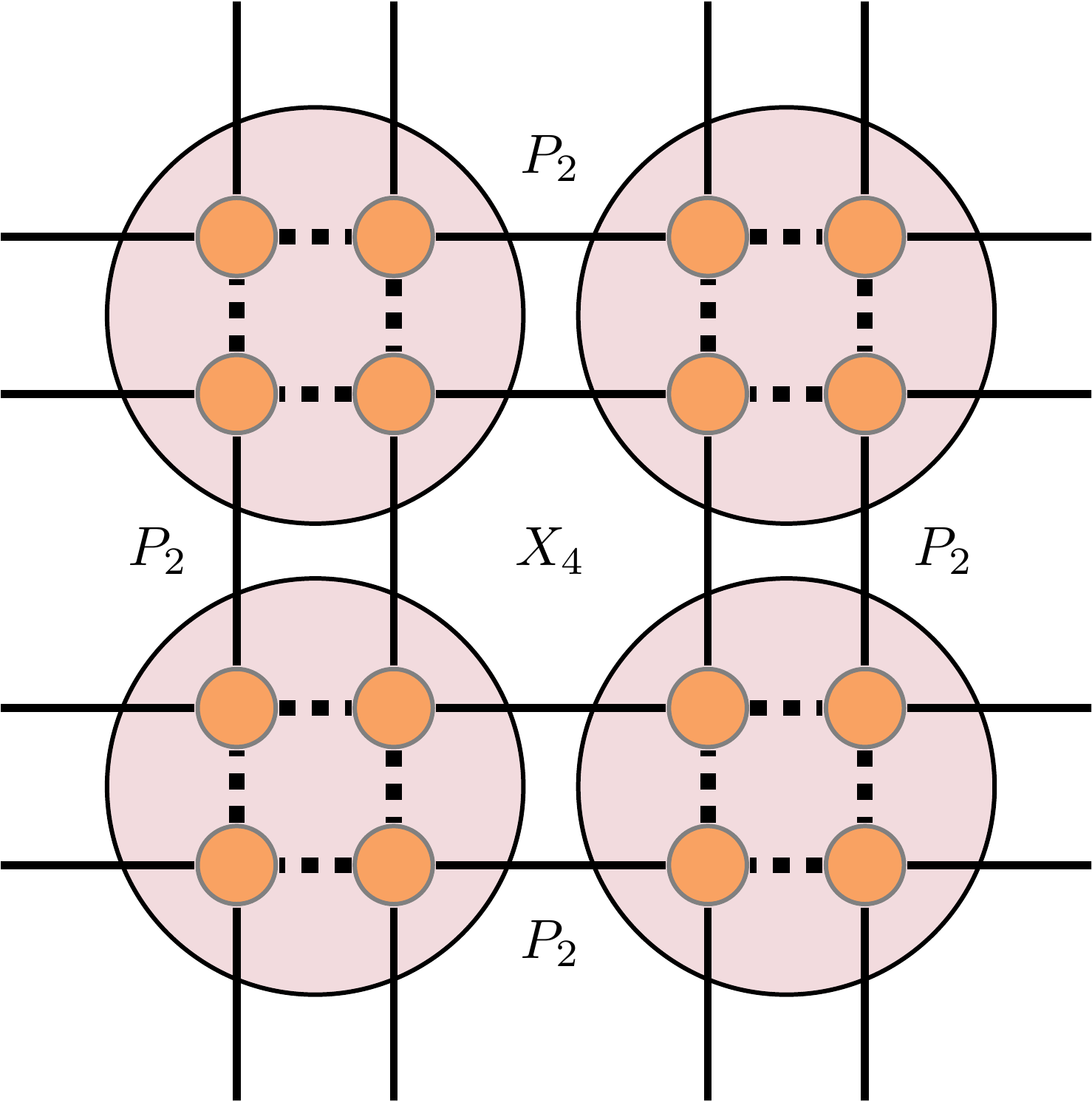}
     \caption{Action of the CZX hamiltonian results in a ground state where the neighboring spins in four different sites (forming a single plaquette) are entangled with each other.}
         \label{fig:czx-hamiltonian}
     \end{subfigure}
        \caption{The CZX state with SPT and an on-site $ Z_2 $ symmetry}
\end{figure}

In this section a model for a SPT phase with a $ Z_2 $ symmetry in a many-body system which admits a particularly clear mapping to the spin-network structure of quantum geometric states. This is the CZX model \cite{Chen2010Local} (named after the unitary operators $ C $, $ Z $ and $ X $  which are used to construct the state). Consider the system shown in \autoref{fig:czx-lattice}. Here each site consists of four spins. The spins between neighboring sites are connected as shown in \autoref{fig:czx-lattice}. The connected spins in four neighboring sites form a \emph{plaquette}. The state we will construct, following \cite{Chen2010Local} will be invariant under the action of $ Z_2 $ symmetry at each site.

\begin{figure}[h]
	\centering
	\begin{subfigure}[t]{0.48\textwidth}
		\centering
		\includegraphics[width=0.3\linewidth]{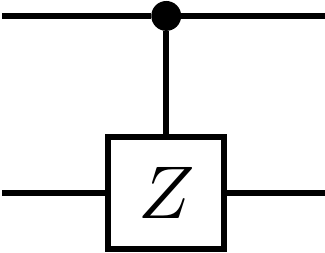}
		\caption{The controlled Z gate. This and other quantum circuits were drawn using the Quantikz package by Alastair Kay \cite{Kay2018Tutorial}}
		\label{fig:controlled-z}
	\end{subfigure}
	\hfill
	\begin{subfigure}[t]{0.48\textwidth}
		\centering
		\includegraphics[width=0.7\textwidth]{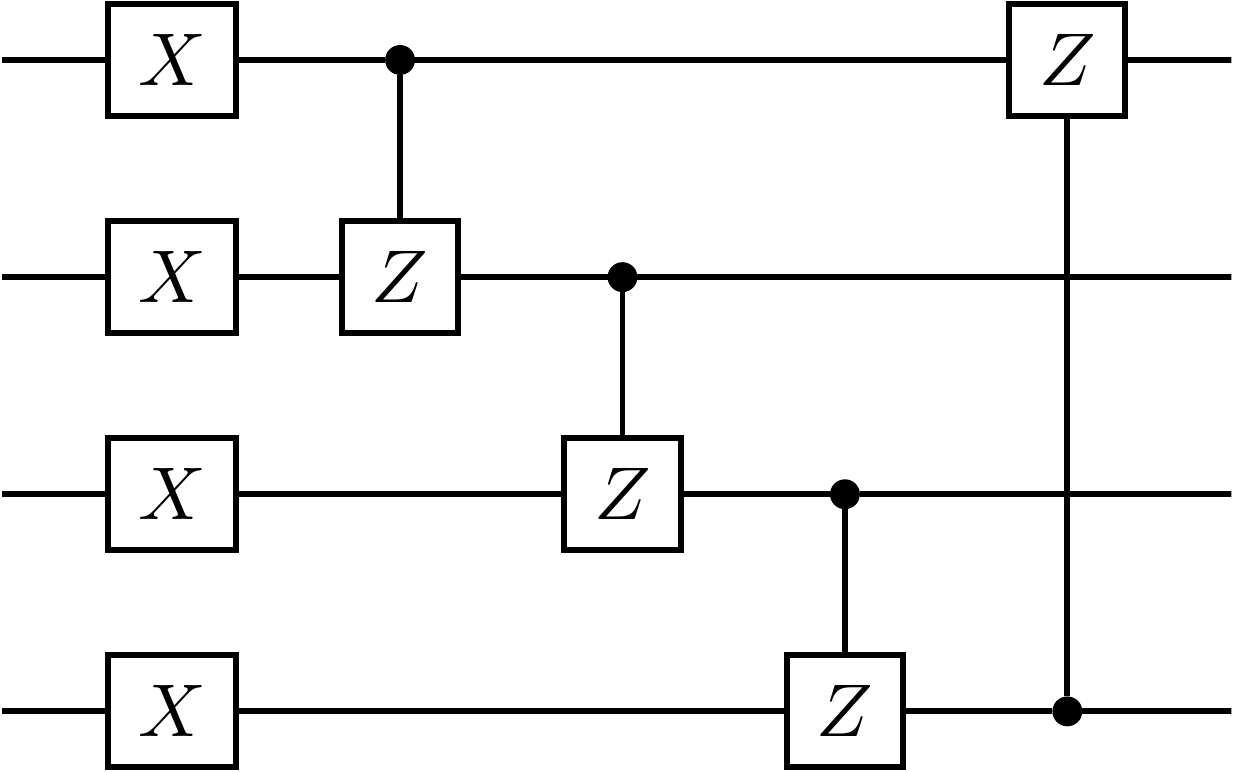}
		\caption{A circuit representation of the operator $ U_{CZX} $}
		\label{fig:czx-circuit}
	\end{subfigure}
\end{figure}

This state is generated by the action of two operators: $ U_X $ and $ U_{CZ} $, which are defined as follows:
\begin{subequations}
	\begin{align}
		U_X & = X_1 \otimes X_2 \otimes X_3 \otimes X_4 \\
		U_{CZ} & = (CZ)_{12} (CZ)_{23} (CZ)_{34} (CZ)_{41}
	\end{align}
\end{subequations}
where $ \{X_i\} $ are the usual Pauli spin flip matrices acting on the four spins of a given site as shown in \autoref{fig:czx-site}. $ \{CZ_{ij}\} $ is the controlled $ Z $ gate acting on two spins, which can be written in operator form as:
\begin{equation}\label{eqn:cz-operator}
	CZ = \ket{00}\bra{00} + \ket{01}\bra{01} + \ket{10}\bra{10} - \ket{11}\bra{11}.
\end{equation}
If the state of the first spin is $ \ket{0} $, the state of the second spin is left unchanged. If first spin is in $ \ket{1} $ state, then the (negative) $ Z $ operator is applied to the second spin. Written as a quantum circuit this operator can be also be expressed in the form shown in \autoref{fig:controlled-z}. The $ U_X $ and $ U_{CZ} $ gates generate a $ Z_2 $ group. This can be seen since both square to $ \mathbbm{1} $ and also commute with each other.

The Hamiltonian for the whole system is given by $ H = \sum H_{p_{i}} $, where $ p_i $  represents the $ i^\text{th} $ plaquette. The Hamiltonian for the $ i^\text{th} $ plaquette is given by:
\begin{equation}\label{eqn:czx-hamiltonian}
	H_{p_{i}}=-\mb{X}_{4} \otimes P_{2}^{u} \otimes P_{2}^{d} \otimes P_{2}^{l} \otimes P_{2}^{r}
\end{equation}
Here $ \mb{X}_4 $ is the four qubit operator given by:
\begin{equation}\label{eqn:czx-entangling-op}
	\mb{X}_4 = \ket{0000}\bra{1111} + \ket{1111}\bra{0000},
\end{equation}
which acts on the four spins in neighboring sites (forming a single plaquette) as shown in \autoref{fig:czx-hamiltonian}. The letter $ \mb{X} $ is used for this operator because it performs the same function as the Pauli $ X $ gate, only acting on groups of four spins at a time.  This operator ensures that the spins within each plaquette are all in the same state - either $ \ket{0} $ or $ \ket{1} $.

The $ P_2^i $ operators, with $i \in \{u,d,l,r\} $, act on the two spins in the neighboring half-plaquettes as shown in \autoref{fig:czx-hamiltonian}. These operators have the form:
\begin{equation}\label{eqn:czx-half-plaquette-op}
	P_2^i = \ket{00}\bra{00} + \ket{11}\bra{11},
\end{equation}
and ensure that each term in the Hamiltonian satisfies the on-site $ Z_2 $ symmetry. All the terms in the Hamiltonian commute with each other:
\begin{equation}\label{eqn:plaquette-commutators}
	[H_{p_i}, H_{p_j}] = 0, \quad \forall i \ne j,
\end{equation}
and consequently the ground state of the total Hamiltonian will simply be the product of the ground state of the $ i^\text{th} $ plaquette. It can easily be checked that this state is given by:
\begin{equation}\label{eqn:plaquette-state}
	\ket{\psi_{p_i}} = \ket{0000} + \ket{1111}
\end{equation}

\begin{figure}[h]
	\centering
	\begin{tabular}{|c|c|}
		\hline
		\addheight{\includegraphics[width=20mm]{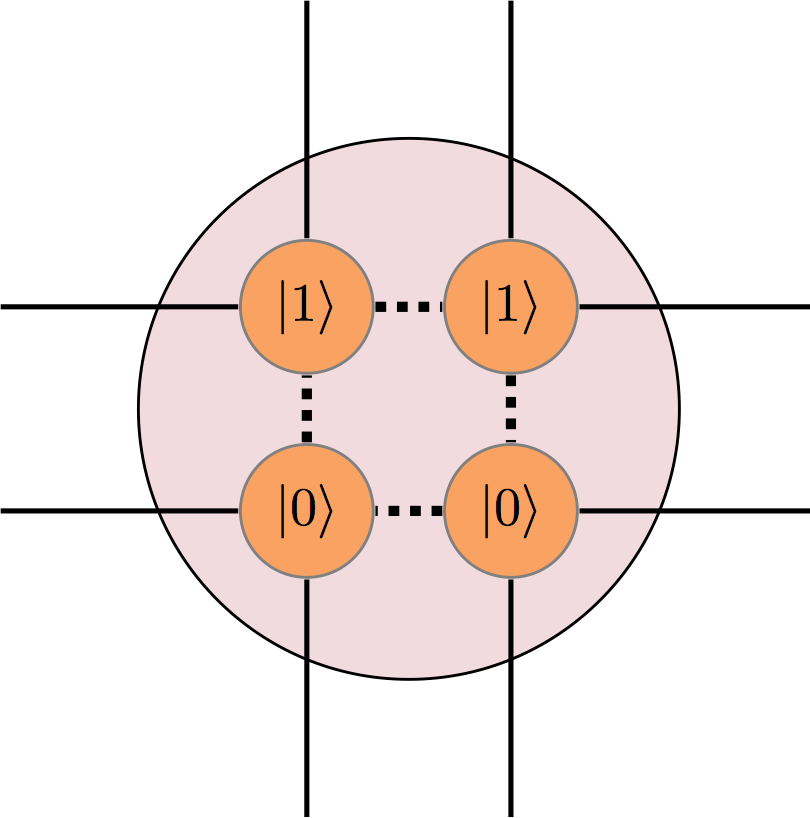}} &
		\addheight{\includegraphics[width=20mm]{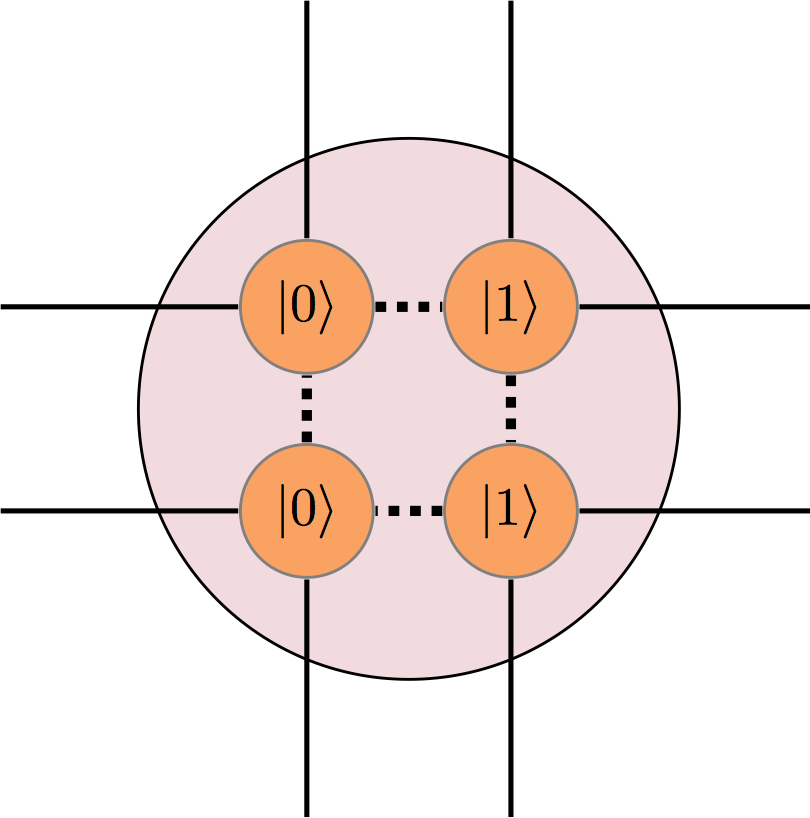}} \\
		\hline
		\addheight{\includegraphics[width=20mm]{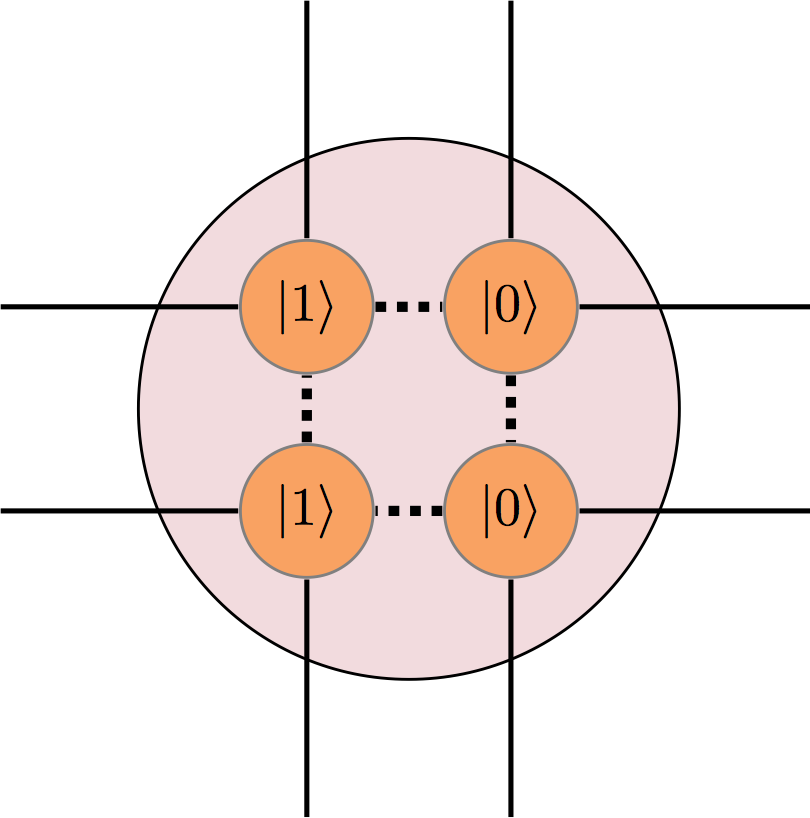}} &
		\addheight{\includegraphics[width=20mm]{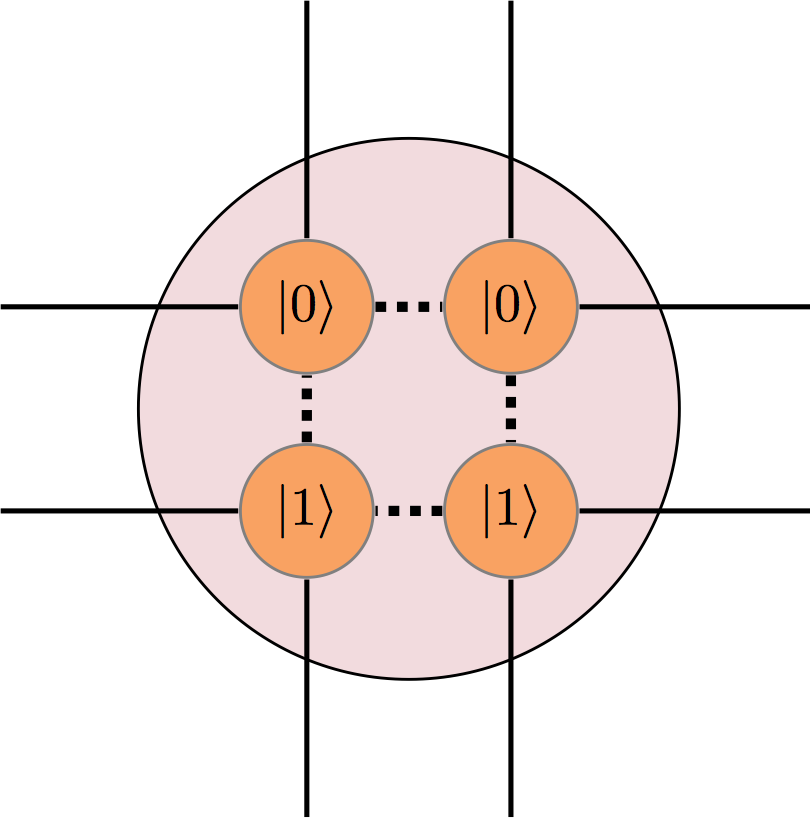}} \\
		\hline
	\end{tabular}
	\caption{The ``code'' subspace of the vertex Hilbert space which is mapped to itself under the action of $ U_{CZ} $ and $ U_X $. E.g. $ U_X \ket{1100} = \ket{0011} $ and $ U_{CZ} \ket{1100} = - \ket{1100} $}
\end{figure}

\begin{figure}[thbp]
	\centering
	\includegraphics[width=0.2\linewidth]{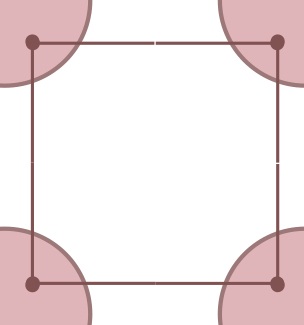}
	\caption{Each plaquette is in an entangled state: $ \ket{0000} + \ket{1111}$}
	\label{fig:czx-entangled}
\end{figure}


\section{SPT Phase of Quantum Geometry}\label{sec:spt-lqg}

In this section we establish the correspondence between the CZX model described in \autoref{sec:spt-czx} and the spin-network states of Loop Quantum Gravity. This mapping is the bridge which allows us to import the technology of symmetry-protected topological phases into the quantum gravity setting, and ultimately to identify the emergence of the arrow of time with a phase transition in a $ Z_2 $ gauge theory on the spin network.

\subsection{Structural Correspondence: CZX Sites and Spin Network Vertices}\label{subsec:czx-lqg-correspondence}

The mapping between the CZX model and the spin network is founded on the structural equivalence of their fundamental building blocks.

Each site in the CZX model consists of four spins, each taking values in $ \mc{H}_{1/2} = \{ \ket{0}, \ket{1} \} $. The on-site Hilbert space is therefore $ (\mc{H}_{1/2})^{\otimes 4} $, which is $ 2^4 = 16 $ dimensional. This is \emph{mathematically identical} to the Hilbert space \eqref{eqn:tetrahedron-hilbert-space} of a 4-valent spin-network vertex where each of the four edges carries the $ j = 1/2 $ representation of $ SU(2) $.

The Gauss constraint \eqref{eqn:physical-wf} restricts the physical states of the spin-network vertex to the $ SU(2) $-invariant subspace, which is two-dimensional and spanned by the states $ \ket{\Psi_1} $ and $ \ket{\Psi_2} $ defined in \eqref{eqn:vol-states}. In the CZX model, the on-site $ Z_2 $ symmetry generated by $ U_{CZX} $ plays an analogous role: it restricts the states of each site to a ``code'' subspace invariant under the combined action of $ U_X $ and $ U_{CZ} $.

\begin{figure}[h]
	\centering
	\begin{subfigure}[t]{0.3\textwidth}
		\centering
		\includegraphics[width=1.0\linewidth]{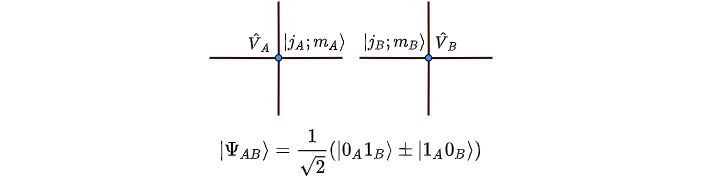}
		\caption{A (bipartite) Bell state formed out of entanglement between two edges of neighboring 4-valent vertices}
		\label{fig:two-edge-engtanglement}
	\end{subfigure}
	\hfill
	\begin{subfigure}[t]{0.3\textwidth}
		\centering
		\includegraphics[width=0.8\linewidth]{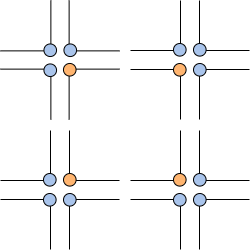}
		\caption{Four 4-valent vertices are shown. In each vertex, one spin is shown in a different color (orange). The four orange spins form a multipartite entangled state as shown in the next figure.}
		\label{fig:four-vertex-state}
	\end{subfigure}
	\hfill
	\begin{subfigure}[t]{0.3\textwidth}
		\centering
		\includegraphics[width=0.8\linewidth]{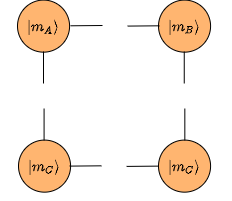}
		\caption{Multipartite entangled state formed out of four spins living on four neighboring vertices.}
		\label{fig:four-edge-entanglement}
	\end{subfigure}
	\caption{Spin network states with bipartite and multipartite entanglement between neighboring vertices}\label{eqn:lqg-entanglement}
\end{figure}

The correspondence extends to the entanglement structure. In the CZX model, the ground state is characterized by entanglement between spins in \emph{neighboring} sites which together form plaquettes (\autoref{fig:czx-entangled}). The plaquette ground state is:
\begin{equation}\label{eqn:czx-plaquette-state}
	\ket{\psi_p} = \ket{0000}_p + \ket{1111}_p
\end{equation}
where the four qubits belong to four different sites. In the spin-network language, this corresponds to a specific pattern of entanglement between edges of neighboring 4-valent vertices. Consider four neighboring vertices, each contributing one edge to a common ``plaquette'' as shown in \autoref{fig:four-edge-entanglement}. The entangled state of these four edges is:
\begin{equation}\label{eqn:sn-plaquette-state}
	\ket{\psi_{\text{plaq}}} = \ket{0000} + \ket{1111}
\end{equation}
which is precisely the GHZ state \eqref{eqn:ghz-state} on four qubits. This is a maximally entangled state which, when projected onto the gauge-invariant sector at each vertex, yields a specific superposition of the intertwiner basis states $ \ket{\Psi_1} $ and $ \ket{\Psi_2} $.

The full CZX ground state is the product of all plaquette states:
\begin{equation}\label{eqn:czx-full-gs}
	\ket{\Psi_{\text{CZX}}} = \prod_p \ket{\psi_p} = \prod_p \left( \ket{0000}_p + \ket{1111}_p \right)
\end{equation}
In the spin-network context, this state describes a quantum geometry in which the entanglement between neighboring vertices is of a very specific, short-range form. As argued in \autoref{sec:topo-order}, short-range entanglement is precisely the condition required for the emergence of a smooth, local macroscopic geometry. The CZX ground state therefore represents a quantum geometry with the correct entanglement structure to give rise to a semiclassical spacetime.

The on-site $ Z_2 $ symmetry of the CZX model acts as a Pauli-$ X $ gate on the two-dimensional code subspace $ \operatorname{span}\{\ket{0000}, \ket{1111}\} $, exchanging the two basis states \cite{Chen2010Local}. The intertwiner subspace at a 4-valent $ j = 1/2 $ spin-network vertex is likewise two-dimensional, spanned by $ \{\ket{\Phi_1}, \ket{\Phi_2}\} $, with the $ Z_2 $ time-reversal action realized by the sign flip of the signed volume operator: $ \hat{Q} \ket{\Phi_{1,2}} = \pm q_0 \ket{\Phi_{1,2}} $. The identification $ \ket{\Phi_1} \leftrightarrow \ket{0000} $, $ \ket{\Phi_2} \leftrightarrow \ket{1111} $ is an isomorphism of $ Z_2 $-invariant qubits: both subspaces carry a $ Z_2 $ automorphism that exchanges the two basis states, and it is via this isomorphism that the topological phase structure of the CZX model -- its phase diagram, topological protection, and edge mode spectrum -- applies directly to the $ Z_2 $ gauge theory on spin networks. The explicit computation verifying these statements is given in \autoref{subsec:czx-intertwiner}.

\subsection{The $ j = 1/2 $ Sector}\label{subsec:j-half-dominance}

The mapping established above relies on the restriction of all spin-network edges to the $ j = 1/2 $ representation of $ SU(2) $. This restriction requires physical justification. We adopt the following as a \textbf{working assumption}: at the Planck scale, the spin-network state is dominated by the $ j = 1/2 $ sector. The physical arguments supporting this assumption are given below; a definitive derivation from first principles remains an open problem in spin-foam dynamics \cite{Dona2020Numerical,Dona2022Asymptotics}.

In the full theory of Loop Quantum Gravity, edges can carry arbitrary representations $ j \in \{1/2, 1, 3/2, 2, \ldots \} $. The area associated with a single edge is given by \eqref{eqn:area-eigenvalue}: $ A_e = 8\pi\gamma l_p^2 \sqrt{j_e(j_e + 1)} $. The $ j = 1/2 $ representation corresponds to the \emph{smallest possible quantum of area} \eqref{eqn:area-minimum} -- the fundamental ``pixel'' of quantum geometry. In the deep Planck regime, the dominant contribution to the spin-network partition function comes from this lowest-spin sector for two reasons:

\begin{enumerate}
	\item \textbf{Amplitude suppression}: In spin-foam models, which provide the dynamics of spin networks, the vertex amplitudes generically suppress higher-spin configurations. The asymptotic behavior of the spin-foam vertex amplitude (such as the EPRL/FK amplitude) for large spins goes as $ A_v \sim j^{-\alpha} $ for some positive exponent $ \alpha $ \cite{Dona2020Numerical,Dona2022Asymptotics}, with the dominant contribution coming from the lowest available spin \cite{Perez2013The-Spin-Foam}. Note, however, that the precise value of $ \alpha $ and whether it is sufficient to ensure strict $ j = 1/2 $ dominance in all spin-foam models remains under active investigation. This is analogous to the dominance of the lowest Landau level in quantum Hall physics, where the lowest energy state governs the macroscopic behavior of the system.

	\item \textbf{Thermal suppression}: Working in Planck units ($\hbar = c = G = 1$, so $l_p=1$), we treat the area eigenvalue as the effective energy cost of an edge. At inverse temperatures $\beta > \beta_c \approx 2/\kappa$ (where $\kappa = 8\pi\gamma$), the thermal Boltzmann weight strongly favours $j=1/2$ edges, since they carry the smallest non-zero area and hence the lowest energy cost. A direct computation of the single-edge partition function shows that the $j=1/2$ contribution exceeds the combined contribution of all higher-$j$ representations for $\beta > \beta_c$ (see \autoref{subsec:czx-intertwiner} and the supplementary calculations). Restoring dimensions, the crossover temperature is $T_c \approx \kappa/(2k_B) = 4\pi\gamma l_p^2/k_B$, which is of order the Planck temperature for $\gamma \sim \mathcal{O}(1)$, so sub-Planck spin networks are dominated by $j=1/2$ edges.

	\item \textbf{Black hole entropy}: The requirement that LQG reproduce the Bekenstein-Hawking entropy $ S_{\rm BH} = A/4 $ selects $ j = 1/2 $ as the dominant representation for the Barbero-Immirzi parameter $ \gamma \approx 0.24 $ \cite{Domagala2004Black-hole,Meissner2004Black-hole}. The microscopic count of $ j = 1/2 $ edge states gives entropy $ S = N \ln 2 $, which matches $ A/4 $ in Planck units for this value of $ \gamma $. Higher-$ j $ contributions are subdominant in this computation.
\end{enumerate}

The $ j = 1/2 $ sector therefore provides the ``background'' around which the quantum geometry is organized. Higher-spin edges represent \emph{excitations} above this background. These excitations introduce corrections to the effective coupling $ K $ in the $ Z_2 $ gauge theory described in \autoref{sec:z2-action} but do not change the universality class of the confinement-deconfinement transition. This is because the $ Z_2 $ structure survives at any spin: time reversal always acts by complex conjugation of the Wigner matrices regardless of $ j $, and the volume operator retains its sign-changing property under this transformation for all representations. What changes with higher $ j $ is the dimension of the intertwiner space (and hence the number of ``flavors'' of the $ Z_2 $ degree of freedom), but the essential dichotomy between positive and negative volume -- the two orientations of time -- persists.

\subsection{The CZX SPT Phase as the Deconfined Phase}\label{subsec:spt-deconfined}

We can now state the central identification of this work. The ground state of the CZX model -- the symmetry-protected topological phase -- corresponds to the \textbf{deconfined phase} of the $ Z_2 $ gauge theory on the spin network.

A crucial distinction must be made explicit. The CZX model as originally defined \cite{Chen2011Symmetry,Chen2010Local} has a \emph{global} $ Z_2 $ symmetry: the same time-reversal operation acts uniformly on all sites. In our spin-network construction, following Chen and Vishwanath \cite{Chen2015Gauging}, this global symmetry is \emph{gauged} -- promoted to a local symmetry by allowing independent $ Z_2 $ transformations at each vertex. The general result, established in \cite{Levin2012Braiding,Haegeman2015Gauging}, is that gauging the protecting symmetry of an SPT phase yields a topological gauge theory in the deconfined phase: for $ G = Z_2 $, the gauged CZX state is the deconfined phase of the $ Z_2 $ lattice gauge theory, equivalent to the toric code ground state. The deconfined phase of our spin-network $ Z_2 $ gauge theory is therefore the \emph{gauged} CZX, not the ungauged SPT. This distinction is consequential for the order parameter: the ungauged SPT is detected by a non-local string order parameter, whereas the gauged (deconfined) phase is detected by the Wilson loop $ W(\gamma) $, which is gauge-invariant by construction and consistent with Elitzur's theorem.

In the SPT phase, the $ Z_2 $ symmetry (time reversal) is unbroken, but the state possesses non-trivial topological order that distinguishes it from a trivial product state. This topological order manifests as follows:
\begin{itemize}
	\item The state cannot be smoothly deformed into a trivial product state by local unitary transformations that respect the $ Z_2 $ symmetry.
	\item The entanglement entropy between a region and its complement exhibits a characteristic pattern determined by the SPT invariant.
	\item When the system has a boundary, protected surface excitations appear: gapless in the ungauged SPT, and a gapped topological order (the all-fermion toric code) in the gauged (deconfined) phase (discussed in the next subsection).
\end{itemize}

These properties correspond precisely to the characteristics of the deconfined phase of the $ Z_2 $ gauge theory:
\begin{itemize}
	\item The deconfined phase cannot be smoothly connected to the confined phase without passing through a phase transition.
	\item The Wilson loop order parameter $ W(\gamma) $ obeys a perimeter law, indicating long-range coherence of the $ Z_2 $ flux.
	\item The topological degeneracy of the ground state on a manifold with non-trivial topology is a hallmark of the deconfined phase.
\end{itemize}

The stability of the SPT phase under local perturbations that respect the $ Z_2 $ symmetry translates directly into the stability of the cosmological arrow of time. A local perturbation of the spin-network state -- such as the excitation of a finite number of vertices to higher-spin representations, or the introduction of matter degrees of freedom at isolated vertices -- cannot destroy the global coherence of the time orientation as long as the perturbation respects the $ Z_2 $ gauge symmetry and the spectral gap remains open.

A comment on dimensions is in order. The CZX model is defined on a two-dimensional square lattice and realizes a $ (2+1) $-dimensional SPT phase. Spin networks, by contrast, are graphs dual to three-dimensional spatial slices, so the effective $ Z_2 $ gauge theory derived in \autoref{sec:z2-action} lives in three spatial dimensions. The correspondence established here is therefore \emph{structural}: the on-site Hilbert space of the CZX model (four spin-$ 1/2 $s per site, \autoref{subsec:czx-intertwiner}) is identical to the intertwiner space of a 4-valent $ j = 1/2 $ spin-network vertex, and the $ Z_2 $ action is the same in both cases. The claim is not that the spin-network state is literally a 2D CZX ground state, but that it realizes the same universality class of $ Z_2 $ gauge theory. Three-dimensional bosonic SPT phases protected by $ Z_2^{\mathcal{T}} $ time-reversal symmetry are classified by $ H^4(Z_2^{\mathcal{T}}, U(1)_{\mathcal{T}}) \cong \mathbb{Z}_2 $ \cite{Vishwanath2013Physics,Kapustin2014Symmetry}, yielding exactly two phases: a trivial phase and a non-trivial phase. The deconfined phase of the 3D $ Z_2 $ gauge theory falls within this classification. A detailed matching of the spin-network phase to the appropriate $ \mathbb{Z}_2 $ entry is left for future work.

It is important to emphasize that we do \emph{not} claim this correspondence holds for arbitrary spin-network graphs. Rather, we expect it to apply to a representative class of regular spatial lattices that dominate the spin-network partition function at low energies. Just as the $j=1/2$ sector dominates because it carries the smallest area eigenvalue, regular lattice topologies (such as cubic or diamond lattices dual to a spatial foliation) are expected to dominate the gravitational path integral because they minimize the vertex-valency entropy and maximize the homogeneity of the spatial slice. Excitations away from regularity — higher-valency vertices, non-uniform edge lengths, or graph defects — act as perturbations that do not change the universality class of the $Z_2$ gauge theory. A systematic derivation of ``regular-lattice dominance'' from spin-foam dynamics remains an interesting open problem.

\subsection{Edge Modes and Matter Degrees of Freedom}\label{subsec:edge-modes}

A striking feature of SPT phases is the existence of protected gapless edge modes at boundaries or defect surfaces. In the CZX model, when the system is defined on a manifold with boundary, the boundary hosts gapless excitations that cannot be gapped out without either breaking the $ Z_2 $ symmetry or introducing topological order at the boundary \cite{Chen2011Symmetry,Chen2010Local}.

In the quantum gravity context, we conjecture that these edge modes correspond to fermionic matter degrees of freedom. The motivation for this conjecture is threefold:

\begin{enumerate}
	\item \textbf{Analogy with string-net models}: In Wen's string-net condensation framework \cite{Wen2002aQuantum,Levin2004String-net}, emergent fermions arise as endpoints of string operators in topologically ordered phases. The CZX model, being an SPT phase with a $ Z_2 $ symmetry, supports analogous excitations at its boundary.

	\item \textbf{The Dirac equation from LQG}: As shown in \autoref{sec:gr-time-reversal}, the sign of the lapse function directly enters the Dirac equation \eqref{eqn:reduced-dirac-eqn} for fermions propagating in a given background geometry. At a domain wall between regions of opposite time orientation (which corresponds to a boundary of the SPT phase), the lapse function changes sign, and the Dirac equation acquires non-trivial boundary conditions. These boundary conditions are expected to give rise to localized fermionic modes, analogous to the Jackiw-Rebbi mechanism for fermion zero modes at domain walls.

	\item \textbf{Spin-statistics}: The edge modes of the CZX model carry projective representations of the $ Z_2 $ symmetry, classified by the group cohomology $ H^2(Z_2, U(1)) \cong Z_2 $ \cite{Chen2011Symmetry}. In the quantum gravity setting, where the $ Z_2 $ corresponds to \emph{anti-unitary} time reversal, the relevant cohomology is that of the anti-unitary group $ Z_2^{\mc{T}} $ acting on $ U(1) $ with the complex-conjugation twist: $ H^2(Z_2^{\mc{T}}, U(1)_{\mc{T}}) \cong Z_2 $. The non-trivial class corresponds to $ \mc{T}^2 = -1 $, which is the defining property of fermionic (half-integer spin) degrees of freedom \cite{Chen2011Symmetry,Kapustin2014Symmetry}. This provides a cohomological rationale for why the boundary excitations of the gauged CZX phase in our spin-network context should be fermionic, though a complete derivation in the LQG setting remains an open problem.
\end{enumerate}

A rigorous demonstration that the edge modes satisfy the Dirac equation in an appropriate semiclassical limit is beyond the scope of the present work. However, the convergence of the three lines of argument above strongly suggests that the SPT structure of quantum geometry provides a natural mechanism for the emergence of fermionic matter from the topological properties of the spin-network state. We leave a detailed analysis for future work.

\section{$ Z_2 $ Action and Time Reversal in Spin Networks}\label{sec:z2-action}

Having established the structural correspondence between spin-network states and the CZX model in the previous section, we now formalize the introduction of a local $ Z_2 $ symmetry on the spin network and analyze its consequences. We will show that the effective theory of this $ Z_2 $ field is a lattice gauge theory whose phase structure naturally provides a mechanism for the emergence of a cosmological arrow of time.

\subsection{The $ Z_2 $ Gauge Field on Spin Networks}\label{subsec:z2-definition}

As established in \autoref{sec:gr-time-reversal}, the sign of the lapse function $ N(t) $ determines the local direction of time evolution. In the tetrad formulation, changing $ N(t) \rightarrow -N(t) $ is equivalent to changing the sign of the determinant of the tetrad field: $ \det({}^4 e) \rightarrow -\det({}^4 e) $. At the level of the spatial geometry encoded by a spin network, this corresponds to a sign flip of the densitized triad field $ \tilde E^a_i $ on the edges of the graph, as shown in \eqref{eqn:momentum-operator}.

We now introduce a $ Z_2 $ gauge field $ \sigma_e \in \{+1, -1\} $ living on the edges of the spin network graph $ \Gamma $. The action of this field on the triad is:
\begin{equation}\label{eqn:z2-triad-action}
	\tilde E^a_i(e) \rightarrow \sigma_e \, \tilde E^a_i(e)
\end{equation}
This transformation has two crucial properties. First, it preserves the spatial 3-metric, since $ h_{ab} = \tilde E^a_i \tilde E^b_j \delta^{ij} $ is quadratic in the triad and therefore invariant under the sign flip. Second, it changes the sign of the determinant of the triad field:
\begin{equation}\label{eqn:z2-det-flip}
	\det({}^3 e) \rightarrow \sigma_e^3 \, \det({}^3 e) = \sigma_e \, \det({}^3 e)
\end{equation}
where the last equality follows because $ \sigma_e^3 = \sigma_e $ for $ \sigma_e \in \{+1, -1\} $. Consequently, the volume of a region $ S $ of the spin network changes sign:
\begin{equation}\label{eqn:z2-volume-flip}
	\text{Vol}(S) = \int_S d^3 x \, \det({}^3 e) \rightarrow \sigma_S \, \text{Vol}(S)
\end{equation}
where $ \sigma_S $ denotes the collective effect of the $ Z_2 $ field in the region $ S $.

At the quantum level, the $ Z_2 $ transformation acts on the angular momentum operators living on the edges via $ \hat J_i \rightarrow \sigma_e \hat J_i $. For the $ j = 1/2 $ sector, the Pauli matrices transform as $ \sigma_k \rightarrow -\sigma_k $, which is precisely the action of time reversal on the spin degrees of freedom. The eigenvalues of the volume operator \eqref{eqn:volume-ops}, which involve the triple product $ \epsilon^{ijk} \hat J_i \hat J_j \hat J_k $, are cubic in the angular momentum operators and therefore change sign under this transformation.

The key insight, following Chen and Vishwanath \cite{Chen2015Gauging}, is that in the tensor network representation of the spin-network state, this $ Z_2 $ transformation can be implemented \emph{locally}. As shown in \autoref{subsec:mps-local-gauge}, time reversal on a subregion of the TNS leaves symmetry flux insertions only on the boundary of that region, with the matrix insertions $ M, M^{-1} $ canceling in the interior. The ``gauging'' of time reversal therefore consists of allowing the $ Z_2 $ field $ \sigma_e $ to take \emph{independent} values on each edge, and summing over all possible configurations of $ \{\sigma_e\} $. This is the content of promoting the global $ Z_2 $ symmetry to a local one.

\subsection{Effective Action for the $ Z_2 $ Field}\label{subsec:z2-effective-action}

Given that the $ Z_2 $ field $ \sigma_e $ lives on the edges of the spin network graph $ \Gamma $, and that gauge-invariant observables must be constructed from products of $ \sigma_e $ around closed loops, the most natural effective action is that of a $ Z_2 $ lattice gauge theory \cite{Wegner1971Duality,Kogut1979An-introduction}:
\begin{equation}\label{eqn:z2-gauge-action}
	S_{Z_2}[\sigma] = -K \sum_{p \in \Gamma^*} \prod_{e \in \partial p} \sigma_e
\end{equation}
where the sum runs over the plaquettes $ p $ of the graph (or equivalently the faces of the dual graph $ \Gamma^* $), and $ K $ is a dimensionless coupling constant. Each term in the sum is the product of the $ Z_2 $ field around a minimal closed loop -- a plaquette -- and is manifestly gauge-invariant under the local $ Z_2 $ transformation $ \sigma_e \rightarrow \tau_v \sigma_e \tau_{v'} $ at the endpoints $ v, v' $ of each edge. Gauge invariance follows explicitly: since $ \tau_v \in \{+1,-1\} $ satisfies $ \tau_v^2 = 1 $, the plaquette product $ \prod_{e \in \partial p} \sigma_e $ is unchanged because each internal vertex appears exactly twice in $ \partial p $, contributing a net factor of $ \tau_v^2 = 1 $.

The coupling constant $ K $ encodes the dynamics of the spin-network state and is, in principle, determined by the intertwiner amplitudes. For spin-network states dominated by the $ j = 1/2 $ sector, $ K $ is related to the relative weight between states with aligned and anti-aligned time orientations at neighboring vertices. In the language of spin-foam amplitudes \cite{Perez2013The-Spin-Foam,Dona2020Numerical,Dona2022Asymptotics}, the coupling $ K $ receives contributions from the vertex amplitude evaluated on configurations with and without $ Z_2 $ flux through the plaquettes.

The partition function for the $ Z_2 $ gauge theory on the spin network is then:
\begin{equation}\label{eqn:z2-partition-function}
	\mc{Z} = \sum_{\{\sigma_e = \pm 1\}} e^{-S_{Z_2}[\sigma]} \, \mc{Z}_{\text{geom}}[\sigma]
\end{equation}
where $ \mc{Z}_{\text{geom}}[\sigma] $ is the contribution from the geometric (spin-network) degrees of freedom in the background of a given $ Z_2 $ configuration. When the backreaction of the $ Z_2 $ field on the geometry can be neglected, the geometric factor is approximately constant and the dynamics is governed entirely by $ S_{Z_2} $.

\subsection{Phase Structure: Confinement and Deconfinement}\label{subsec:z2-phases}

The $ Z_2 $ lattice gauge theory in three spatial dimensions is one of the best studied models in mathematical physics \cite{Wegner1971Duality,Kogut1979An-introduction,Fradkin1979Phase}. Its phase structure is well established and contains two distinct phases separated by a phase transition at a critical coupling $ K_c $:

\begin{enumerate}
	\item \textbf{Confined phase} ($ K < K_c $): The $ Z_2 $ flux is disordered. The Wilson loop observable:
	\begin{equation}\label{eqn:wilson-loop-z2}
		W(\gamma) = \left\langle \prod_{e \in \gamma} \sigma_e \right\rangle
	\end{equation}
	obeys an \emph{area law}: $ W(\gamma) \sim e^{-c \cdot \text{Area}(\gamma)} $, where $ \text{Area}(\gamma) $ is the minimal area enclosed by the closed path $ \gamma $ on the graph. In this phase, the $ Z_2 $ flux through any macroscopic surface fluctuates wildly and there is no coherent assignment of time orientation over large distances. This is the \emph{pre-geometric phase}: a state of quantum gravitational ``foam'' in which the notion of a smooth spacetime with a well-defined arrow of time does not exist.

	\item \textbf{Deconfined phase} ($ K > K_c $): The $ Z_2 $ flux is ordered. The Wilson loop obeys a \emph{perimeter law}: $ W(\gamma) \sim e^{-c' \cdot \text{Perimeter}(\gamma)} $. In this phase, the time orientation is coherent over macroscopic distances -- a measurement of the local arrow of time at widely separated vertices of the spin network will yield the same result with a probability that approaches unity as the coupling increases. This is the phase corresponding to our observed semiclassical spacetime, in which the arrow of time is uniform on cosmological scales.
\end{enumerate}

A crucial remark is in order regarding Elitzur's theorem \cite{Elitzur1975Impossibility}. This theorem states that in a theory with a local gauge symmetry, no \emph{local} order parameter charged under the gauge group can acquire a non-zero expectation value. In our context, this means that $ \expect{\sigma_e} = 0 $ in both phases. This is \emph{not} a contradiction with our claim that the arrow of time emerges in the deconfined phase. The relevant order parameter is not the local field $ \sigma_e $ but the Wilson loop $ W(\gamma) $ defined in \eqref{eqn:wilson-loop-z2}, which is a \emph{non-local}, gauge-invariant observable. The transition between the confined and deconfined phases is a genuine phase transition detected by the change in the scaling behavior of $ W(\gamma) $ from area law to perimeter law, entirely consistent with Elitzur's theorem.

This distinction has a direct physical interpretation. We should not ask whether a single vertex of the spin network ``points forward or backward in time'' -- this is a gauge-dependent question with no physical meaning, just as asking for the value of the vector potential at a point has no physical meaning in electrodynamics. What \emph{is} physically meaningful is the \emph{relative} time orientation between vertices connected by a macroscopic path, which is precisely what the Wilson loop measures. In the deconfined phase, the Wilson loop tells us that widely separated vertices share the same time orientation, giving rise to a macroscopic arrow of time.

It is worth noting that the $ Z_2 $ lattice gauge theory possesses a deeper algebraic structure. As shown by Bais and Propitius \cite{Propitius1996Discrete}, discrete gauge theories in general admit a description in terms of quantum doubles (Hopf algebras), which provide a unified framework for understanding their excitation spectrum, fusion rules and braiding statistics. This Hopf algebraic structure of the $ Z_2 $ gauge theory on spin networks is expected to play an important role in a more complete treatment of the dynamics and will be explored in future work\footnote{The significance of Hopf algebras for spin network dynamics was recognized early on by Markopoulou \cite{Markopoulou1997Dual,Markopoulou2000Quantum}, though this perspective has not been widely pursued in the LQG community.}.

\subsection{The Cosmological Phase Transition}\label{subsec:cosmological-transition}

We are now in a position to state the central claim of this paper. The emergence of the cosmological arrow of time is identified with the \textbf{confinement-deconfinement transition} of the $ Z_2 $ gauge theory living on the primordial spin network.

In the deep Planck regime -- near the ``Big Bang'' -- the effective coupling $ K $ is small and the system is in the \emph{confined phase}. The $ Z_2 $ flux is maximally disordered: the time orientation fluctuates randomly from vertex to vertex, and there is no coherent notion of ``past'' and ``future''. In this phase the Wilson loop expectation value decays exponentially with the area enclosed, indicating that the relative time orientation between distant vertices is completely uncorrelated. The quantum state of geometry in this phase resembles a quantum gravitational foam with no semiclassical interpretation.

As the spin network evolves -- which in the language of spin foams corresponds to the expansion of the universe -- the effective coupling $ K $ grows\footnote{The precise mechanism by which $ K $ increases during cosmological evolution requires a detailed analysis of the spin-foam vertex amplitudes, which we leave for future work. However, the basic physical picture is that as the number of vertices in the spin network increases (corresponding to the expansion of space), the effective temperature of the $ Z_2 $ gauge theory decreases, driving the system towards the ordered phase.}. At the critical value $ K = K_c $, the system undergoes a phase transition into the \emph{deconfined phase}. Beyond this point the $ Z_2 $ flux becomes ordered: the Wilson loops obey a perimeter law, and the time orientation becomes coherent over arbitrarily large distances. This is the moment at which a macroscopic, cosmological arrow of time emerges.

Several features of this picture are physically attractive:

\begin{itemize}
	\item \textbf{Universality}: The confined-deconfined transition in the $Z_2$ gauge theory is second-order and belongs to the Ising universality class.\footnote{For the three-dimensional quantum $Z_2$ gauge theory at $T=0$ (relevant to the spin-foam path integral), Kramers--Wannier duality maps the transition to the three-dimensional quantum Ising model. For a purely classical three-dimensional treatment the dual exponents are those of the two-dimensional classical Ising model.} The critical exponents are universal and do not depend on the microscopic details of the spin-network state. This suggests that the emergence of the arrow of time is a robust phenomenon, insensitive to the precise form of the quantum gravity dynamics at the Planck scale.

	\item \textbf{Topological protection}: In the deconfined phase, the coherent time orientation is topologically protected. Local perturbations -- such as the creation of a small region with reversed time orientation -- are energetically suppressed and cannot destabilize the global arrow of time. This can be understood in the language of quantum error correction \cite{Dennis2002Topological}: the deconfined phase of the $ Z_2 $ gauge theory is equivalent to the three-dimensional toric code \cite{Dennis2002Topological}, which is a topological quantum error correcting code. Local ``errors'' (vertices with reversed time orientation) are automatically corrected by the topological order, preventing the formation of macroscopic domains of reversed time.

	\item \textbf{Domain walls}: The theory naturally accommodates the possibility of domain walls separating regions of opposite time orientation. Such domain walls would correspond to surfaces across which the $ Z_2 $ flux is concentrated, and their dynamics would be governed by an effective tension proportional to $ K - K_c $. In the deep deconfined phase ($ K \gg K_c $), these domain walls are heavy and confined, making regions of reversed time extremely rare and short-lived. However, near the phase transition ($ K \approx K_c $), domain walls proliferate, and the arrow of time becomes ill-defined -- as expected in the deep Planck regime.

	\item \textbf{Compatibility with the SPT framework}: As we argued in \autoref{sec:spt-lqg}, the spin-network state in the deconfined phase is in a symmetry-protected topological (SPT) phase -- specifically, the CZX state. The SPT order ensures that the state cannot be smoothly deformed into a trivial product state without either breaking the $ Z_2 $ symmetry or closing the spectral gap. This provides additional stability to the arrow of time beyond what is provided by the gauge theory alone.
\end{itemize}

To summarize: the arrow of time in our universe is not the result of a special initial condition or a boundary condition imposed by hand. It is the natural consequence of a phase transition in a $ Z_2 $ gauge theory living on the fundamental spin-network structure of quantum geometry. The confined phase represents the pre-geometric epoch where no arrow of time exists; the deconfined phase represents our observed semiclassical universe with its uniform cosmological arrow. The transition between these phases is a \emph{confinement-deconfinement transition} detected by the Wilson loop order parameter, fully consistent with Elitzur's theorem and the gauge-theoretic nature of the $ Z_2 $ field.

\section{Discussion}\label{sec:discussion}

As explained in \autoref{sec:gr-time-reversal} it is only in the tetrad (Palatini) formulation of general relativity that a change in the sign of the lapse function has any physical effects. Changing the sign of the lapse $ N(t) \rightarrow -N(t) $ causes the determinant of the \emph{tetrad} field to change sign $ \det ({}^4 e) \rightarrow - \det ({}^4 e) $. Spin networks are objects which, as such, live on purely spatial hypersurfaces and have no knowledge of the lapse function or other quantities which require information about the spacetime at other time slices. Therefore one might ask whether it is makes sense to speak about time reversal at the level of spin networks at all and even if it does, how can one possibly connect something which requires knowledge of time evolution (change of the sign of the lapse) to something which does not possess such knowledge (spin networks encoding the geometry of spatial hypersurfaces)? we will address both of the concerns in turn.

\subsection{Time Reversal in Spin Networks}

The first criticism is the following. Spin networks are purely spatial objects. The ``time evolution'' of spin networks is generated by the Hamiltonian constraint operator of LQG and results in four-dimensional simplicial complexes known as ``spin foams''. If one wishes to talk about time reversal then would not spin foams, rather than spin networks, be the appropriate arena?

It is true that a more complete discussion of time reversal in LQG should take place at the level of spin foams. However, firstly, such an analysis would be significantly more involved than the discussion presented in this work. Second, consider a simple system such as a single spin $ \ket{\uparrow} $ particle moving in ordinary flat space. Applying time reversal to this system simply flips the spin of the particle $ \ket{\downarrow} $. One can do so without having any knowledge of the past or future trajectory of the particle in question. Whether or not the entire worldline of the particle is symmetric under time reversal is a separate question and the possible types of trajectories (symmetric or anti-symmetric under time reversal) are used to classify different types of order in condensed matter systems such as in spin hall insulators.

In the same vein, one can always write down the results of applying time-reversal to any spin-network. One simply has to flip all the spins labeling the edges or, if one wishes, flip only the spins in a given finite subset of the graph network where we want time-reversal to act. Whether the spin-foam, resulting from the time evolution of the time reversed spin network state, is time symmetric or not is a separate question\footnote{However, one would expect that spin-networks which violate time-reversal symmetry would also generate spin-foam histories which are not time-reversal invariant.}. Thus, examining the properties of spin networks under the action of time-reversal is a completely valid line of investigation.

\subsection{Time Reversal, Spatial Volume and the Lapse}

The second criticism concerns being able to connect the sign of the lapse to spin networks. The lapse is a quantity which tells us how spatial slices at different times evolve into each other. Spin networks on the other hand are purely spatial objects with knowledge only of the spatial geometry in a given time slice. Why should one know about the other? It turns out that this criticism already has its resolution built into the structure of spin networks.

When we apply time-reversal to a portion of the spin network we flip the spins living on the edges which fall in that given region. Doing so is equivalent to changing the sign of the triad field: $ e_a^i \rightarrow - e_a^i $, because as explained in the discussion in the text centered around \eqref{eqn:momentum-operator}, the quantum operator corresponding to the triad field is the $ SU(2) $ angular momentum operator $ \hat J_i $.

Now if we change the sign of the triad field on each edge $e$, $E_e \rightarrow \sigma_e E_e$, this results in changing the sign of the local determinant $ \det({}^3 e)_v $ at each vertex as well\footnote{For an $ N \times N $ matrix $ \mb{A} $, negating all matrix elements gives $ \det(-\mb{A}) = (-1)^N \det(\mb{A}) $}. While the global volume scale is set by the 4-metric determinant, the relative orientation of the spatial frame is encoded in this per-edge $Z_2$ scaling. This implies that the oriented volume of the classical geometry described by the given spin network also changes sign locally:

\begin{equation}\label{eqn:volume-sign}
	\text{Vol}_v = \int_{v} d^3 x \, \det({}^3 e) \rightarrow \sigma_v \text{Vol}_v
\end{equation}
where $\sigma_v$ is the product of $Z_2$ fluxes meeting at vertex $v$.

Note, that this only happens for the triad and not for the tetrad! Therefore if we implement time-reversal on a spin-network in the manner described in this work the resulting \emph{four-dimensional} volume element will also undergo a change in sign $ \det({}^4 e) \rightarrow - \det({}^4 e) $. Since the determinant of the tetrad and triad fields are related to each other\footnote{This follows from the form \eqref{eqn:4-metric-adm} of the 4-metric $ g_{\mu\nu} $ written in terms of the lapse, shift and the 3-metric $ h_{ab} $, from which one can deduce that $ \sqrt{\det(g)} = N \sqrt{\det(h)} $.} by a factor of the lapse function $ \det({}^4 e) = N(t) \det({}^3 e) $, changing the sign of the determinant of the triad field is \emph{equivalent} to changing the sign of the lapse function.


\subsection{Elitzur's Theorem and the Nature of the Order Parameter}

A natural objection to the framework presented in this paper is that Elitzur's theorem \cite{Elitzur1975Impossibility} forbids the spontaneous breaking of local gauge symmetries. Since our $ Z_2 $ time-reversal symmetry has been promoted to a local gauge symmetry via the Chen-Vishwanath procedure, one might worry that no order parameter can distinguish between the two phases we have described.

This objection is addressed by the analysis in \autoref{subsec:z2-phases}. The key point is that the relevant order parameter is \emph{not} the local field $ \sigma_e $ (whose expectation value indeed vanishes in all phases, as Elitzur's theorem demands) but the Wilson loop $ W(\gamma) = \expect{\prod_{e \in \gamma} \sigma_e} $, which is a non-local, gauge-invariant observable. The confined and deconfined phases are sharply distinguished by the scaling behavior of $ W(\gamma) $: area law in the confined phase, perimeter law in the deconfined phase. This distinction is a genuine phase transition, fully consistent with Elitzur's theorem. For a careful analysis of gauge-invariant order parameters and the subtleties of the confinement-Higgs transition, see \cite{Greensite2020The-Higgs}.

It is also worth emphasizing that the Chen-Vishwanath ``gauging'' procedure is distinct from the standard Yang-Mills gauging. In the Yang-Mills case, promoting a global symmetry to a local one introduces new dynamical degrees of freedom (the gauge field) and associated gauge redundancy from the outset. In the TNS framework, the gauging consists of allowing local application of the symmetry to subregions of the tensor network, with the symmetry flux lines (domain walls) playing the role of the gauge field configurations. The sum over all possible domain wall configurations is the analogue of the functional integral over gauge field configurations. This distinction is important because it clarifies that the $ Z_2 $ gauge symmetry in our context arises from the structure of the tensor network representation rather than from a fundamental gauge principle, and Elitzur's theorem, while still applicable, constrains only the local observables.

\subsection{Quantum Error Correction and Stability of the Arrow of Time}

The identification of the deconfined phase with a topological quantum error correcting code \cite{Dennis2002Topological,Mildenberger2024Probing,Homeier2023Quantum,Jahn2021Holographic,Steinberg2023Holographic} provides an appealing physical picture for the stability of the cosmological arrow of time. In the deconfined phase, the $ Z_2 $ gauge theory on the spin network is equivalent to Kitaev's toric code (strictly speaking, to the three-dimensional generalisation thereof \cite{Dennis2002Topological}), which is known to be a topological quantum memory: quantum information stored in its ground state is protected against local errors (bit flips and phase flips) up to a critical error rate.

Translated into the language of quantum geometry, this means that the coherent time orientation of the universe is protected against local ``errors'' -- isolated vertices or small regions where the local arrow of time might be reversed due to quantum fluctuations. As long as the density of such errors remains below a critical threshold (set by the coupling $ K $ being sufficiently above $ K_c $), the topological order of the deconfined phase ensures that these errors are corrected and the global arrow of time is preserved.

This provides a satisfying resolution to the question of why we do not observe pockets of reversed time in our universe: such pockets are the analogue of uncorrectable errors in the toric code, and in the deep deconfined phase they are exponentially suppressed.

\subsection{Algebraic Structure and Future Directions}

The $ Z_2 $ lattice gauge theory on the spin network possesses a rich algebraic structure that we have only touched upon in this work. As shown by Bais and Propitius \cite{Propitius1996Discrete}, discrete gauge theories admit a description in terms of quantum doubles -- Hopf algebras whose representation theory classifies the excitation spectrum, fusion rules and braiding statistics of the theory. The quantum double of $ Z_2 $ is a well-studied mathematical object, and its application to the gauge theory on spin networks is expected to yield additional insights into the dynamics of the confinement-deconfinement transition and the nature of the excitations above the ground state.

A particularly intriguing direction concerns the relationship between the Hopf algebraic structure of the $ Z_2 $ gauge theory and the dynamics of spin networks themselves. The evolution of spin networks under the Hamiltonian constraint of LQG can be formulated in terms of local moves (Pachner moves) on the underlying graph, and these moves have been shown to possess a Hopf algebraic structure \cite{Markopoulou1997Dual,Markopoulou2000Quantum}. Whether the Hopf algebra governing the spin network dynamics is compatible with or constrained by the quantum double structure of the $ Z_2 $ gauge theory is an open question with potentially deep implications for the consistency of the framework presented here.

Other directions for future work include:
\begin{itemize}
	\item A detailed derivation of the coupling constant $ K $ from spin-foam vertex amplitudes, which would allow the critical coupling $ K_c $ to be related to physical quantities such as the Planck temperature or the cosmological constant.
	\item An extension of the analysis to higher-spin sectors and an investigation of how higher-spin fluctuations modify the phase diagram of the $ Z_2 $ gauge theory.
	\item A rigorous analysis of whether the surface excitations of the SPT phase satisfy the Dirac equation in the semiclassical limit, which would establish the conjecture that fermionic matter emerges from the topological structure of quantum geometry.
	\item An investigation of the spin-foam history generated by the confinement-deconfinement transition, which would provide a four-dimensional description of the cosmological phase transition that gives rise to the arrow of time. The coarse-graining flow of spin-foam intertwiners \cite{Dittrich2016Coarse} provides a natural framework for studying the phase structure of the effective $ Z_2 $ gauge theory across scales.
	\item An exploration of whether the confinement-deconfinement transition can be understood in the framework of emergent spacetime from entanglement \cite{Cao2018Bulk}, and how the relational notion of time \cite{Hohn2021The-Trinity} interfaces with the arrow of time arising from the gauge-theoretic phase transition described here.
\end{itemize}

\section*{Acknowledgments}

The author wishes to acknowledge the Inter-University Centre for Astronomy and Astrophysics (IUCAA), Pune, India, where this work was initiated during a Visiting Associateship in 2018.

The author used Claude (Anthropic) as an AI writing assistant during the 2026 revision of this manuscript. Its use was limited to formalization of theoretical arguments, LaTeX editing, bibliography management, and error correction. All scientific content, physical arguments, and conclusions are the author's own. A detailed account of AI-assisted contributions, with reference to the publicly available git commit history, accompanies the cover letter.


\printbibliography[title={Bibliography}]

\appendix

\renewcommand{\theequation}{\Alph{section}.\arabic{equation}}

\numberwithin{equation}{section}

\section{Time Reversal of Quantum Systems}\label{sec:time-reversal}

In classical mechanics the action of time-reversal symmetry on physical observables can be defined as shown in \autoref{tbl:trs-classical}


Under time-reversal the Poisson brackets change sign:
\begin{equation}\label{eqn:classical-poisson}
\left\{ x_i, p_j \right\} = \delta_{ij} \rightarrow \left\{ x'_i, p'_j \right\} = -\delta_{ij}
\end{equation}
where $ \vect{x'} = \mc{T} \vect{x} = \vect{x} $, $ \vect{p'} = \mc{T} \vect{p} = - \vect{p} $ and $ \mc{T} $ is the generator of time-reversal. However, the classical equations of motion remain invariant:
\begin{equation}\label{eqn:classical-eom}
\vect{\dot{x}'} = \frac{\partial H}{\partial \vect{p'}}; \quad \vect{\dot{p}'} = - \frac{\partial H}{\partial \vect{x'}}
\end{equation}
In quantum mechanics the situation is quite different and this difference can ultimately be traced back to the appearance of the imaginary $ i = \sqrt{-1} $ in the canonical commutation relations and the Schr\"{o}dinger equation. If, proceeding by analogy with the classical case, we implement time reversal for quantum operators in the expected way:
\begin{equation}\label{eqn:trs-quantum}
\vect{\hat{x}} \rightarrow \vect{\hat{x}}; \quad \vect{\hat{p}} \rightarrow -\vect{\hat{p}}
\end{equation}
the quantum commutators pick up a negative sign, similar to what happens in the classical case \eqref{eqn:classical-poisson}:
\begin{equation}\label{eqn:quantum-poisson}
\left[ \hat{x}_i, \hat{p}_j \right] = i \hbar \delta_{ij} \rightarrow \left[\hat{x}'_i, \hat{p}'_j \right] = - i \hbar \delta_{ij}
\end{equation}

\begin{figure}[htbp]
	\centering
	\begin{tabular}{|p{5cm}|p{6cm}|}
		\toprule
		Physical Observable & Under time reversal \\
		\midrule
		Position & \( \vect{x} \rightarrow \vect{x}\) \\
		Time & \( t \rightarrow -t\) \\
		Linear Momentum & \(\vect{p} \rightarrow -\vect{p} = d\vect{x}/d(-t) = - d\vect{x}/dt\) \\
		Angular Momentum & \( \vect{l} = \vect{x} \times \vect{p} \rightarrow -\vect{l} \)  \\
		Energy/Hamiltonian & \( H(\vect{x},\vect{p}) \rightarrow H(\vect{x}, -\vect{p}) \)  \\
		Electric Field & \( E^i \rightarrow -E^i\) \\
		Magnetic Field & \( B^i \rightarrow - B^i \) \\
		\bottomrule
	\end{tabular}
	\caption{Action of time-reversal symmetry on classical observables}
	\label{tbl:trs-classical}
\end{figure}

However, the equation of motion is no longer invariant:
\begin{equation}\label{eqn:trs-schrodinger}
i \hbar \frac{\partial \psi}{\partial t} =  \hat{H} \psi \rightarrow i \hbar \frac{\partial \psi}{\partial t'} =  - \hat{H} \psi
\end{equation}
where $ t' = -t $. This can be resolved, however, if following Wigner we define the quantum time reversal operator $ \mc{T} $ to act as follows on any quantum state as follows:
\begin{equation}\label{eqn:wigner-time-reversal}
\mc{T} ( c_1 \psi_1 + c_2 \psi_2 ) = c_1^* \mc{T} \psi_1 + c_2^* \mc{T} \psi_2
\end{equation}
Such an operator is called \emph{anti-linear} or \emph{anti-unitary}. Acting with this operator on both sides of the Schr\"{o}dinger equation we obtain:
\begin{equation}\label{eqn:trs-schrodinger-2}
\mc{T} \left( i \hbar \frac{\partial \psi}{\partial t} \right) = i \hbar \frac{\partial (\mc{T} \psi)}{\partial t'} = H (\mc{T} \psi)
\end{equation}
where in the second step we have used the fact that $ i \rightarrow -i $ under the action of $ \mc{T} $ and that $ t' = -t $. Consequently, if $ \psi $ satisfies the Schr\"{o}dinger equation, then $ \mc{T} \psi $ satisfies the time-reversed Schr\"{o}dinger equation. A second benefit of this definition is that under $ \mc{T} $, the quantum commutators no longer change sign:
\begin{equation}\label{eqn:quantum-poisson-2}
\mc{T} \left(\left[ \hat{x}_i, \hat{p}_j \right]\right) \mc{T}^{-1} = \mc{T} i \mc{T}^{-1} \hbar \delta_{ij} \rightarrow \left[\hat{x}'_i, \hat{p}'_j \right]= i \hbar \delta_{ij}
\end{equation}
where we have used the fact that an operator $ \mc{O} $ transforms as $ U \mc{O} U^{-1} $ under the action of any symmetry $ U $. As a sidenote let us mention that while the Wigner definition of time-reversal leads to an invariant equation of motion \emph{and} invariant commutators in the quantum case, the definition of time-reversal in the classical case leaves the equation of motion invariant, but not the Poisson brackets. This dichotomy between the classical and quantum cases deserves further study, but we postpone that for a later occasion.

\section{ADM Splitting and Tetrads}\label{sec:adm-tetrad}

Recall that in the ADM formulation the four-dimensional spacetime metric can be written in terms of a lapse function $ N(x^\mu) $ and a spatial shift vector field $ N^i(\vec{x}) $:
\begin{equation}\label{eqn:adm-metric}
	{}^4 g_{\mu\nu}  = (- N^2 + N^i N_i) dt^2 + 2 N_i dt dx^i + h_{ij} dx^i dx^j
\end{equation}
where $ i \in \{1,2,3\} $, $ \mu,\nu \in \{0,1,2,3\} $, $ N(x^\mu) $ is the lapse function, $ N^i(\vec{x}) $ is the shift vector field with $ \vec{x} \equiv (x,y,z) $ being a purely spatial co-ordinate; and $ h_{ij} $ is the internal metric of the constant time surfaces. This metric can be expressed in terms of a tetrad (``vielbein'' or local Lorentz frame) as:
\[ g_{\mu\nu} = e_\mu^I e_\nu^J \eta_{IJ} \]
where $ \eta_{IJ} = \text{diag}(-1,1,1,1) $ is the flat Minkowski metric. In the so-called time gauge the time-time component of the above metric \eqref{eqn:adm-metric} can be used to determine the tetrad components for $ e_0^I $:
\begin{equation}\label{eqn:timelike-tetrad}
	e_0^0 = N(x^\mu) dt; \quad e_0^i = N^i(\vec{x}) dx_i
\end{equation}
with $ i \in \{1,2,3\} $, and all other components of the tetrad set to zero. There is no summation over repeated indices in this expression. The remaining components of the tetrad field require knowledge of the internal metric and also a choice of local Lorentz frame. We can always choose a set of coordinates, locally, in which the spatial metric becomes diagonal:
\begin{equation}\label{eqn:diag-metric}
	h_{ij} \equiv \text{diag}(h_1, h_2, h_3)
\end{equation}
The tetrad components for $ e_i^I $, $ i \in \{1,2,3\} $ then can be chosen to be:
\begin{equation}\label{eqn:diag-tetrad-metric}
	e_i^i = \sqrt{h_i},
\end{equation}
with the rest being set to zero. When the spatial metric is flat and the shift vector is set to zero, the only non-zero components of the tetrad field will be given by:
\[ e_0^0 = N(t) dt; \quad e_i^0 = d x^i \]

\section{Time Reversal in Classical GR}\label{sec:gr-time-reversal}

In \cite{Christodoulou2012How-to-detect,Rovelli2012Discrete} the question of discrete symmetries in LQG is considered and it is shown that in the presence of fermions the sign of the lapse function can, in principle, be detected. This can be seen as follows. Consider a geometry where the shift vector is identically zero, the lapse vector is a function only of the time coordinate and the spatial metric is the flat Euclidean metric. In this case the tetrad components become \footnote{For a reminder of how the lapse function can be related to tetrad components we refer the reader to \autoref{sec:adm-tetrad}}:
\begin{equation}\label{eqn:lapse-tetrad}
e_0^0 = N(t) dt; \quad e_i^0 = d x^i,
\end{equation}

Now, when fermions are present both the tetrad field and the gravitational spin connection $ \omega_\mu^{IJ} $ play  a role in determining their dynamics via the Dirac equation:
\begin{equation}\label{eqn:massive-dirac-eqn}
\gamma_I e_\mu^I \left[ \partial^\mu + \frac{i}{2} \omega^\mu_{KL} \tau^{KL} \right] \psi + m \id \psi = 0,
\end{equation}
where $ \{\gamma_I\} $ are the Dirac matrices, $ \tau^{KL} $ are the generators of the Lorentz group and $ \id $ is the identity matrix included here for clarity. The spin connection for the given geometry \eqref{eqn:lapse-tetrad} vanishes. Assuming that $ \psi $ is a function only of time, the above expression reduces to:
\begin{equation}\label{eqn:reduced-dirac-eqn}
\gamma_0 N(t) \partial_t \psi + m \id \psi = 0.
\end{equation}
This expression has a straightforward solution given by:
\begin{equation}\label{eqn:dirac-solution}
\psi(t) = e^{i m f(t) } \psi_0
\end{equation}
where the phase $ f(t) $ is given by:
\begin{equation}\label{eqn:dirac-phase}
f(t) = \int_{t_0}^{t} N(t) dt
\end{equation}
Since the phase in \eqref{eqn:dirac-solution} is a function of the lapse one can, in principle, design an interference experiment involving two fermions with different masses whose result would depend on the sign of the lapse function. Alternatively one can describe the behavior of the fermion field in a region of negative lapse as evolving ``backwards in time''. Since, in principle, nothing forbids the lapse from taking on negative values this leads us to the possibility of the existence of ``pockets'' of geometry where time ``runs backwards'' as compared to the surrounding regions.

In terms of the action principle, the ambiguity in the sign of the lapse can be seen as arising due to an ambiguity in the determinant of the tetrad matrix. The usual Einstein-Hilbert action in the metric formalism is:
\begin{equation}\label{eqn:eh-metric-action}
S_{EH} = \frac{1}{\kappa} \int d^4x \sqrt{-g}~R + \int d^4 x \sqrt{-g}~\mc{L}_{matter},
\end{equation}
where $ \kappa = 8 \pi G_N $, $ R $ is the Ricci scalar and $ g = \det(g) $ is the determinant of the 4-metric. Expressed in terms of the tetrad and connection variables \cite[Sec 4.5.3]{Vaid2016LQG-for-the-Bewildered} the action becomes:
\begin{align}
S_{P}[e, \omega] & =\frac{1}{2 \kappa} \int d^{4} x \star\left(e^{I} \wedge e^{J}\right) \wedge F^{K L} \epsilon_{I J K L} \nonumber \\
& = \frac{1}{4 \kappa} \int d^{4} x \epsilon^{\mu \nu \alpha \beta} \epsilon_{I J K L} e_{\mu}^{I} e_{\nu}^{J} F_{\alpha \beta}^{K L} \nonumber \\
& = \frac{1}{4 \kappa} \int d^{4} x (\det {}^4 e) R[e]
\end{align}
where $ F_{\alpha \beta}^{K L} $ is the curvature of the spin-connection $ \omega_\mu^{IJ} $ and $ \det({}^4 e) $ is the determinant of the \emph{tetrad} field. From the form of this action and the relation \eqref{eqn:lapse-tetrad}, which implies that $ \det e \propto \text{sign}(N(t))$, we can see that the sign of the tetrad action $ S_P $ depends on the sign of the lapse. The sign of the lapse, however, has no effect on the sign of the determinant of the 4-metric and therefore does not affect the action given in \eqref{eqn:eh-metric-action}. One can see this more concretely by noting that in terms of the ADM lapse and shift fields, the 4-metric can be expressed as follows \cite[Eq (4.14)]{Vaid2016LQG-for-the-Bewildered}:
\begin{equation}\label{eqn:4-metric-adm}
g_{\mu\nu} = \begin{pmatrix}
-N^2 + N^a N_a & \mb{N} \\
\mb{N}^T	   & h_{ab}
\end{pmatrix}
\end{equation}
where $ \mb{N} $ is the spatial shift vector field, $ N(t) $ is the lapse function and $ h_{ab} $ is the metric on the spatial hypersurfaces. Since only square of the lapse occurs in the 4-metric, the sign of the lapse does not affect its determinant $ \sqrt{-\det{g}} $. Therefore, it is only in the tetrad-connection formalism that a \emph{local} change in the sense of time-evolution and represented by a change in the sign of the lapse can have an effect on the dynamics.

\section{Quantum States of Geometry}\label{sec:quantum-geometry}

The key point of this paper is that time-reversal symmetry can be made into a local gauge field acting on quantum states of spacetime geometry. These states of quantum geometry, known as ``spin networks'', originate in an approach towards a theory of quantum gravity known as \emph{loop quantum gravity} \cite{Rovelli2014Covariant, Vaid2016LQG-for-the-Bewildered}. Here we outline the key steps leading from classical gravity to its quantization in terms of spin-networks. The notation rules are as follows: lower case greek letters are $ 3+1 $ dimensional spacetime indices $ \alpha, \beta, \ldots \in \{0,1,2,3\} $, upper case latin letters are $ \mf{so}(3,1) $ Lie-algebra indices $ I, J, \ldots \in \{0,1,2,3\} $, lower case latin letters from the beginning of the alphabet are three-dimensional spatial indices $ a,b,c \in \{1,2,3\} $ and those from the middle of the alphabet are $ \mf{su}(2) $ Lie algebra valued indices $ i,j,k \in \{1,2,3\} $.

We will divide the process into several parts: the first part describes the path from the usual metric theory of gravity to the connection formalism and the form of the ADM constraints; the second describes the quantization strategy and how we obtain graph states as the candidate quantum states of geometry; the third explains how to impose the Gauss constraint to obtain gauge invariant states of quantum geometry; and the fourth elaborates on how the intertwiner (or vertex) degrees of freedom come about from the imposition of gauge invariance.

\subsection*{From metric to connection variables}
	\begin{enumerate}[label=\arabic*.]
		\item Starting with the usual Einstein-Hilbert action in the 1\supersc{st} order formalism:
		\begin{equation}
		S_{EH} = \frac{1}{\kappa} \int d^4x \sqrt{-g}~R[g,\Gamma],
		\end{equation}
		where $ \kappa = 8 \pi G_N $, $ R $ is the Ricci scalar and $ g = \det(g) $ is the determinant of the 4-metric, we perform a canonical transformation \cite{Ashtekar1986New-Variables,Ashtekar1987New-Hamiltonian} from the four-metric and Christoffel connection being our dynamical variables $ \{g_{\mu\nu}, \Gamma_{\alpha\beta}^\delta \} $ to a new set of variables $ \{ e_\mu^I, A_\mu^{IJ} \} $ (known as the Ashtekar or Ashetkar-Barbero variables), where $ e_\mu^I $ is a tetrad field (choice of local Lorentz frame) and $ A_\mu^{IJ} $ is a $ \mf{so}(3,1) $ Lorentz gauge connection.
		\item In the new variables the Einstein-Hilbert action can be written in the form:
		\begin{align}\label{eqn:eh-tetrad-action}
		S_{P}[e, A] & =\frac{1}{2 \kappa} \int d^{4} x \star\left(e^{I} \wedge e^{J}\right) \wedge F^{K L} \epsilon_{I J K L} \nonumber \\
		& = \frac{1}{4 \kappa} \int d^{4} x \epsilon^{\mu \nu \alpha \beta} \epsilon_{I J K L} e_{\mu}^{I} e_{\nu}^{J} F_{\alpha \beta}^{K L}
		\end{align}
		where $ F_{\alpha \beta}^{K L} $ is the curvature of the spin-connection $ A_\mu^{IJ} $.
		\item Next we perform a $ 3+1 $ split of spacetime following the usual ADM (Arnowitt-Deser-Misner) prescription which yields the following expression for the Einstein-Hilbert-Palatini action:
		\begin{equation}\label{eqn:eh-tetrad-adm}
		S_{P}= \int d t \int_M d^{3} x\left(E^\mu_{IJ} \dot A_\mu^{IJ} - N \mathcal{H} + N_{a} \mathcal{C}^{a} + \Lambda_{IJ} \mc{G}^{IJ} \right)
		\end{equation}
		where $ A_\mu^{IJ} $ is the configuration variable, $ \dot A_\mu^{IJ} $ is the generalized ``velocity'' and:
		\begin{equation}\label{eqn:adm-momentum}
		E^\mu_{IJ} = \frac{1}{2} \epsilon_{IJKL} \epsilon^{abc} e_a^K e_b^L
		\end{equation}
		is the momentum conjugate to $ A $; $ \epsilon^{abc} $ is the anti-symmetric tensor living only on the spatial slices. The inner integral is over the spatial 3-manifolds $ M $ which are the ``leaves'' of our $ 3+1 $ decomposition.
		\item One can now perform a Legendre transformation to obtain the Hamiltonian for general relativity in connection variables:
		\begin{equation}\label{eqn:tetrad-hamiltonian}
		H_{LQG} = N \mathcal{H} - N_{a} \mathcal{C}^{a} - \Lambda_{IJ} \mc{G}^{IJ}
		\end{equation}
		Note that, since $ N $, $ N^a $ and $ \Lambda_i $ in \eqref{eqn:eh-tetrad-adm} are all Lagrange multipliers, the associated terms $ \mc{H} $, $ \mc{C}^a $ and $ \mc{G}^i $ are all constraints, \ie they are functions which have to vanish on the space of solutions. $ N $ is known as ``lapse'' function, $ N^a $ is the ``shift'' vector field. In contrast to the usual ADM prescription, in the connection formulation we get an additional constraint $ \Lambda_{IJ} \mc{G}^{IJ} $, where $ \mc{G}^{IJ} $ has the form:
		\begin{equation}\label{eqn:gauss-constraint}
		\mc{G}_{IJ} = \mc{D}_a E^a_{IJ} \equiv 0
		\end{equation}
		and is known as the ``Gauss'' constraint.
		\item Here the classical variables are a Lorentz $ \mf{so}(3,1) $ connection $ \omega_\mu^{IJ} $ and an ``electric field'' $ E^\mu_{IJ} $ of the Yang-Mills type. When restricted to the spatial manifold these variables turn into a $ \mf{su(2)} $ connection $ A_a^i $ and a densitized triad field $ \tilde E^a_i = \sqrt{\det(h)} e^a_i$, where $ h_{ab} $ is the 3-metric on $ M $ and $ e^a_i $ is an orthonomal vector field spanning the local tangent space $ T_p (M) $ at any point of M. The Gauss constraint \eqref{eqn:gauss-constraint} can then be written in the form:
		\begin{subequations}
			\begin{align}
			\mc{G}_i = \mc{D}_a \tilde E^a_{i} & = 0 \label{eqn:gauss-constraint-v2} \\
			\Rightarrow \oint_{S^2} E^a_i n_a d^2 x & = 0 \label{eqn:gauss-constraint-v3}
			\end{align}
		\end{subequations}
		where in the second line we have integrated the differential form of the Gauss constraint over a closed two-dimensional surface of spherical topology with $ n_a $ being the normal to the area element at any point on the surface. This, then, becomes the form of ``Gauss' Law'' which says that - in the absence of any charged matter - the total electric flux through any closed surface is zero. We shall see that this statement is crucial for constructing the physical states of quantum geometry.
	\end{enumerate}

\subsection*{From connection variables to graph states}
	\begin{enumerate}[label=\arabic*.]
		\item Quantization proceeds by using the intuition gained from quantizing gauge theories of the Yang-Mills types. In Yang-Mills theories, rather constructing operators living in $ L^2(M) $ (where $ M $ is the spatial manifold on which our quantum fields live) one constructs gauge-invariant observables of the Wilson type and promotes these to operators in the quantum theory. For the connection variable this means using as the classical variable the holonomies along a Wilson loop:
		\begin{equation}\label{eqn:wilson-loop}
		h(A) = \int_\gamma dx^\mu A_\mu^I(x) \tau_I		\end{equation}
		\item The key difference between loop quantum gravity and other theories is that in LQG the background spacetime is \emph{itself} a quantum entity and therefore while performing the quantization we do not assume the presence of a pre-existing background geometry. Thus, while constructing the quantum phase space we should be careful to ensure that we work only with background independent operators. We can do so by \cite[Sec 3.2]{Ashtekar2004Background} using holonomies to probe the background geometry, however, since we need to probe the geometry over an extended region and not only along a single curve, instead of individual curves we use arbitrary graphs with a holonomy $ h_e(A) $ attached to each edge. The quantum configuration space associated with a given graph $ \Gamma $, with $ L $ edges becomes:
		\begin{equation}\label{eqn:lqg-config-space}
		\Psi_\Gamma(A) = \psi(h_{e_1}(A), \ldots, h_{e_L}(A))
		\end{equation}
		where $ \psi $ is function on $ L $ copies of the gauge group $ G $ of the theory which in our case happens to be $ SU(2) $.
		\item Such functions are known as ``cylindrical'' functions. We will write the space of cylindrical functions \wrt a graph $ \Gamma $ as $ \Cyl_\Gamma $. An inner-product on $ \Cyl_\Gamma $ can be given using the Haar measure on $ SU(2) $:
		\begin{equation}\label{eqn:lqg-inner-product}
		\innerp{\Psi_1}{\Psi_2}_\Gamma := \int_{G^L} d\mu_L \bar\psi_1 \psi_2
		\end{equation}
		\item Now, states defined on a single graph will naturally represent only a small sector of all possible states on the space of connections $ \mc{A} $. As shown in \cite{Ashtekar1994A-Manifestly} the above measure can be extended to a space of ``generalized connections'' $ \bar{\mc{A}} $ which consists of states living on \emph{arbitrary} graphs.
		\item Using the inner product \eqref{eqn:lqg-inner-product} the Hilbert space for a single graph $ \Gamma $ becomes:
		\begin{equation}\label{eqn:graph-hilbert-space-v1}
		\mc{H}_\Gamma = L^2(SU(2)^L),
		\end{equation}
		\ie the space of square integrable functions living on $ L $ copies of $ SU(2) $.
		\item we can use the Peter-Weyl theorem to perform a sort of Fourier transform on states living in $ \mc{H}_\Gamma $ and write them in terms of spin-labels $ \{j_{e_i}\} $ associated with each edge rather than with function on the group manifold:
		\begin{namedthm*}{Peter-Weyl Theorem}
			The matrix elements of the Wigner representation matrices of a compact Lie group are orthogonal \wrt the Haar measure on the group:
			\begin{equation}\label{eqn:peter-weyl}
			\int d\mu(g) ~\overline{D^{j'}_{m'n'}(g)}~D^j_{mn}(g)  = \frac{1}{d_j} \delta^{jj'} \delta_{mm'} \delta_{nn'},
			\end{equation}
		\end{namedthm*}
		where $ d\mu(g) $ is the Haar measure, $ d_j = (2j + 1) $ is the dimension of the $ j^{\text{th}} $ representation of $ SU(2) $ and $ D^j_{mn}(g) $ representation matrix of the element $ g \in G $. This theorem implies that any square integrable function $ f(g) $ on $ G $ can be expressed as a linear sum over the group representation matrices:
		\begin{equation}\label{eqn:group-fourier}
		f(g) = \sum_{j;mn} f^{mn}_j D^j_{mn}(g)
		\end{equation}
		Each representation matrix is $ (2j+1) \times (2j+1) $ dimensional and can therefore be thought of \cite[Sec 5.2]{Rovelli2014Covariant} as a map from the Hilbert space $ \mc{H}_j $ of a spin $ j $ particle to itself:
		\begin{equation}\label{eqn:su2-map}
		D^j : \mc{H}_j \rightarrow \mc{H}_j
		\end{equation}
		Each such matrix can therefore also be viewed as an element of $ \mc{H}_j \otimes \mc{H}_j $. Consequently the space $ L^2(G) $ can be decomposed as a direct sum over all possible representations of $ SU(2) $:
		\begin{equation}\label{eqn:su2-decomp}
		L^2(G) = \oplus_j (\mc{H}_j \otimes \mc{H}_j)
		\end{equation}
		\textbf{Note:} this implies that with each edge there are associated \emph{two} copies of the Hilbert space of a spin $ j $ particle. This fact will be crucial going forwards.
		\item Each graph $ \Gamma $ has $ L $ edges with a total Hilbert space given by \eqref{eqn:graph-hilbert-space-v1}. Using \eqref{eqn:su2-decomp} this space can then be written as a tensor product over the the Hilbert space of each edge:
		\begin{equation}\label{eqn:graph-hilbert-space-v2}
		\mc{H}_\Gamma = L^2(SU(2)^N) = \otimes_e \left[ \oplus_{j_e} (\mc{H}_{j_e} \otimes \mc{H}_{j_e} )\right]
		\end{equation}
		Let $ e_{vv'} $ be the (directed) edge connecting the two vertices labeled $ v $ and $ v' $ with $ v $ being the source of the edge and $ v' $ the target. We associate one of the two Hilbert spaces living on $ e_{vv'} $ with $ v $ and the other with $ v' $. The Hilbert space of a given edge can thus be written as:
		\begin{equation}\label{eqn:edge-hilbert-space-v1}
		\mc{H}_{e} = \oplus_{j_e} (\mc{H}^{e,v}_{j_e} \otimes \mc{H}^{e,v'}_{j_e}),
		\end{equation}
		keeping in mind that each $ j $ is associated with a particular edge.
		Equivalently we can label the source vertex of the edge $ e $ as $ v \equiv v_{e+} $ and the target vertex as $ v' \equiv v_{e-} $. Then the above equation becomes:
		\begin{equation}\label{eqn:edge-hilbert-space-v2}
		\mc{H}_{e} = \oplus_{j_e} (\mc{H}^{v_{e+}}_{j_e} \otimes \mc{H}^{v_{e-}}_{j_e})
		\end{equation}
		\item Using this notation we can rewrite the Peter-Weyl decomposition a little differently by grouping together the Hilbert spaces associated with each vertex rather than the Hilbert spaces associated with each edge. The Hilbert space associated with a single \emph{vertex}, becomes:
		\begin{equation}\label{eqn:vertex-hilbert-space}
		\mc{H}_v = \underset{\{j_e\}}{\bigoplus} \left[ \underset{{e^+\in E^+_v}}{\bigotimes} \mc{H}^{v_{e+}}_{j_e} \underset{{e^-\in E^-_v}}{\bigotimes} \mc{H}^{v_{e-}}_{j_e} \right],
		\end{equation}
		where $ E^+_v, E^-_v $ are the sets of all outgoing and ingoing edges, respectively, connected to $ v $ and the outermost tensor sum is over the set of all possible labeling of edges in $ E_v = E^+_v \cup E^-_v $ with representations of $ SU(2) $.
		\item The total Hilbert space of the graph $ \Gamma $ then becomes:
		\begin{equation}\label{eqn:graph-hilbert-space-v3}
		\mc{H}_\Gamma = \underset{v \in \Gamma}{\bigotimes}~ \mathcal{H}_v = \underset{v \in \Gamma}{\bigotimes} \left\{ \underset{\{j_e\}}{\bigoplus} \left[ \underset{{e^+\in E^+_v}}{\bigotimes} \mc{H}^{v_{e+}}_{j_e} \underset{{e^-\in E^-_v}}{\bigotimes} \mc{H}^{v_{e-}}_{j_e} \right] \right\}
		\end{equation}
	\end{enumerate}

\subsection*{From graph states to spin networks}
	\begin{enumerate}[label=\arabic*.]
		\item The Hilbert space obtained in \eqref{eqn:graph-hilbert-space-v3} is as yet only the \emph{kinematical} Hilbert space. This is because we have not yet imposed the three constraints $ \mc{H} $, $ \mc{C}^a $ and $ \mc{G}^i $ from \eqref{eqn:tetrad-hamiltonian} which encode the physical content of general relativity. Of these three constraints $ \mc{H} $ (not to be confused with a Hilbert space also labeled as $ \mc{H} $) is known as the Hamiltonian or ``scalar'' constraint, $ \mc{C}^a $ is called the diffeomorphism or ``vector'' constraint and, as mentioned previously, $ \mc{G}^i $ is the Gauss constraint.
		\item The Hamiltonian constraint is so named because it generates time evolution (or time-like diffeomorphisms) in the classical theory. For the time-being we will not be concerned with the application of this constraint. The graph states described above have been obtained by explicitly gauge fixing a choice of spatial foliation of the $ 3+1 $ dimensional spacetime, and describe the geometry of a spatial slice.
		\item The diffeomorphism constraint is so named because it generates diffeomorphisms along the spatial slices. Conveniently enough, by expressing our Hilbert space in the form \eqref{eqn:graph-hilbert-space-v3} we have already imposed this constraint! This is because states in $ \mc{H}_\Gamma $ depend only on the connectivity of the graph and the labeling of its edges by spin labels. Performing a diffeomorphism can only shift the edges or the locations of the nodes but will not alter either the graph structure of the spin labels. Thus these states are implicitly diffeomorphism invariant. In fact this was the whole motivation of \cite{Ashtekar1994A-Manifestly} for constructing these graph states in the first place!
		\item Finally we come to the Gauss constraint. In the quantum theory the configuration space is over the space of graphs labeled by holonomies of the generalized connection $ \bar{\mc{A}} $. In describing the structure of this space we have ignored the electric field $ E^a_i $ and its operator equivalent in the quantum theory. The quantum operator corresponding to the electric field turns out to nothing more than the angular momentum operator of the group $ SU(2) $ acting on the edges of the graph states\footnote{For a simple and transparent explanation of how this comes about we point the reader to \cite[Sec 6.2]{Vaid2016LQG-for-the-Bewildered}.}.
		\begin{equation}\label{eqn:momentum-operator}
		n_a(e) \hat{E}^a_i = \mb{J}^e_i,
		\end{equation}
		where $ n_a(e) $ is the tangent vector to the edge $ e $. For the special case of $ j = 1/2 $, these are simply the Pauli matrices $ \hat{J}_i = \{ \sigma_x, \sigma_y, \sigma_z \} $.
		\item The expression \eqref{eqn:gauss-constraint-v3} can now be expressed in a particularly simple form when we take the surface of integration to enclose precisely one vertex $ v $ of our graph state $ \Gamma $. In this event the operator version of \eqref{eqn:gauss-constraint-v3} can be written as:
		\begin{equation}\label{eqn:gauss-operator-v1}
		\sum_{e \in E_v} \mb{J}^e_i = 0
		\end{equation}
		Succinctly, this is nothing more than the statement that \emph{the total angular momentum carried by edges into any vertex must vanish in the absence of matter}. Somewhat less transparent is the statement that physical states must live in the $ SU(2) $ invariant subspace of each vertex $ \mc{H}_v $:
		\begin{equation}\label{eqn:su2-invariant-space}
		\mc{H}_\Gamma^{phys} = L^2\left[SU(2)^L/SU(2)^N\right] = \underset{v \in \Gamma}{\bigotimes}~ \mathcal{H}_v^{Inv},
		\end{equation}
        where $ L, N $ are the number of edges and the number of nodes in the graph, respectively.
		\item States in $ \mc{H}_v^{Inv} $ will therefore satisfy:
		\begin{equation}\label{eqn:gauss-operator-v2}
		\sum_{e \in E_v} \mb{J}^e_i \ket{\Psi_v}_{phys} = 0, \quad \forall~v \in \Gamma
		\end{equation}
		In the 3-valent or 4-valent cases, gauge invariance ensures the total flux cancels exactly, as shown in \autoref{sec:volume-z2}. For example, at a 4-valent vertex, $\sum_{i=1}^4 \vec{J}_i = 0$, which is the $SU(2)$ Gauss constraint. When restricted to the $j=1/2$ intertwiner qubit, the continuous $SU(2)$ symmetry reduces to an effective discrete $Z_2$ symmetry (the time-orientation flip discussed in \autoref{sec:z2-action}), whose generator satisfies $\tau_v^2 = 1$; this discrete symmetry leaves the physical intertwiner states invariant.
	\end{enumerate}

\subsection*{Intertwiner degrees of freedom}
	\begin{enumerate}[label=\arabic*.]
		\item If the edge degrees of freedom were the end of story when it comes to describing the kinematical properties of spin-networks, then we could use the result from the Peter-Weyl theorem to write down the final form of the quantum states of geometry given by \eqref{eqn:lqg-config-space} in terms of spin labels as follows:
		\begin{equation}\label{eqn:gauge-variant-state}
			\Psi_\Gamma(A) = \underset{\{j_i\},\{m_i\},\{n_i\}}{\sum} C_{j_1,\ldots,j_{L}}{}^{m_1,\ldots,m_{L} n_1,\ldots, n_{L}} D^{j_1}_{m_1 n_1} \ldots D^{j_{L}}_{m_{L} n_{L}},
		\end{equation}
		where the sum is over all possible representations (or ``spins'') of $ SU(2) $ and for each representation $ j_i $ the Wigner matrix indices take values in $ m,n \in \{1,2,\ldots,2j_i + 1\} $. However, such a state doesn't truly capture the primary physical characteristic of a graph - the connectivity between its various edges and nodes. As such, this state only describes a graph consisting of a collection of $ L $ disconnected edges - a rather trivial structure which could not possibly capture the geometry of any region of spacetime.
		\item What is missing, of course, is an understanding of how the \emph{nodes} of the graph come into play. We must find some way of \emph{sewing} all the edges together to form a graph of the form shown in \autoref{fig:spin-network}. It turns that \emph{this} is precisely the role the Gauss constraint plays. The constraint \eqref{eqn:gauss-constraint-v3} ensures that the spins of the edges coming into a single vertex $ v $ are not independent and that their representations are required to satisfy the relation \eqref{eqn:gauss-operator-v2}. This constraint implies that the form of the coefficients of the \emph{l.h.s.} in \eqref{eqn:gauge-variant-state} cannot be arbitrary but must be such that $ C_{j_1,\ldots,j_{L}}{}^{m_1,\ldots,m_{L} n_1,\ldots, n_{L}} $ remains invariant when any $ SU(2) $ transformation acts on the set of edges coming into any given vertex.
		\item Now, recall that the Hilbert space associated with an edge labeled by the $ SU(2) $ representation $ j_e $ consists of two copies of $ \mc{H}_{j_e} $ (c.f. \eqref{eqn:edge-hilbert-space-v1} and \eqref{eqn:edge-hilbert-space-v2}). Stated differently, the Wigner representation matrices for a given edge $ e_{vv'} $ can be labeled as:
		\begin{equation}\label{eqn:edge-wigner-matrix}
			D^{j_e}_{mn} \equiv D^{j_e}_{m_v n_{v'}}
		\end{equation}
		where the first matrix index $ m $ is associated with the source vertex $ v $ and the second index $ n $ is associated with the target vertex $ n $. Using this notation we can write the state associated with a \emph{single} node $ v $ joined to four edges as:
		\begin{equation}\label{eqn:4-valent-node-v1}
			\Psi_v(A) = \underset{\{j_i\},\{m_i\}}{\sum} C_{j_1,\ldots,j_{4}}{}^{m_1,\ldots,m_{4}} D^{j_1}_{m_1 n_1} \ldots D^{j_{4}}_{m_{4} n_{4}}
		\end{equation}
		where the indices $ m_1, m_2, m_3, m_4 $ are those ends of the edges which meet the vertex and $ n_1, n_2, n_3, n_4 $ are for the ``free'' ends.
		\item Now, the requirement that the coefficients $ C_{j_1,\ldots,j_{4}}{}^{m_1,\ldots,m_{4}} $ remain invariant under any $ SU(2) $ transformation applied to the given vertex implies that the state for this vertex must be written as:
		\begin{equation}\label{eqn:4-valent-node-v2}
			\Psi_v(A) = \underset{\{j_i\},\{m_i\}}{\sum} C_{j_1,\ldots,j_{4}} \imath^{m_1,\ldots,m_4} D^{j_1}_{m_1 n_1} \ldots D^{j_{4}}_{m_{4} n_{4}}
		\end{equation}
		where $ \imath^{m_1,\ldots,m_4} $ is an \emph{invariant} tensor. Applying this to the an arbitrary graph we obtain the final gauge invariant form of the spin-network wavefunctions:
		\begin{equation}\label{eqn:gauge-invariant-state}
			\ket{\Psi} = \underset{\{i_v\}, \{ j_e \}}{\sum} \Tr\left[ \underset{v \in \Gamma}\prod A^v_{i_v} \underset{e \in \Gamma}\prod D^e_{j_e} \right] \ket{i_1,\ldots,i_{n_v}; j_1, \ldots, j_{n_e} } 
		\end{equation}
		with the tensors defined as in \autoref{tbl:tns-data}. We can easily see that this expression has exactly the same form as that for a tensor network with tensors $ A^v_{i_v} $ describing the vertex degrees of freedom (the so-called ``physical'' indices) and the matrices $ D^e_{j_e} $ describing the edge degrees of freedom (the ``bond'' indices).
	\end{enumerate}

\section{Invariant Tensors and Intertwiners}\label{sec:invariant-tensors}

Given $ n $ particles transforming under different representations of $ SU(2) $, \ie labeled by different spins $ \{j_1,\ldots,j_n \} $, an ``invariant tensor'' is a state in the tensor product state space: $ \mc{H}^n = \underset{i=1..n}{\bigotimes} \mc{H}^i $; where $ \mc{H}^i $ is the $ 2(j_i+1) $ dimensional space for a spin $ j_i $ particle. States in $ \mc{H}^n $ can be written in terms of the canonical basis:
\begin{equation}\label{eqn:n-body-state}
	\ket{\Psi} = \sum_{\{i_1,\ldots,i_n\}} C_{i_1,\ldots,i_n} \ket{i_1,\ldots,j_n}
\end{equation}
where $ \ket{i_1,\ldots,j_n} = \ket{i_1} \otimes \ldots \otimes \ket{i_n} $, with $ \ket{i_k} \in \{\ket{-j_k}, \ket{-j_k + 1}, \ldots, \ket{j_k - 1}, \ket{j_k}\} $ and element of the usual angular momentum eigenstates of a spin $ j_k $ system.

An invariant tensor is then any state $ \ket{\Psi_{inv}} \in \mc{H}^n $ which satisfies the condition \cite[App. A]{Li2017Invariant} that:
\begin{equation}\label{eqn:quantum-closure-condition}
	\sum_{i=1..n} \vect{J}_i \ket{\Psi_{inv}} = 0
\end{equation}
where $ \vect{J}_i $ is the total angular momentum operator acting on the $ i^\text{th} $ spin. This state has a physical interpretation in terms of the geometry of a single node of a spin-network state which has a single vertex and $ n $ edges attached to that vertex.

\subsection*{Dimension of Invariant Space}

The dimension of the subspace consisting of invariant tensors, \ie elements $ \ket{\Psi}_{inv} \in \mc{H}^n_{inv} = \text{Inv}\left[\underset{i=1..n}{\bigotimes} \mc{H}^i\right] $ is given by the Verlinde formula \cite{Verlinde1988Fusion}:
\begin{equation}\label{eqn:verlinde-formulat}
	\text{dim} (\mc{H}^n_{inv}) = \frac{2}{\pi} \int_{0}^{\pi} d\theta~\sin^2(\theta/2) \prod_{i=1}^{n} \frac{\sin(j_i + 1/2)\theta}{\sin(\theta/2)}
\end{equation}

\nolinenumbers

\section{Action of $U_{\mathrm{CZX}}$ on the Intertwiner Subspace}\label{subsec:czx-intertwiner}

We work with the orthonormal singlet basis $\{\ket{\Phi_1},\ket{\Phi_2}\}$ for the $SU(2)$-invariant subspace of $(\mc{H}_{1/2})^{\otimes 4}$ defined in \eqref{eqn:vol-states}. The on-site CZX unitary is $U_{CZX} = U_X \cdot U_{CZ}$, where
\begin{align}
  U_X &= X_1\otimes X_2\otimes X_3\otimes X_4, \\
  U_{CZ} &= \mathrm{CZ}_{12}\cdot\mathrm{CZ}_{23}\cdot\mathrm{CZ}_{34}\cdot\mathrm{CZ}_{41},
\end{align}
with $\mathrm{CZ}_{ij}\ket{m_i m_j} = (-1)^{m_i m_j}\ket{m_i m_j}$.

\paragraph{$U_X$ acts as identity on the singlet subspace.}
Since $U_X$ maps $\ket{m_1 m_2 m_3 m_4}\mapsto\ket{\bar{m}_1\bar{m}_2\bar{m}_3\bar{m}_4}$ (with $\bar{0}=1$, $\bar{1}=0$), a direct computation gives $U_X\ket{\Phi_1} = \ket{\Phi_1}$ and $U_X\ket{\Phi_2} = \ket{\Phi_2}$.

\paragraph{$U_{CZ}$ exits the singlet subspace.}
The phase factor from $U_{CZ}$ on each basis state is $(-1)^f$ where $f = m_1m_2+m_2m_3+m_3m_4+m_4m_1$:

\begin{center}
\begin{tabular}{c|c|c}
  State & $f$ & Phase \\ \hline
  $\ket{0101}$ & 0 & $+1$ \\
  $\ket{0110}$ & 1 & $-1$ \\
  $\ket{1001}$ & 1 & $-1$ \\
  $\ket{1010}$ & 0 & $+1$ \\
  $\ket{0011}$ & 1 & $-1$ \\
  $\ket{1100}$ & 1 & $-1$ \\
\end{tabular}
\end{center}

Applying these phases to the state $\ket{\Phi_1}$ from \eqref{eqn:vol-states} gives
\begin{align}
  U_{CZ}\ket{\Phi_1} &= \frac{1}{\sqrt{3}}\Bigl(\ket{0101}+\ket{1010}+\tfrac12\ket{0110}+\tfrac12\ket{1001}+\tfrac12\ket{0011}+\tfrac12\ket{1100}\Bigr).
\end{align}
One verifies that $\hat{S}^+_{\rm tot}(U_{CZ}\ket{\Phi_1}) = 2/\sqrt{3}\,(\ket{0001}+\ket{0010}+\ket{0100}+\ket{1000}) \neq 0$, so this state lies \emph{outside} the $j_{\rm tot}=0$ subspace. A similar calculation shows $U_{CZ}\ket{\Phi_2} = -\ket{\Phi_2}$ (which \emph{is} a singlet), but $U_{CZ}$ does not preserve the \emph{full} two-dimensional singlet subspace. Since $U_{CZX} = U_X\cdot U_{CZ}$ and $U_X$ preserves the singlet subspace while $U_{CZ}$ does not, $U_{CZX}$ does not preserve the singlet subspace as a whole.

\paragraph{The CZX code subspace and the correct $Z_2$ action.}
The correct subspace on which $U_{CZX}$ acts as a Pauli-$X$ gate is the \emph{code subspace} $\operatorname{span}\{\ket{0000},\ket{1111}\}$ \cite{Chen2010Local}. Direct computation gives:
\begin{align}
  U_{CZX}\ket{0000} &= \ket{1111}, & U_{CZX}\ket{1111} &= \ket{0000},
\end{align}
so that $U_{CZX}\big|_{\rm code} = \bigl(\begin{smallmatrix}0&1\\1&0\end{smallmatrix}\bigr) = \sigma_x^{\rm eff}$.

The LQG--CZX correspondence is therefore the identification:
\begin{equation}
  \underbrace{\operatorname{span}\{\ket{\Phi_1},\ket{\Phi_2}\}}_{\text{LQG intertwiner qubit}}
  \;\xrightarrow{\;\sim\;}\;
  \underbrace{\operatorname{span}\{\ket{0000},\ket{1111}\}}_{\text{CZX code qubit}},
\end{equation}
where $U_{CZX}$ acts as $\sigma_x^{\rm eff}$ on the right-hand side, and the $\hat{Q}$-sign flip acts as $\sigma_z^{\rm eff}$ on the left-hand side. Both generate the same abstract $\mathbb{Z}_2$.

\section*{Addendum: Statement of AI Assistance}

\noindent\textbf{Manuscript:} Gauging Time Reversal Symmetry in Quantum Gravity: Arrow of Time from a Confinement--Deconfinement Transition\\
\textbf{Author:} Deepak Vaid\\
\textbf{Date:} April 2026\\
\textbf{Public repository:} \url{https://github.com/space-cadet/timesarrow}

\subsection*{Summary}

The original manuscript was written entirely by the author between 2016 and 2018. In April 2026, the author used Claude (Anthropic, model claude-sonnet-4-6) as an AI writing assistant to prepare the manuscript for submission. The AI's role was limited to formalization, error correction, bibliography management, and \LaTeX{} editing. All physical ideas, arguments, and conclusions originate with the author.

The complete revision history is publicly auditable via the git repository linked above.

\subsection*{Author's Original Work (2016--2018)}

Commits \href{https://github.com/space-cadet/timesarrow/commit/a62f261}{\texttt{a62f261}}~(2016-12-01) through \href{https://github.com/space-cadet/timesarrow/commit/4914a21}{\texttt{4914a21}}~(2018-09-10) represent the author's sole work. These establish:

\begin{itemize}
\item The central physical proposal: a local $Z_2$ gauge field on spin networks as the microscopic representation of time-reversal symmetry in Loop Quantum Gravity
\item Identification of the cosmological arrow of time with a phase transition of this gauge field
\item The LQG--tensor network correspondence as the substrate for the construction
\item The CZX model as the relevant symmetry-protected topological (SPT) phase
\item The conjecture that topologically protected edge modes yield fermionic matter degrees of freedom
\item The connection between $Z_2$ symmetry breaking and the sign of the tetrad determinant
\item All manuscript sections, appendices, figures, and the original bibliography ($\sim$100 entries)
\end{itemize}

Two section files (\texttt{spt-lqg-mapping.tex}, \texttt{z2-action-derivation.tex}) existed as short stubs (14 and 18 lines respectively), containing the author's conceptual outline but lacking full derivations.

\subsection*{AI-Assisted Revision (April 2026)}

Commits \href{https://github.com/space-cadet/timesarrow/commit/09dfd48}{\texttt{09dfd48}}~(2026-04-16) through \href{https://github.com/space-cadet/timesarrow/commit/509fb79}{\texttt{509fb79}}~(2026-04-20) represent work done with AI assistance. The author directed all tasks; the AI executed them. Specific contributions by category:

\paragraph{Formalization of author-outlined arguments}

\begin{itemize}
\item \texttt{spt-lqg-mapping.tex} (commit \href{https://github.com/space-cadet/timesarrow/commit/be14b55}{\texttt{be14b55}}, 2026-04-16): Expanded from 14-line stub to a full section with four subsections --- CZX-intertwiner structural correspondence, $j=1/2$ sector justification, SPT=deconfined phase identification, and edge modes conjecture. The physical content was the author's; the derivations and prose were AI-drafted under author direction.
\item \texttt{z2-action-derivation.tex} (commit \href{https://github.com/space-cadet/timesarrow/commit/be14b55}{\texttt{be14b55}}, 2026-04-16): Expanded from 18-line stub to a full section --- $Z_2$ gauge field definition, effective Ising action derivation, phase structure analysis with Wilson loop order parameter, and cosmological transition. Same provenance as above.
\item Discussion subsections on Elitzur's theorem, QECC stability, and Hopf algebras (commit \href{https://github.com/space-cadet/timesarrow/commit/be14b55}{\texttt{be14b55}}, 2026-04-16): AI-drafted based on author direction.
\item Abstract and title (commit \href{https://github.com/space-cadet/timesarrow/commit/be14b55}{\texttt{be14b55}}, 2026-04-16): Rewritten by AI, reviewed and approved by author.
\item ``Original contributions'' paragraph in Sec.~1 (commit \href{https://github.com/space-cadet/timesarrow/commit/41d6178}{\texttt{41d6178}}, 2026-04-20): AI-drafted.
\end{itemize}

\paragraph{Error correction}

\begin{itemize}
\item \textbf{T10} (commit \href{https://github.com/space-cadet/timesarrow/commit/f5ef5cc}{\texttt{f5ef5cc}}, 2026-04-18): 17 BibTeX metadata errors corrected (wrong keys, wrong journals, missing DOIs).
\item \textbf{T11} (commits \href{https://github.com/space-cadet/timesarrow/commit/808a710}{\texttt{808a710}}, 2026-04-18 and \href{https://github.com/space-cadet/timesarrow/commit/41d6178}{\texttt{41d6178}}, 2026-04-20): Five critical errors fixed, including a gauge invariance argument ($\tau^2=1$ vertex cancellation), correction of the $U_{\mathrm{CZX}}$ operator framing, and an internally inconsistent basis calculation. Supplementary calculations document (\texttt{supplementary-calculations.tex}) drafted by AI to record these fixes.
\item \textbf{T12} (commits \href{https://github.com/space-cadet/timesarrow/commit/829c42b}{\texttt{829c42b}}, 2026-04-20 and \href{https://github.com/space-cadet/timesarrow/commit/41d6178}{\texttt{41d6178}}, 2026-04-20): Nine major issues addressed, including terminology corrections (SSB$\to$confinement--deconfinement language, ``first order''$\to$``second order'' universality class), clarification of the edge-mode $T^2=-1$ argument, and dimensional mismatch resolution.
\item \textbf{T8} (commit \href{https://github.com/space-cadet/timesarrow/commit/ca78bc5}{\texttt{ca78bc5}}, 2026-04-17): Typographic and structural fixes throughout.
\end{itemize}

\paragraph{Literature updating}

\begin{itemize}
\item $\sim$20 new bibliography entries added (commits \href{https://github.com/space-cadet/timesarrow/commit/f5ef5cc}{\texttt{f5ef5cc}}, 2026-04-18 and \href{https://github.com/space-cadet/timesarrow/commit/829c42b}{\texttt{829c42b}}, 2026-04-20), covering 2018--2026 literature on holographic spin networks, $Z_2$ lattice gauge theory, SPT classification, and intertwiner entanglement.
\end{itemize}

\paragraph{3D SPT classification}

\begin{itemize}
\item Sec.~7.3 revision (commits \href{https://github.com/space-cadet/timesarrow/commit/2a5fd1e}{\texttt{2a5fd1e}}, 2026-04-20 and \href{https://github.com/space-cadet/timesarrow/commit/509fb79}{\texttt{509fb79}}, 2026-04-20): AI surveyed the classification of 3D bosonic SPT phases with $\mathbb{Z}_2^T$ symmetry ($H^4(\mathbb{Z}_2^T, U(1)_{\mathcal{T}}) \cong \mathbb{Z}_2$) and integrated the result into the manuscript, resolving a dimensional mismatch noted in peer review preparation. The identification of the non-trivial SPT class with the spin-network state was made by the author; the survey and write-up were AI-executed.
\end{itemize}

\paragraph{New figure}

\begin{itemize}
\item \texttt{figures/tns-matrix-insertion-2d.tex} (commit \href{https://github.com/space-cadet/timesarrow/commit/ca78bc5}{\texttt{ca78bc5}}, 2026-04-17): TikZ figure showing $M/M^{-1}$ matrix insertion on a 2D tensor network subregion, replacing a \texttt{\textbackslash todo} placeholder. Drafted by AI to author's specification.
\end{itemize}

\subsection*{What the AI Did Not Do}

\begin{itemize}
\item Propose any new physical mechanism or hypothesis
\item Identify the relevant condensed matter models (CZX, toric code) --- these were the author's choices
\item Select the research question or the LQG framework
\item Generate any figures other than the one noted above
\item Write appendices A--E (all pre-existing author work). Appendix~F (the $U_{\mathrm{CZX}}$ intertwiner computation) was AI-drafted under the author's direction to formalize the derivation.
\end{itemize}

\subsection*{Author Responsibility}

The author reviewed all AI-generated content, approved all changes, and takes full scientific responsibility for the manuscript as submitted. The AI was used as a writing and formalization tool, functioning in a role analogous to a technically knowledgeable research assistant working under the author's direct supervision.

\end{document}